\documentclass[a4paper,11pt]{article}
\usepackage{jheppub} 
\usepackage{lineno}

\usepackage{newtxtext}
\usepackage{newtxmath}
\usepackage{mathrsfs}

\usepackage{graphicx}
\usepackage{subfigure}
\usepackage{amsmath, bm}
\usepackage{float}
\usepackage{multirow}
\usepackage{slashed}
\usepackage{xcolor}
\usepackage{mathtools,braket}
\usepackage{color,physics,tensor}

\usepackage{natbib} 
\setcitestyle{numbers}
\newcommand{\im}{{\rm Im}}

\newcommand{\be}{\begin{eqnarray}}
\newcommand{\ee}{\end{eqnarray}}
\newcommand{\ba}{\begin{array}}
\newcommand{\ea}{\end{array}}
\newcommand{\bea}{\begin{eqnarray}}
\newcommand{\eea
}{\end{eqnarray}}
\newcommand{\bi}{\begin{itemize}}
\newcommand{\ei}{\end{itemize}}
\newcommand{\nn}{{\nonumber}}

\title{\boldmath Non-diagonal DVCS and transition GPDs: \\ A unified framework for spinless hadron case $\gamma^*\pi\to\gamma\pi\pi$}

\author[a,1]{Sangyeong Son \note{Corresponding author.}}
\author[a,b,c]{and Kirill M. Semenov-Tian-Shansky}
\affiliation[a]{Department of Physics, Kyungpook National University, Daegu 41566, Republic of Korea}
\affiliation[b]{NRC ``Kurchatov Institute'' - PNPI, Gatchina 188300, Russia}
\affiliation[c]{Higher School of Economics, National Research University, St. Petersburg  194100, Russia}

\emailAdd{thstkd3754@gmail.com}
\emailAdd{ksemenov@knu.ac.kr}

\abstract{
Hadron-to-two-hadron transition generalized parton distributions (GPDs) extend the concept 
of hadron-to-resonance transition GPDs and provide a unified description of non-diagonal hard exclusive reactions in the generalized Bjorken limit. We present the formalism for the case of spinless hadrons addressing the non-diagonal deeply virtual Compton scattering 
$\gamma^*\pi\to\gamma\pi\pi$ 
in terms of 
$\pi\to\pi\pi$ 
transition GPDs, which generalize GPDs for   
$\pi \to f_0, \, \rho, \, f_2, \, \cdots$ 
transitions. We work out the basic properties of 
$\pi\to\pi\pi$ 
transition GPDs and establish the soft pion theorems 
at the $2\pi$ production threshold.  We construct the partial wave expansion of  
$\pi\to\pi\pi$ 
transition GPDs in the two-pion decay angles and employ the dispersive approach to constrain $\pi\to\pi\pi$ 
transition GPDs in terms of 
$\pi\pi$-scattering phases with help of the Omn\`{e}s representation. We estimate the 
$e^-\pi^+ \to e^- \gamma \pi^+ \pi^0$ 
cross section in the kinematics of the JLab@12~GeV incorporating the isolated 
$\rho(770)$ 
resonance state and work out the angular distributions of the cross section, specifying 
the observables sensitive to the polarization states of the produced 
$\rho(770)$ 
resonance. 
}

\begin{document}
\maketitle
\flushbottom

\section{Introduction}
\label{sec:Intro}
Hard exclusive inelastic reactions, such as deeply virtual Compton scattering (DVCS) and deeply virtual meson production (DVMP), 
are pivotal for exploring the complex internal structure of hadrons in terms of the fundamental degrees of freedom 
of Quantum Chromodynamics (QCD).
With the help of QCD collinear factorization theorems \cite{Collins:1996fb, Collins:1998be},
these processes can be factorized, separating the hard and soft regimes of interaction. The non-perturbative
dynamics of hadron constituents  is 
encoded in terms of generalized parton distributions (GPDs)~\cite{Muller:1994ses, Radyushkin:1997ki, Ji:1996ek, Ji:1996nm, Ji:1998pc}.
Over the past few decades, significant developments in the GPD framework have been achieved; see refs.~\cite{Goeke:2001tz, Diehl:2003ny, Belitsky:2005qn, Diehl:2023nmm} for a review.

GPDs, defined through hadronic matrix elements  of  bilocal QCD string  light-cone  operators, are convenient tools for investigating the internal structure of QCD composite systems. The Mellin moments of GPDs provide access to generalized form factors (FFs) that parametrize the hadronic matrix elements of local QCD operator. 
In particular, the second Mellin moment of GPD yields the so-called gravitational form factors (GFFs), 
which are defined by the matrix elements of the QCD energy-momentum tensor~\cite{Kobzarev:1962wt, Pagels:1966zza}, thereby enabling the exploration of the mechanical properties of hadrons.
A prominent example of such studies
is Ji's spin sum rule \cite{Ji:1996ek},  which  quantifies the total angular momentum carried 
by quarks and gluons within a hadron.
In addition to angular momentum, the key mechanical characteristics of hadrons include mechanical radii, pressure distribution, and internal stresses acting on quarks,  all of which  are closely related to the so-called $D$-term FF~\cite{Polyakov:2002yz, Polyakov:2018zvc, Lorce:2018egm}.
The empirical extraction of the $D$-term FF using DVCS data from the CLAS collaboration has  recently been performed to determine the pressure and shear force distributions, as well as the mechanical radius 
of the proton~\cite{Burkert:2018bqq, Burkert:2021ith, Burkert:2023atx}.
Moreover,  
the Fourier  transform  of GPDs into the impact parameter space provides the transverse spatial distribution of partons with a given longitudinal momentum fraction, revealing the ($2+1$)-dimensional tomographic images of the hadron's interior~\cite{Burkardt:2000za, Ralston:2001xs, Burkardt:2002ks}.

Non-diagonal DVCS and DVMP reactions involving nucleon-to-resonance transitions,
such as $N\to \Delta, N^*$~\cite{Frankfurt:1998jq, Frankfurt:1999xe},
extend the study of  the  internal QCD dynamics of hadrons from the ground state nucleon to resonance excitations, 
giving rise to the concept of transition GPDs.
Analogous to the diagonal case, the mechanical properties of resonances can be explored 
through the non-diagonal hadronic matrix elements of the QCD energy-momentum tensor, parametrized by transition GFFs \cite{Ozdem:2019pkg,Polyakov:2020rzq,Azizi:2020jog,Ozdem:2022zig,Kim:2022bwn, Kim:2023xvw}. 
Transition GPDs may also provide access to  the spatial distribution of active partons in course of the resonance formation. 
The recent white paper~\cite{Diehl:2024bmd} presents a comprehensive overview of current research on non-diagonal hard exclusive reactions and the investigation of baryon resonance structure within the framework of transition GPDs.

In recent years, significant advancements have been reported in both the theoretical and experimental understanding of non-diagonal hard exclusive reactions, which can be described using transition GPDs.
The CLAS collaboration has performed the first measurement of the beam-spin asymmetry (BSA) for the hard exclusive reaction of $\gamma^* p \to \pi^-\Delta^{++}\to \pi^- p\pi^{+}$, which is expected to provide access to the $p\to\Delta^{++}$ transition GPDs~\cite{CLAS:2023akb}. On the theoretical side, 
the DVMP process $\gamma^* p\to\pi^-\Delta^{++}$ has been analyzed in terms of the $p\to\Delta^{++}$ transition GPDs ~\cite{Kroll:2022roq}.
The cross section was estimated accounting for contributions of
both the helicity nonflip and transversity $p\to\Delta^{++}$ transition GPDs. 
The non-diagonal DVCS process $e^- N\to e^- \gamma \pi N$ has also been studied within the framework of $N\to R$ transition  GPDs,  
accounting for the  lowest-lying $\pi N$ resonances,  $R = \Delta(1232)$, $P_{11}(1440)$, $D_{13}(1520)$, and $S_{11}(1535)$ \cite{Semenov-Tian-Shansky:2023bsy}.
In particular, observables such as the BSA and differential cross sections for $e^- N\to e^- \gamma \pi N'$ were estimated in the first and second $\pi N$ resonance regions, 
specifying their distributions in the invariant mass and the $\pi N$ decay angles in the resonance rest frame. 
Moreover, in the strange sector, the DVMP of kaons was investigated using the 
$p\to\Lambda$ 
GPDs~\cite{Kroll:2019wug}.

An important next step in the development of the formalism of transition GPDs involves exploring GPDs for transitions from a single hadron into a two-hadron system. In particular, the 
$N \to \pi N$ 
transition GPDs introduced in refs.~\cite{Polyakov:1998sz,Polyakov:2006dd} have emerged as  intriguing theoretical objects. Near the pion production threshold, they can be constrained in a model-independent manner through the chiral dynamics using soft-pion theorems \cite{Pobylitsa:2001cz,Chen:2003jm,Guichon:2003ah,Kivel:2004bb}. 
In the 
$\pi N$ 
resonance region, 
$N \to \pi N$  
transition GPDs provide a unified description of 
$N \to R$ 
transitions.

However, to date, the formalism of hadron-to-two-hadron transition GPDs has only been developed in a preliminary and sketchy manner.
Due to the many-body nature of the corresponding hard exclusive reactions, these transition GPDs depend on additional variables describing the final state of the two-hadron system. The Lorentz structure also becomes more complex than in the diagonal case. 
Therefore, before addressing the more  intricate scenario involving 
spin-$\frac{1}{2}$
hadrons, it is more practical to first examine the spinless case, with pions serving as the most natural candidates.

In this paper we consider the non-diagonal 
$\gamma^*\pi \to \gamma\pi\pi$ 
DVCS  using 
$\pi\to\pi\pi$ 
transition GPDs. We  explore the basic properties of 
$\pi \to \pi \pi$ 
transition GPDs, with particular emphasis on the choice of 
kinematic variables that describe the produced  
$2\pi$ 
system.
Additionally, we derive constraints for 
$\pi\to\pi\pi$ 
GPDs based on soft pion theorems. We also construct a framework to utilize 
$\pi \to \pi \pi$ 
transition GPDs across the entire $2\pi$-resonance region. In the vicinity of the 
$\rho(770)$ resonance, we analyze the  angular distribution of the pion decay for the 
$e^-\pi^+\to e^-\gamma\pi^+\pi^0$ 
cross section. We then generalize our findings and construct the partial-wave (PW) expansions of  
$\pi\to\pi\pi$ 
GPDs in terms of the two-pion decay angles. To develop a realistic dependence of
$\pi\to\pi\pi$ 
transition GPDs on the invariant mass of the $2\pi$ system, we follow the strategy formulated in refs.~\cite{Polyakov:1998ze, Lehmann-Dronke:1999vvq, Lehmann-Dronke:2000hlo} 
for the $2\pi$ generalized distribution amplitudes (GDAs) 
and make use of the Omnès-type $N$-subtracted dispersion 
relation to constrain 
$\pi \to \pi \pi$  
transition GPDs in terms of $\pi\pi$-scattering phases.
Furthermore, we apply the Froissart-Gribov (FG) projection technique to the 
$\pi\to\pi\pi$ 
Compton FFs to  characterize excitation of a $2\pi$-resonance through the cross channel spin-$J$ probe induced by the non-local QCD string light-cone operator.

This paper is organized as follows. In section~\ref{sec:2}, we work out the kinematics of the $e^-\pi\to e^-\gamma\pi\pi$ reaction, 
define the $\pi\to\pi\pi$ transition GPDs and discuss their basic properties. 
In section~\ref{sec:3}, we focus on the reaction $e^-\pi\to e^-\gamma R \to e^-\gamma \pi \pi$ in the vicinity of $\rho(770)$ resonance using $\pi\to\rho$ transition GPDs. 
In section~\ref{Sec_Resonance_region}, we construct the double PW expansion of $\pi\to\pi\pi$ GPDs 
in terms of the two-pion decay angles and present the dispersive technique that provides the dependence of 
$\pi\to\pi\pi$ GPDs on the invariant mass of the produced 
$2\pi$ system through the Omnès representation.
We then discuss the application of the FG projection to the $\pi\to\pi\pi$ Compton FFs. Our phenomenological models for $\pi\to\pi\pi$  transition GPDs and numerical results for the typical kinematics of the JLab@12~GeV are presented in section~\ref{sec:results}, and, finally, section~\ref{sec:summary}  provides our summary and outlook.

\section{\texorpdfstring{\bm{$e^-\pi \to e^-\gamma\pi \pi $} within the framework of GPDs}{e^-\pi^+\to e^-\gamma\pi^+\pi^0 in description of generalized parton distributions}}
\label{sec:2}
In this study, as a toy model of a non-diagonal hard exclusive reaction involving only spinless hadrons, 
we address the non-diagonal DVCS off a pion with a transition from 
the target pion into a two-pion system:
\begin{eqnarray}
    \gamma^*(q) + \pi(k) \to \gamma(q') + \pi(k_1) + \pi(k_2). 
    \label{DVCS_pi_to_2pi}
\end{eqnarray}
We analyze this process by adapting the collinear factorization mechanism in the generalized Bjorken kinematics, in which 
\begin{eqnarray}
    s = (k+q)^2, \; s_1 = (q'+k_1)^2, \;  Q^2 \to \infty; \quad x_B = \frac{Q^2}{2 k\cdot q} \; \; \text{and} \; \; \frac{s_1}{s} \, - \, \text{fixed;} \nonumber \\
    t \equiv \Delta^2 = (q-q')^2, \; t' = (k_2 - k)^2,\;~\text{and}~\;W_{\pi\pi}^2 = (k_1+k_2)^2~\text{of hadronic mass scale}.
    \label{Bjorken_limit}
\end{eqnarray}
Here, $Q^2 \equiv -q^2 > 0$ stands for the photon virtuality, $x_B$ is the Bjorken variable,
and the invariant $s$ is expressed as
\begin{eqnarray}
    s = m_\pi^2 + Q^2\left(\frac{1-x_B}{x_B}\right),
\end{eqnarray}
with $m_\pi$ denoting the pion mass.
The status of collinear factorization theorems for DVCS and DVMP processes, like (\ref{DVCS_pi_to_2pi}), with a transition from a target hadron to a low invariant mass two-hadron system, is essentially the same as for the usual DVCS \cite{Collins:1998be} and DVMP \cite{Collins:1996fb}, as argued e.g. in refs.~\cite{Goeke:2001tz, Diehl:2003ny, Polyakov:2006dd}.

The DVCS process~(\ref{DVCS_pi_to_2pi}), accompanied by the background
non-diagonal Bethe-Heitler (BH) process, can be investigated through the photon electroproduction reaction involving the $\pi\to\pi\pi$ transition on the hadron side:
\begin{eqnarray}
    e^-(l) + \pi(k) \to e^-(l') + \gamma(q') + \pi(k_1) + \pi(k_2).
    \label{Electroproduction_of_2pi}
\end{eqnarray}
The lepton side of the reaction~(\ref{Electroproduction_of_2pi}) is completely analogous to that of the usual diagonal exclusive
electroproduction reactions. 

\begin{figure}[H]
 \begin{center}
\epsfig{figure= 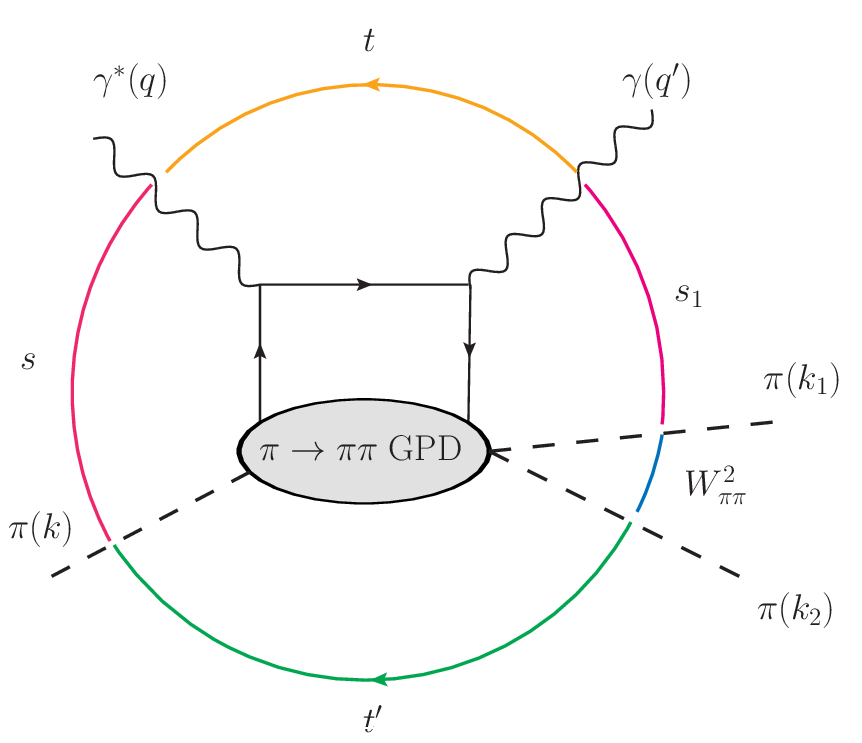, height=6cm}
 \end{center}
 \caption{Handbag diagram for non-diagonal $\pi \to \pi \pi$ DVCS (\ref{DVCS_pi_to_2pi}). Diagram with crossed real and virtual photons is not shown. Colored arcs
 show a choice of invariant kinematical variables (\ref{Bjorken_limit}) describing a $2 \to 3$ scattering process.}
\label{Fig_Diagram_ND_DVCS_pions}
\end{figure}

\begin{figure}[h]
    \centering
    \includegraphics[width=0.7\linewidth]{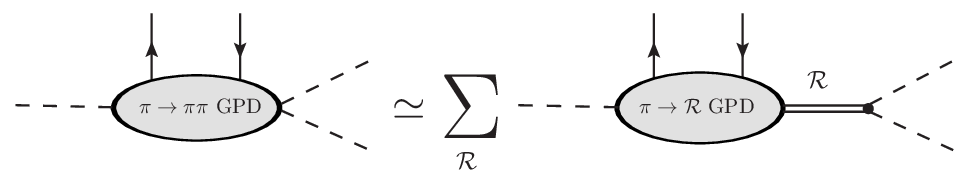}
    \caption{General $\pi\to\pi\pi$ transition GPD as a tool for a unified description of the $2 \pi$ resonance spectrum for $W_{\pi \pi} \sim M_{\cal R}$.}
    \label{fig:transitionGPD}
\end{figure}

Figure~\ref{Fig_Diagram_ND_DVCS_pions} depicts the handbag diagram for the DVCS process~(\ref{DVCS_pi_to_2pi}) in the framework of the collinear factorization, providing a description
of the reaction in terms of the $\pi\to\pi\pi$ transition GPDs.
Non-diagonal $\pi\to\pi\pi$ transition GPDs are introduced
as matrix elements of the  QCD string quark  operator between the initial $\pi$ and the final $2\pi$ states:
\begin{eqnarray}
    &&\big\langle \pi(k_1) \pi(k_2) \big| \bar{\psi}(0)[0,z]\psi(z) \big| \pi(k) \big\rangle,
\end{eqnarray}
defined on the light-cone 
$z^2 = 0$, 
where the Wilson line 
$[0, z]$ 
ensures the gauge invariance of the quark operator. 
Within kinematics corresponding to the generalized Bjorken limit (\ref{Bjorken_limit}), with invariant mass of the final state $\pi \pi$ system, $W_{\pi \pi}$, being of order of $2\pi$ resonance mass, $M_{\cal R}$, the 
$\pi\to\pi\pi$ 
GPDs generalize the concept of pion-to-resonance transition GPDs, 
see figure~\ref{fig:transitionGPD}. 
We argue that these objects provide a unified description of the reaction 
(\ref{DVCS_pi_to_2pi}) 
across the entire $2\pi$-resonance region, representing a step forward in describing non-diagonal hard exclusive reactions compared to simpler frameworks that account only for transition from the target pion to isolated resonances.

The reaction~(\ref{DVCS_pi_to_2pi}) 
is, in principle, experimentally accessible through 
the Sullivan-type process~\cite{Sullivan:1971kd}:
\begin{eqnarray}
e^- + p \to e^- + n + \pi^+ \to e^- + n + \gamma + \pi^+ + \pi^0.
\label{Sullivan}
\end{eqnarray}
Within the kinematic region where the absolute value of the momentum transfer between the two nucleons is  sufficiently  small, the
reaction 
(\ref{Sullivan}) 
is argued to proceed  through a one-pion exchange,  and the amplitude can be factorized into two subprocesses: 
$p\to n\pi^+$ 
and 
$e^-\pi^+\to e^-\gamma\pi^+\pi^0$~\cite{Qin:2017lcd, Amrath:2008vx, Chavez:2022tkf},
with the target 
$\pi^+$ 
being  nearly  on-shell. The validity of  these approximations is discussed  in detail in 
refs.~\cite{Qin:2017lcd, Amrath:2008vx}.
Therefore, in the following discussion, the target pion in 
(\ref{Electroproduction_of_2pi}) 
is assumed to be on the mass shell.

\subsection{Kinematics}
\label{subsec:2_1}

In this subsection, we introduce two complementary sets of kinematic variables for the description of the DVCS subprocess $\gamma^*\pi\to\gamma\pi\pi$.
In general, a description of a $2\to 3$ process requires five kinematic variables. A convenient choice, as shown in figure~\ref{Fig_Diagram_ND_DVCS_pions},
includes the following variables:
\begin{eqnarray}
    && 
        s=(k+q)^2; \quad s_1=(q'+k_1)^2; \quad s_2 = (k_1+k_2)^2 \equiv W_{\pi\pi}^2; \quad t_1 = t \equiv \Delta^2 ; \quad t_2 = (k_2-k)^2 \equiv t'. \nn \\ &&
     \label{Set_invatiants}
\end{eqnarray}
Here, 
$W_{\pi\pi}^2=(k_1+k_2)^2 \equiv p_{\pi\pi}^2$ 
is the invariant mass of the two pions in the final state 
with 
$p_{\pi\pi} = k_1 + k_2$, 
and 
$\Delta=q-q'$ 
is the momentum transfer between the initial and final hadronic states. 

On the other hand, describing the process 
$\gamma^*\pi\to\gamma\pi\pi$ 
in terms of the 
$2\pi$ 
decay angles allows for a connection with the conventional tools of the PW analysis used in hadron resonance spectroscopy.
There are several ways to introduce a set of angles for describing a $2 \to 3$ reaction, see, e.g., section~V of ref.~\cite{Byckling1973}. 
We choose to use the polar and azimuthal angles of the $2\pi$ decay in the so-called helicity frame, with the polar axis aligned along $\vec{q}'$ 
in the center-of-mass (c.m.) system of 
$2\pi$ 
\cite{Byckling1973}:
\begin{eqnarray}
    \cos\theta_\pi^* = \frac{\Vec{q}'\cdot \Vec{k}_2}{|\Vec{q}'| |\Vec{k}_2|}\biggl|_{\Vec{k}_1 = -\Vec{k}_2}; \qquad 
    \cos\varphi_\pi^* = \frac{(\Vec{q}'\times\Vec{k})\cdot(\Vec{q}'\times\Vec{k}_2)}{|\Vec{q}'\times\Vec{k}| |\Vec{q}'\times\Vec{k}_2|}\biggl|_{\Vec{k}_1 = -\Vec{k}_2}. \label{pi-angles}
\end{eqnarray}
The $2\pi$ decay angles 
$\Omega_\pi^* = (\theta_\pi^*, \varphi_\pi^*)$  
provide an alternative  to the invariants $s_1$ and $t'$ and are incorporated into the set of kinematic variables:
\begin{eqnarray}
    s; \quad t; \quad W_{\pi\pi}^2; \quad \Omega_\pi^* = (\theta_\pi^*, \varphi_\pi^*).
    \label{Set_angles}
\end{eqnarray}

\begin{figure*}[t]
\centering
\includegraphics[page=1,width=0.7\textwidth]{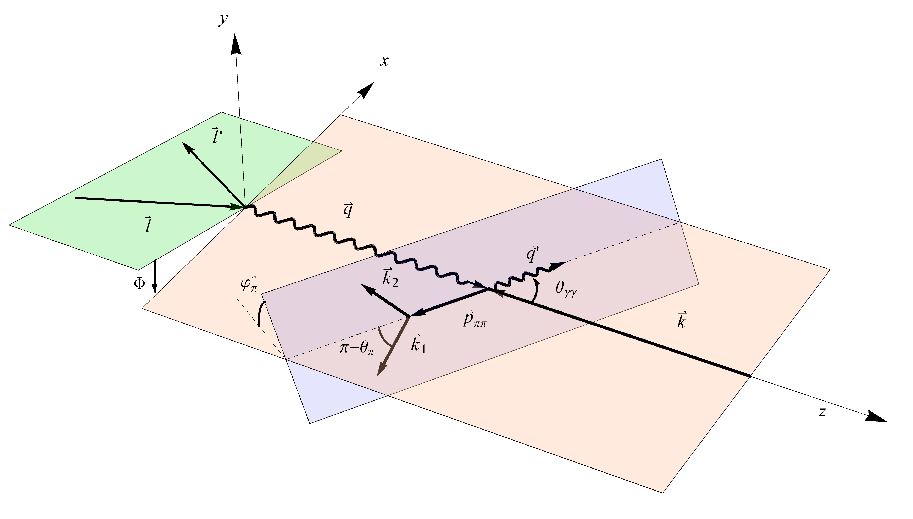}
\caption{Kinematic planes of the $e^-\pi\to e^-\gamma R \to e^-\gamma \pi\pi$ reaction (\ref{Electroproduction_of_2pi}) in the $\gamma^*(q) \pi(k)$ c.m. frame. 
$z$-axis is chosen along $\vec{q}$; the $xz$ plane corresponds to the production plane, where the hard subprocess $\gamma^*(q) \pi(k) \to \gamma(q') R(p_{\pi \pi})$ occurs. $\theta_{\gamma \gamma}$ is the scattering angle of the $\gamma^* \pi \to R \gamma$ subprocess in the  $\gamma^*(q) \pi(k)$ c.m. frame.  The leptonic and decay planes form angles $\Phi$ and $\varphi_\pi^*$ with the production plane. The angle between the production and decay planes, $\varphi_\pi^*$, is not affected by the boost along $\vec{q}'$ and hence is the same as in the $\pi(k_1) \pi(k_2)$ c.m. frame.  }
\label{fig:kin_plane}
\end{figure*}

Figure~\ref{fig:kin_plane} presents the kinematics of the reaction~(\ref{Electroproduction_of_2pi}) 
in the 
$\gamma^*\pi$ 
c.m. frame. 
$\theta_{\gamma \gamma}$ 
is the scattering angle of the 
$\gamma^* \pi \to R \gamma$ 
subprocess in the  
$\gamma^*(q) \pi(k)$ 
c.m. frame.
The leptonic and decay planes form angles 
$\Phi$ and $\varphi_\pi^*$ 
with the production plane. Note that the angle between the production and decay planes, 
$\varphi_\pi^*$, 
is not affected by the boost along 
$\vec{q}'$ 
and hence is the same as in the 
$\pi(k_1) \pi(k_2)$ 
c.m. frame.
The polar decay angle 
$\theta_\pi$ 
shown in figure~\ref{fig:kin_plane} 
differs from the polar decay angle 
$\theta_\pi^*$ 
defined in the 
$2\pi$
c.m. frame, 
as it is modified by the boost along
$\vec{q}'$, 
that connects the 
$\gamma^*\pi$ 
c.m. frame and the 
$2 \pi$ 
c.m. frame.

In order to establish a link between the sets of kinematic variables 
(\ref{Set_invatiants}) 
and 
(\ref{Set_angles}),
we now express the invariants 
$s_1$ and $t'$ 
in terms of the decay angles of the final state $2\pi$ system. 
The expressions for 
$\cos\theta_\pi^*$ and $\sin^2\theta_\pi^*$ 
are given by (see eqs.~(V.1.9) and (V.1.10) of 
ref.~\cite{Byckling1973}):
\begin{eqnarray}
    \cos\theta_\pi^* = \frac{G_2\begin{pmatrix}
        k_1 + k_2, q' \\ k_1 + k_2, k_1
    \end{pmatrix}}{[\Delta_2(k_1+k_2, q')\Delta_2(k_1+k_2, k_2)]^{1/2}},
    \label{cos_theta_pi}
\end{eqnarray}
and 
\begin{eqnarray}
    \sin^2\theta_\pi^* =
     \frac{W_{\pi\pi}^2\Delta_3(k_1+k_2, q', k_2)}{\Delta_2(k_1+k_2, q')\Delta_2(k_1+k_2, k_2)}, \label{sin_theta_pi}
\end{eqnarray}
where $G_n$ and $\Delta_n$ denote  the generic and symmetric $n\times n$ Gram determinants, 
given by 
eqs.~(\ref{GramDetG}) 
and 
(\ref{GramDetDelta}), 
respectively. The Gram determinants in the denominator of 
eq.~(\ref{cos_theta_pi}) 
can be computed using the Källén kinematic triangular function 
(\ref{TriangLambda}):
\begin{eqnarray}
    [\Delta_2(k_1+k_2, q')\Delta_2(k_1+k_2, k_2)]^\frac{1}{2} &=& \lambda^{1/2}(s, W_{\pi\pi}^2, 0)\lambda^{1/2}(W_{\pi\pi}^2, m_\pi^2, m_\pi^2) \nonumber \\
    &=& \frac{1}{4}W_{\pi\pi}(s-W_{\pi\pi}^2)\sqrt{W_{\pi\pi}^2-4m_\pi^2}.
\end{eqnarray}
The invariant $s_1$ is then  expressed as 
\begin{eqnarray}
    s_1 = \frac{s-W_{\pi\pi}^2+2m_\pi^2}{2} + \cos\theta_\pi^* \frac{(s-W_{\pi\pi}^2)\sqrt{W_{\pi\pi}^2-4m_\pi^2}}{2W_{\pi\pi}}. \label{s_1}
\end{eqnarray}
It is important to note that $s_1$ is a large variable of  the order of $s$ in the generalized Bjorken limit 
(\ref{Bjorken_limit}).

In the chiral limit ($m_\pi = 0$), the expressions for $\cos\theta_\pi^*$ and $\sin^2\theta_\pi^*$ read:
\begin{eqnarray}
    \cos\theta_\pi^* \Big|_{m_\pi = 0} = -\frac{s-2s_1-W_{\pi\pi}^2}{s-W_{\pi\pi}^2}, 
    \quad
    \sin^2\theta_\pi^* \Big|_{m_\pi = 0} = \frac{4s_1(s-s_1-W_{\pi\pi}^2)}{(s-W_{\pi\pi}^2)},
\end{eqnarray}
and $s_1$ reduces to
\begin{eqnarray}
    s_1 \Big|_{m_\pi = 0} = \frac{s-W_{\pi\pi}^2}{2}(1+\cos\theta_\pi^*).
\end{eqnarray}

Now we turn to the invariant $t'$. The cosine and sine functions of the helicity frame azimuthal angle 
$\varphi_\pi^*$ 
can be expressed using the Gram determinants (see eq.~(V.8.6) of ref.~\cite{Byckling1973}):
\begin{eqnarray}
    \cos\varphi_\pi^* = \frac{G_3\begin{pmatrix}
        k+q, q', k \\ k+q, q', k_2
    \end{pmatrix}}{[\Delta_3(k+q, q',k)\Delta_3(k+q, q', k_2)]^{1/2}}, \label{cos_phi_pi} 
\end{eqnarray}
and
\begin{eqnarray}
    \sin^2\varphi_\pi^* = \frac{\Delta_2(k+q, q')\Delta_4(k+q, q', k, k_2)}{\Delta_3(k+q, q', k)\Delta_3(k+q, q', k_2)}. \label{sin_phi_pi}
\end{eqnarray}
Employing eq.~(\ref{s_1}), we establish the relation
\begin{eqnarray}
    \Delta_3(k+q, q', k_2)^\frac{1}{2}
    &=& \frac{\sin\theta_\pi^*}{4}(s-W_{\pi\pi}^2)\sqrt{W_{\pi\pi}^2-4m_\pi^2}.
\end{eqnarray}
The Gram determinant $G_3$ in the numerator of eq.~(\ref{cos_phi_pi}) reads
\begin{eqnarray}
    G_3\begin{pmatrix}
        k+q, q', k \\ k+q, q', k_2
    \end{pmatrix}
    = \frac{1}{4}\begin{vmatrix}
        2s & s-W_{\pi\pi}^2 & s-s_1+m_\pi^2 \\
        s-W_{\pi\pi}^2 & 0 & s-s_1-W_{\pi\pi}^2+m_\pi^2 \\
        s+Q^2+m_\pi^2 & Q^2-W_{\pi\pi}^2+\Delta^2 & 2m_\pi^2 - t'
    \end{vmatrix},
\end{eqnarray}
which enables us to express the invariant $t'$   in terms of the pion decay angles $\theta_\pi^*$ and $\varphi_\pi^*$. 
Using 
eqs.~(\ref{cos_phi_pi}) 
and~(\ref{s_1}), 
we derive the following expression for the invariant 
$t'$:
\begin{eqnarray}
    t' &=& R_{0,0}(\Delta^2,W_{\pi \pi}^2) \nn   \\  
    &&\mbox{} + \sqrt{1-\frac{4m_\pi^2}{W_{\pi\pi}^2}}\left( R_{1,0}(x_B,\Delta^2,W_{\pi \pi}^2) \cos\theta_\pi^* + R_{1,1}(x_B,\Delta^2,W_{\pi \pi}^2) \sin\theta_\pi^*\cos\varphi_\pi^*\right), \label{t'}  
\end{eqnarray}
where 
\begin{eqnarray}
    R_{0,0}(\Delta^2,W_{\pi \pi}^2) = -\frac{W_{\pi\pi}^2-\Delta^2-3m_\pi^2}{2},
\end{eqnarray}
and the  other two  coefficients are approximately given in the Bjorken limit by
\begin{eqnarray}
    R_{1,0}(x_B,\Delta^2,W_{\pi \pi}^2) &=& -\frac{1}{2}\left[\left(\frac{1+x_B}{1-x_B}\right)W_{\pi\pi}^2 + \Delta^2 -m_\pi^2 \right] +\mathcal{O}\biggl(\frac{1}{Q^2}\biggr), \nonumber \\
    R_{1,1}(x_B,\Delta^2,W_{\pi \pi}^2) &=& \frac{W_{\pi\pi}}{1-x_B}\sqrt{(1-x_B)(x_B m_\pi^2 - \Delta^2) - x_B W_{\pi\pi}^2}+\mathcal{O}\biggl(\frac{1}{Q^2}\biggr). \label{coefficients_t'}
\end{eqnarray}
Hence,
$t'$ 
is a ``small'' variable of  the order of 
$Q^0$ 
in the Bjorken limit. The full expressions for 
$R_{1,0}$ 
and 
$R_{1,1}$ 
can be read off from 
eq.~(\ref{tprime_through_angles}).

We also provide a useful expression for the Gram determinant $\Delta_4$,
\begin{eqnarray}
    \Delta_4(k+q, q', k, k_2) \equiv - \Big[\varepsilon_{\alpha\beta\gamma\delta}(k+q)^\alpha q'^\beta k^\gamma k_2^\delta \Big]^2 
    \label{GramDet_4},
\end{eqnarray}
which can be written as
\begin{eqnarray}
\Delta_4(k+q, q', k, k_2) &=& \frac{1}{16}\sin^2\theta_\pi^*\sin^2\varphi_\pi^* (W_{\pi\pi}^2-4m_\pi^2)\nonumber \\
&& \mbox{} \times 
G(
2s-W_{\pi\pi}^2, 
-Q^2, 
-\Delta^2-2Q^2, 
s, 
0, 
m_\pi^2
),
\end{eqnarray}
where 
$\varepsilon_{\alpha\beta\gamma\delta}$ 
is the totally antisymmetric tensor%
\footnote{We adopt the convention $\varepsilon_{0123}=-\varepsilon^{0123} = +1$ and use the shorthand notation $\varepsilon^{\alpha\beta\gamma\delta} p_{1 \alpha} p_{2 \beta} p_{3 \gamma} p_{4 \delta} \equiv \varepsilon(p_1,p_2,p_3,p_4)$. },
and 
$G(x,y,z,u,v,w)$ 
is the kinematic tetrahedral function, 
given in
eq.~(\ref{CayleyDet}). 
Note that the physical domain of the reaction~(\ref{DVCS_pi_to_2pi})
corresponds to the condition
\begin{eqnarray}
    \Delta_4(k+q, q', k, k_2) \leq 0.
\end{eqnarray}
In the chiral limit ($m_\pi = 0$), this yields the inequality
\begin{eqnarray}
    x_B \leq \frac{-\Delta^2}{W_{\pi\pi}^2-\Delta^2}.
\end{eqnarray}

We conclude with the choice of eight variables characterizing the kinematics of $2\to 4$ reaction (\ref{Electroproduction_of_2pi}):
\begin{eqnarray}
    && Q^2, \quad y \equiv \frac{k\cdot q}{k \cdot l}, \quad \Phi, \quad x_B, \quad \Delta^2, \quad W_{\pi\pi}^2, \quad \Omega_\pi^* = (\theta_\pi^*, \varphi_\pi^*),
\end{eqnarray}
where $y$ is the lepton energy loss and the angle $\Phi$ between the leptonic and production planes is depicted in figure~\ref{fig:kin_plane}.
Therefore, the sevenfold differential cross section of the reaction 
(\ref{Electroproduction_of_2pi})
is expressed as
\begin{eqnarray}
        \frac{d^7\sigma}{dx_B dQ^2 d\Delta^2 d\Phi dW_{\pi\pi}^2 d\Omega_\pi^*} &=& \frac{1}{256(2\pi)^7}\frac{x_B y^2}{Q^4 \sqrt{1+\frac{4m_\pi^2 x_B^2}{Q^2}}}\sqrt{1-\frac{4m_\pi^2}{W_{\pi\pi}^2}} \nonumber \\
    && \mbox{} \times \bar{\sum_i}\sum_f \left|\mathcal{M}(e^-\pi^+\to e^-\gamma\pi^+\pi^0)\right|^2,
    \label{CS_7fold}
\end{eqnarray}
where we sum (average) over the final (initial) particle polarizations, 
and $d \Omega_\pi^* \equiv d \cos \theta_\pi^* d \varphi_\pi^*$ is the solid angle of the $2\pi$ decay.

\subsection{Parametrizations of $\pi\to\pi\pi$ transition GPDs}

In the generalized Bjorken limit, we describe the non-diagonal DVCS process~(\ref{DVCS_pi_to_2pi}) 
in terms of $\pi\to\pi\pi$ transition GPDs.

The leading twist-$2$ unpolarized $\pi\to\pi\pi$ transition GPDs are defined
as the Fourier transforms of the non-diagonal matrix elements of the  isoscalar and isovector quark light-cone 
($n^2=0$) 
operators, 
$\hat{O}_S$ and $\hat{O}_{V}^d$, 
\begin{eqnarray}
    \begin{Bmatrix}
        \hat{O}_S \\ \hat{O}^d_V \\
    \end{Bmatrix}
    \biggl(-\frac{\lambda n}{2}, \frac{\lambda n}{2}\biggr) = \Bar{\psi}\biggl(-\frac{\lambda n}{2}\biggr)\slashed{n}
    \begin{Bmatrix}
        1 \\ \tau^d \\
    \end{Bmatrix}
    \psi\biggl(\frac{\lambda n}{2}\biggr),
    \label{Def_OSV}
\end{eqnarray}
where $\tau^d$  ($d=1,\,2,\,3$) are the Pauli matrices. 
Note that we  choose the light-cone gauge $A\cdot n=0$, which reduces  the Wilson line in (\ref{Def_OSV}) 
to unity.

Due to the isospin invariance, the parametrizations of the
unpolarized 
$\pi\to\pi\pi$ 
transition GPDs, 
$(H^{\pi\to\pi\pi}_S, H^{\pi\to\pi\pi}_{V,I})$ 
corresponding to 
$(\hat{O}_S, \hat{O}_{V}^d)$,
take the following form:
\begin{eqnarray}
      &&  \frac{1}{2} \int \frac{d \lambda}{ 2 \pi} e^{i \lambda x (\bar{P} \cdot n)} \big\langle \pi^b(k_1)\pi^c(k_2) \big| 
    \begin{Bmatrix}
        \hat{O}_S \\ \hat{O}^d_V \\
    \end{Bmatrix}
    \biggl(-\frac{\lambda n}{2}, \frac{\lambda n}{2}\biggr)\big| \pi^a(k) \big\rangle \nonumber \\
    && \mbox{} = \frac{i\varepsilon(n, \Bar{P}, \Delta, k_1)}{f_\pi^3}
    \begin{Bmatrix}
        i\varepsilon^{abc}H^{\pi\to\pi\pi}_{S}
             \\
        \sum\limits_{I = 0}^{2} P^{I, bc}_{da} H^{\pi\to\pi\pi}_{V,I}
            \end{Bmatrix}
    (x, \xi, t;  \cdots 
        ).
    \label{Def_Hunpolarized_pi_to2pi}
\end{eqnarray}
Here, 
$\Bar{P} \equiv \frac{k + p_{\pi\pi}}{2}$  
represents  the average hadron momentum,
$\varepsilon^{abc}$ 
is the  
SU$(2)$  
isospin  totally antisymmetric  tensor, and 
$f_\pi \simeq 92.4~\mathrm{MeV}$ 
is the pion decay constant. Our conventions for the light-cone vectors and
the Sudakov decomposition of the relevant momenta are summarized in appendix~\ref{App_Sudakov}.
The projection operators 
$P^{I, bc}_{da}$ 
for the isospin states 
$I = 0, 1, 2$ 
of the final state pions are provided in appendix~\ref{App_projectors}.
The GPDs, 
$H^{\pi\to\pi\pi}_S$ 
and 
$H^{\pi\to\pi\pi}_{V,I}$,
are functions of the usual GPD variables, i.e., the parton longitudinal momentum fraction $x$, the skewness parameter
$\xi= - \frac{\Delta \cdot n}{2 (\bar{P} \cdot n)}$, and the invariant momentum transfer $t$.
They also depend on  the factorization scale $\mu^2$ 
(not shown explicitly), as well as three 
 additional variables  that describe the final state $2 \pi$ system, 
that we denote by ellipsis $(\cdots)$ in the GPDs, i.e., 
 $H^{\pi\to\pi\pi}(x,\xi,t;\cdots)$,
 and discuss further below.

Similarly, the leading twist-$2$ polarized $\pi\to\pi\pi$ transition GPDs 
are defined from the non-diagonal matrix elements of the axial-vector bilocal quark isoscalar and isovector operators, 
$\hat{O}_{5S}$ and $\hat{O}_{5V}^d$, 
\begin{eqnarray}
    \begin{Bmatrix}
        \hat{O}_{5S} \\ \hat{O}^{d}_{5V} \\
    \end{Bmatrix}
    \biggl(-\frac{\lambda n}{2}, \frac{\lambda n}{2}\biggr) = \Bar{\psi}\biggl(-\frac{\lambda n}{2}\biggr)\slashed{n}\gamma_5
    \begin{Bmatrix}
        1 \\ \tau^d \\
    \end{Bmatrix}
    \psi\biggl(\frac{\lambda n}{2}\biggr).
    \label{Def_OSV5}
\end{eqnarray}
Then, the parametrizations of the
 leading twist-$2$ polarized isoscalar and isovector GPDs ($\tilde{H}^{\pi\to\pi\pi}_S$, $\tilde{H}^{\pi\to\pi\pi}_{V,I}$),
 corresponding to $(\hat{O}_{5S}, \hat{O}_{5V}^d)$,
read:
\begin{eqnarray}
                &&  \frac{1}{2} \int \frac{d \lambda}{ 2 \pi} e^{i \lambda x (\bar{P} \cdot n)} \big\langle \pi^b(k_1)\pi^c(k_2) \big|
        \begin{Bmatrix}
            \hat{O}_{5S} \\ \hat{O}^{d}_{5V} \\
        \end{Bmatrix}
        \biggl(-\frac{\lambda n}{2}, \frac{\lambda n}{2}\biggr) \big|\pi^a(k)\big\rangle \nonumber \\
        && \mbox{} = \frac{i}{f_\pi}
    \begin{Bmatrix}
        i\varepsilon^{abc}\tilde{H}^{\pi\to\pi\pi}_{S} \\
        \sum\limits_{I = 0}^{2} P^{I, bc}_{da} \tilde{H}^{\pi\to\pi\pi}_{V,I}
                \\
    \end{Bmatrix}
    (x, \xi, t; \cdots).
        \label{Def_Hpolarized_pi_to2pi}
\end{eqnarray}

In addition to the usual  GPD variables 
$(x,\xi,t)$, 
three extra variables,  denoted by ellipsis 
$(\cdots)$ 
in the GPDs, are required to describe the final state two-hadron system.  The first of these is the invariant mass of the produced two-hadron system (pion pair), 
$W_{\pi \pi}^2 \equiv (k_1 + k_2)^2$. 
For the remaining two variables, there are two alternative sets of choices, each more appropriate depending on the specific context:

\begin{itemize}
\item The first possibility involves using the decay angles 
$\Omega_{\pi}^* = (\theta_\pi^*, \varphi_\pi^*)$,
defined in (\ref{pi-angles}) within the c.m. frame of the $2\pi$ system.
This approach is particularly well-suited for PW analysis and dispersive methods, as discussed in sections~\ref{sec:3} and \ref{Sec_Resonance_region}.
\item The second option, which greatly simplifies the manifestation of the polynomiality property of the Mellin moments of transition GPDs
\footnote{See refs.~\cite{Goeke:2001tz, Polyakov:2006dd}, where the non-diagonal $N\to \pi N$ transition GPDs are discussed.},
 consists of using the invariant momentum transfer variable 
$t' \equiv (k_2-k)^2$ 
and the light-cone momentum fraction 
$\alpha$, 
which describes how the longitudinal momentum of the final hadron state is distributed between the produced pions:
\begin{eqnarray}
    \alpha= \frac{k_1\cdot n}{p_{\pi\pi}\cdot n} = \frac{k_1 \cdot n}{1-\xi}; 
    \qquad
    1-\alpha = \frac{k_2 \cdot n}{p_{\pi\pi}\cdot n} = \frac{k_2\cdot n}{1-\xi}.
    \end{eqnarray}
\end{itemize}

In order to establish a link between the two sets of variables, we use  
(\ref{Lc_vector_covariant}) 
to express 
$\alpha$ through the invariants to the leading order (LO) in $1/Q$:
\begin{eqnarray}
    \alpha = \frac{s_1 -m_\pi^2}{s-W_{\pi\pi}^2} + \mathcal{O}(1/Q^2). \label{alpha_s}
\end{eqnarray}
We thus confirm that in the generalized Bjokren limit~(\ref{Bjorken_limit}), the value of $\alpha$ is indeed fixed.
Now, with the help of (\ref{s_1}), $\alpha$ 
can be written as a function of the cosine of the pion polar decay angle in the helicity frame:
\begin{eqnarray}
    \alpha = \frac{1}{2}\left(1 - \sqrt{1-\frac{4m_\pi^2}{W_{\pi\pi}^2}}\cos\theta_\pi^* \right) + \mathcal{O}(1/Q^2). \label{alpha}
\end{eqnarray}
The expression of the invariant variable $t'$ in terms of 
the decay angles is given by eq.~(\ref{t'}) with $x_B$ expressed from (\ref{skewedness}).

In order to  illustrate the  manifestation of the polynomiality property of $\pi \to \pi \pi$  GPDs,  we consider the Mellin moments
of the unpolarized isoscalar GPD $H_S^{\pi \to \pi \pi}(x,\xi,t; W_{\pi \pi}^2,t',\alpha)$. 
The $(N+1)$-th Mellin moments of $H_S^{\pi \to \pi \pi}$ is related to  the generalized FFs of the transition matrix element of the local 
twist-2 operator:
\begin{eqnarray}
    \mathscr{O}_{S}^{\mu\mu_1 \cdots \mu_N} = \mathbf{S}\Bar{\psi}\gamma^{\mu}i\overset{\leftrightarrow}{D}\vphantom{D}^{\mu_1}\cdots i\overset{\leftrightarrow}{D}\vphantom{D}^{\mu_N}\psi,
\end{eqnarray}
where $\mathbf{S}$  
denotes symmetrization in all uncontracted Lorentz indices and subtraction  
of trace  terms,  and $\overset{\leftrightarrow}{D}\vphantom{D}^\mu  \equiv (\overset{\rightarrow}{D}\vphantom{D}^\mu - \overset{\leftarrow}{D}\vphantom{D}^\mu)/2$ is the covariant derivative. The FF decomposition of the $\pi \to \pi\pi$ transition matrix element of $\mathscr{O}_S^{\mu\mu_1\cdots\mu_N}$ can be presented as
\begin{eqnarray}
    \big\langle \pi^b(k_1)\pi^c(k_2) \big| \mathscr{O}_{S}^{\mu \mu_1\cdots \mu_N}(0) \big| \pi^a(k)\big\rangle &=& \mathbf{S} \frac{i\varepsilon^{abc}}{f_\pi^3} i\varepsilon(\mu, \Bar{P},\Delta, k_1) \sum_{\substack{i,j = 0 \\ i + j \leq N}}^N \mathcal{A}_{N+1, i, j}(t; W_{\pi\pi}^2,t') \nonumber \\
    && \mbox{} \times \Delta^{\mu_1}\cdots\Delta^{\mu_i} \Bar{P}^{\mu_{i+1}}\cdots \Bar{P}^{\mu_{i+j}} k_1^{\mu_{i+j+1}}\cdots k_1^{\mu_{N}}, \label{FF_decomposition}
\end{eqnarray}
where the corresponding generalized transition FFs are functions of $t$, $t'$, and $W_{\pi\pi}^2$.
By contracting the matrix elements in~(\ref{FF_decomposition}) with $n_\mu n_{\mu_1}\cdots n_{\mu_N}$, 
one obtains the sum rule for the ($N+1$)-th Mellin moment of the isoscalar unpolarized $\pi\to\pi\pi$ GPD:
\begin{eqnarray}
    &&\int_{-1}^1 dx x^N H^{\pi\to\pi\pi}_S(x,\xi,t;W_{\pi\pi}^2, \alpha, t') \nonumber \\
    && = \sum_{\substack{i,j = 0 \\ i+j \leq N}}^N (-2\xi)^i[(1-\xi)\alpha]^{N-i-j}\mathcal{A}_{N+1, i,j}(t; W_{\pi\pi}^2,t'),
    \label{polynomiality}
\end{eqnarray}
which yields polynomials in $\xi$ and $\alpha$ of order $N$.
Note  that the time  reversal invariance does not impose constraints on
transition GPDs, and both even and odd powers of $\xi$ appear in 
(\ref{polynomiality}).

We also stress that switching to the variables $\Omega_\pi^*$ using (\ref{alpha}) 
and (\ref{t'}) introduces non-polynomial factors in  $\xi$, originating from (\ref{coefficients_t'}). 
As a result, while the Lorentz invariance — fundamental to the polynomiality property of GPDs
— remains preserved, the polynomiality property itself becomes less straightforward with this choice of variables for transition GPDs.

The invariance under charge conjugation allows us to establish
symmetry properties of $\pi\to\pi\pi$ GPDs with respect to 
$x$.
The (isoscalar) isovector (un)polarized $\pi\to\pi\pi$ GPDs are even functions of $x$:
\begin{eqnarray}
     H^{\pi\to\pi\pi}_S(x,\xi,t;W_{\pi\pi}^2, \; \ldots
         ) &=&  H^{\pi\to\pi\pi}_S(-x,\xi,t;W_{\pi\pi}^2, \; \ldots
     ), \nn \\
     \tilde{H}^{\pi\to\pi\pi}_{V,I}(x,\xi,t;W_{\pi\pi}^2, \; \ldots
     )
     &=& \tilde{H}^{\pi\to\pi\pi}_{V,I}(-x,\xi,t;W_{\pi\pi}^2, \; \ldots
               ),    
\end{eqnarray}
where $(\ldots)$ in the transition GPD arguments stand for the variables $\{ \alpha, \,t'\}$ or $\{\theta_\pi^*, \,\varphi_\pi^*\}$, describing the $2\pi$ system.
Analogously, the (isovector) isoscalar (un)polarized $\pi\to\pi\pi$ GPDs are odd functions of $x$:
\begin{eqnarray}
    \tilde{H}^{\pi\to\pi\pi}_S(x,\xi,t;W_{\pi\pi}^2, \;
    \ldots
    )
    &=& - \tilde{H}^{\pi\to\pi\pi}_S(-x,\xi,t;W_{\pi\pi}^2, \; \ldots
       ), \nn  \\  
    H^{\pi\to\pi\pi}_{V,I}(x,\xi,t;W_{\pi\pi}^2,
    \, \ldots
      )
    &=& -H^{\pi\to\pi\pi}_{V,I}(-x,\xi,t;W_{\pi\pi}^2, \, \ldots   
    ).
\end{eqnarray}

Relying on the isospin invariance, we work out
combinations of transition GPDs corresponding to  physical process. 
For instance, in the case of $\pi^+\to\pi^+\pi^0$, we obtain
\begin{eqnarray}
    H^{\pi^+\to\pi^+\pi^0} &=& \frac{4}{9}H^{\pi^+\to\pi^+\pi^0}_u + \frac{1}{9}H^{\pi^+\to\pi^+\pi^0}_d \nonumber \\
    &=& \frac{5}{9}H^{\pi\to\pi\pi}_S + \frac{1}{6}\big(H^{\pi\to\pi\pi}_{V,2} - H^{\pi\to\pi\pi}_{V,1}\big). \label{H_pi-2pi}
\end{eqnarray}
To derive these relations, we make use of the identities 
\begin{eqnarray}
    H^{\pi^\pm\to\pi^\pm\pi^0}_{u+d} = \pm 2H^{\pi\to\pi\pi}_S,
    \quad
    H^{\pi^\pm\to\pi^\pm\pi^0}_{u-d} = H^{\pi\to\pi\pi}_{V,2} - H^{\pi\to\pi\pi}_{V,1}.
\end{eqnarray}
The isospin symmetry relations among the $\pi\to\pi\pi$ transition 
GPDs for the  $u-$and $d-$quark flavors are given by
\begin{eqnarray}
    H^{\pi^+\to\pi^+\pi^0}_u = -H^{\pi^-\to\pi^-\pi^0}_d, 
    \quad
    H^{\pi^+\to\pi^+\pi^0}_d = -H^{\pi^-\to\pi^-\pi^0}_u.
\end{eqnarray}
Similar relations are established for other charge eigenstates of pions, and they also hold for the polarized transition GPDs.

\subsection{$\pi \to \pi\pi$ transition GPDs at the $2\pi$ threshold}
\label{Sub_sec_Threshold}

The exact $2 \pi$ threshold kinematics of the
reaction (\ref{DVCS_pi_to_2pi}),
with two pions produced at rest in the $2 \pi$ c.m. frame, corresponds to $W_{\pi \pi}^2= 4m_\pi^2$.
Using eqs.~(\ref{s_1}) and (\ref{t'}), one can check that this corresponds to the following threshold values of the invariants: 
\begin{eqnarray}
    s_{1 \, \mathrm{th}} = \frac{s-2m_\pi^2}{2}, \qquad    \label{s_1_th}
    t'_{\mathrm{th}} = \frac{\Delta^2-2m_\pi^2}{2}  \label{t'_th},
\end{eqnarray}
and, from eq.~(\ref{alpha}), we get%
\footnote{Our definition of threshold kinematics is consistent with the results of ref.~\cite{Polyakov:2006dd},
adapted for the case of equal masses of the final state hadrons ($\pi\pi$ instead of $\pi N$).}
$
\alpha_{\rm th}= \frac{1}{2}. 
$

Following the approach outlined in refs.~\cite{Pobylitsa:2001cz, Kivel:2004bb, Polyakov:1998ze, Polyakov:2006dd, Polyakov:1999gs}, we impose constraints on 
$\pi\to\pi\pi$ GPDs in the near-threshold region using the soft pion theorem, which 
is based on the Partial  Conservation of Axial Current (PCAC) hypothesis,
see {e.g.} \cite{Adler1968}.
According to this theorem,  the  matrix elements of the bilocal quark operators $\hat{\mathcal{O}} = \{ \hat{O}_{S}, \hat{O}_{5S}, \hat{O}_{V}^d, \hat{O}_{5V}^d \}$, 
which define $\pi \to \pi \pi$ transition GPDs, are
reduced to the pion matrix elements of their chiral rotation:
\begin{eqnarray}
    \big\langle\pi^b(k_1)\pi^c(k_2)\big| \hat{\mathcal{O}} \big| \pi^a(k) \big\rangle 
    \Big|_{ 2\pi \, \text{threshold}}
    &=& -\frac{i}{f_\pi}\big\langle \pi^c(k_2) \big| \big[Q^b_5, \hat{\mathcal{O}}\big] \big| \pi^a(k) \big\rangle,
    \label{SPT}
   \end{eqnarray}
where $Q_5^a$ is the axial charge operator,
\begin{eqnarray}
    Q_5^a = \int d^3x J_{5 \, 0}^a(x),
\end{eqnarray}
with the axial current $J_{5 \, \mu}^a (x) = \Bar{\psi}(x)\gamma_\mu\gamma_5\frac{\tau^a}{2}\psi(x)$.
The required commutators are computed as follows:
\begin{eqnarray}
&&
    \left[ Q^a_5, \hat{O}_S(0,z) \right] = 0;  \qquad \qquad \qquad \ \ \ 
    \left[ Q^a_5, \hat{O}_{5S}(0,z) \right] = 0; \nonumber \\ &&
    \left[ Q^a_5, \hat{O}^b_V(0,z) \right]= i\varepsilon^{abc}\hat{O}^c_{5V}(0,z); \quad
    \left[ Q^a_5, \hat{O}^b_{5V}(0,z) \right]= i\varepsilon^{abc}\hat{O}^c_{V}(0,z).
\end{eqnarray}
This leads us to express the polarized $\pi\to\pi\pi$ transition GPDs in terms of the unpolarized  isovector pion GPD $H^{\pi}_V$, 
which is defined in the isospin invariant form through the pion matrix element of the vector light-cone quark operator  (\ref{Def_OSV}) \cite{Polyakov:1999gs}:
\begin{eqnarray}
        \frac{1}{2}\int\frac{d\lambda}{2\pi}e^{ix_\pi\lambda n\cdot \Bar{P}_\pi}\big\langle\pi^b(p'_\pi)\big|
    \begin{Bmatrix}
        \hat{O}_S \\ \hat{O}^c_V \\
    \end{Bmatrix}
    \biggl(-\frac{\lambda n}{2}, \frac{\lambda n}{2}\biggr)\big| \pi^a(p_\pi)\big\rangle = \left(\Bar{P}_\pi\cdot n \right)
    \begin{Bmatrix}
        \delta^{ab}H^\pi_S(x_\pi,\zeta,t_\pi) \\
        i\varepsilon^{abc}H^\pi_V(x_\pi,\zeta,t_\pi)\\
    \end{Bmatrix}
    , \nonumber \\
\end{eqnarray}
where $\Bar{P}_\pi \equiv \frac{p'_\pi+p_\pi}{2}$ is the average pion momentum;  $\zeta$ is the skewness variable defined with respect to the momentum transfer between the initial and final pions; and $t_\pi=(p_\pi'-p_\pi)^2$ is the invariant momentum transfer.

Hence, at the $2\pi$ threshold, the isovector polarized $\pi\to\pi\pi$ GPDs reduce to 
\begin{eqnarray}
    \tilde{H}^{\pi\to\pi\pi}_{V,2}(x,\xi,t;4m_\pi^2, \alpha_{\rm th}, t'_{\rm th}) &=& H^{\pi}_V(x_\pi, \zeta, t'_{\rm th}) \theta\left( 1-|x_\pi| \right),  \nn \\
    \tilde{H}^{\pi\to\pi\pi}_{V,1}(x,\xi,t;4m_\pi^2,
  \alpha_{\rm th}, t'_{\rm th}) &=& -H^{\pi}_V(x_\pi, \zeta, t'_{\rm th}) \theta\left( 1-|x_\pi| \right),  \nn \\
    \tilde{H}^{\pi\to\pi\pi}_{V,0}(x,\xi,t;4m_\pi^2,
  \alpha_{\rm th}, t'_{\rm th}  
    ) &=& -4H^{\pi}_V(x_\pi, \zeta, t'_{\rm th}) \theta\left( 1-|x_\pi| \right), 
\end{eqnarray}
where $\theta(x)$ is the unit step function; 
and the average parton momentum fraction $x_\pi$ and the skewness variable $\zeta$ are expressed as
\begin{eqnarray}
    x_\pi = \frac{2x}{2-\alpha_{\rm th} (1-\xi)};  \qquad 
    \zeta = \frac{(k-k_2)\cdot n}{(k+k_2)\cdot n} = \frac{2\xi+\alpha_{\rm th}(1-\xi)}{2-\alpha_{\rm th}(1-\xi)}.
\end{eqnarray}

On the other hand, the isoscalar polarized $\pi\to\pi\pi$ GPD, as well as the unpolarized $\pi\to\pi\pi$ GPDs, reduces to zero in this limit:
\begin{eqnarray}
    \tilde{H}^{\pi\to\pi\pi}_{S}(x,\xi,t;4m_\pi^2, \alpha_{\rm th}, t'_{\rm th}) = 0; \quad
    H^{\pi\to\pi\pi}_{S,V}(x, \xi, t; 4m_\pi^2, \alpha_{\rm th}, t'_{\rm th}) = 0.
\end{eqnarray}

It is also instructive to work out the $2\pi$ threshold kinematics in terms of the $2\pi$ decay angles, $\theta_\pi^*$ and $\varphi_\pi^*$. 
Substituting the threshold value eq.~(\ref{s_1_th}) into the expression for $\cos^2\theta_\pi^*$ we get for the numerator of eq.~(\ref{cos_theta_pi}),
\begin{eqnarray}
    \left( G_2\begin{pmatrix}
        k_1 + k_2, q' \\ k_1 + k_2, k_2
    \end{pmatrix} \right)^2 \Biggr|_{s_1= s_{1 \, \mathrm{th}}} = \frac{1}{16}W_{\pi\pi}^4(W_{\pi\pi}^2 - 4m_\pi^2)^2. \label{cos_theta_pi_numerator}
\end{eqnarray}
Therefore, after substituting $s_1 = s_{1 \,\mathrm{th}}$ it is safe to set $W_{\pi\pi}$ to the $2\pi$ threshold value in eq.~(\ref{cos_theta_pi}), since the threshold results in the first order zero in its denominator,
\begin{eqnarray}
    \Delta_2(k_1+k_2, q')\Delta_2(k_1+k_2, k_2) = \frac{1}{16}W_{\pi\pi}^2 (s-W_{\pi\pi}^2)^2(W_{\pi\pi}^2-4m_\pi^2),
\end{eqnarray}
while the numerator eq.~(\ref{cos_theta_pi_numerator}) has the second order zero for $W_{\pi\pi} = 2m_\pi$. We conclude that with that sequence of taking limits (first $s_1\to s_{1 \, \mathrm{th}}$ and then $W_{\pi\pi}\to 2 m_\pi$) we obtain the prescription
\begin{eqnarray}
    \cos^2\theta_\pi^* \Big|_\mathrm{th} = 0 \quad \mathrm{and} \quad \sin^2\theta_\pi^* \Big|_\mathrm{th} = 1.
\label{Threshold_theta}    
\end{eqnarray}
Similarly, for the azimuthal angle $\varphi_\pi^*$, we obtain
\begin{eqnarray}
    \cos^2\varphi_\pi^* \Big|_\mathrm{th} = 0 \quad \mathrm{and} \quad \sin^2\varphi_\pi^* \Big|_\mathrm{th} = 1,
    \label{Threshold_phi}
\end{eqnarray}
by taking the limit $t' \to t_{\mathrm{th}}$ first, and then $W_{\pi\pi} \to 2 m_\pi$.

\section{\texorpdfstring{Contribution of the $\rho$-meson resonance}{Contribution of the \rho-meson resonance}}
\label{sec:3}
In this section, we address  the $e^-\pi\to e^-\gamma\pi\pi$ reaction in the vicinity of an intermediate $2\pi$-resonance state. 
We focus on the $\rho(770)$, a well-defined resonance state that is  accurately described by the Breit-Wigner formula.
The subsequent decay of the $\rho(770)$ into two pions is described using the standard effective $\rho\pi\pi$  Lagrangian (\ref{L_rho_pipi}). 
This  allows us to work out the $2\pi$ decay angular dependence of the $\pi\to\pi\pi$ transition GPDs 
in the case of pseudoscalar-to-vector meson transition, providing insight into the structure 
of the double PW expansion of $\pi\to\pi\pi$  transition  GPDs. 
 
We assume that for 
$W_{\pi \pi} \sim m_\rho$, 
the 
$e^-\pi\to e^-\gamma\pi\pi$ 
reaction is dominated by the effect of 
$\pi\to\rho$ transition. We introduce the parametrization for the
$\pi \to \rho$ transition GPDs and compute the amplitude of 
$e^-\pi\to e^-\gamma\rho$ process, accounting for both the BH and non-diagonal DVCS subprocess contributions.
Subsequently, we combine this amplitude with the $\rho\to\pi\pi$ decay amplitude to 
describe 
the $e^-\pi\to e^-\gamma\rho\to e^-\gamma\pi\pi$ process.

\subsection{Parametrization of $\pi\to\rho$ transition GPDs}

We introduce the leading twist-$2$ unpolarized and polarized $\pi\to\rho$ transition GPDs. 
The usual counting of degrees of freedom gives the number of independent helicity amplitudes for $\pi \to \rho$ transition induced by the quark-antiquark operator:
\be
\underbrace{\frac{1}{2}}_{P-\text{invariance}} \times \underbrace{2 \times 2}_{\text{quark helicities}} \times \underbrace{3}_{\rho-\text{meson helicities}} =6.
\ee
Thus, there are a total of $6$ quark twist-$2$ GPDs for the $\pi \to \rho$ transition: $3$ unpolarized and polarized GPDs, and 
$3$ transversity GPDs  (which we do not consider here).

The unpolarized  isoscalar and isovector  $\pi\to\rho$ transition GPDs 
are defined as the Fourier transforms of the matrix elements of light-cone quark operators (\ref{Def_OSV}) and
can be parametrized as 
\begin{eqnarray}
        && 
        \frac{1}{2} \int \frac{d \lambda}{ 2 \pi} e^{i \lambda x (\bar{P} \cdot n)}
        \big\langle\rho^b(p_{\pi\pi}, \lambda_\rho) \big|
    \begin{Bmatrix}
        \hat{O}_S \\ \hat{O}^c_V \\
    \end{Bmatrix}
    \biggl(-\frac{\lambda n}{2}, \frac{\lambda n}{2}\biggr)\big| \pi^a(k) \big\rangle  \nonumber \\
    && = \frac{\varepsilon\left(n, \mathcal{E}^*(p_{\pi\pi},\lambda_\rho), \Bar{P}, \Delta\right)}{m_\rho}
    \begin{Bmatrix}
        \delta^{ab}H_S^{\pi\to\rho}(x,\xi,t) \\
        i\varepsilon^{abc} H^{\pi\to\rho}_V(x,\xi,t) \\
    \end{Bmatrix}
    , \label{pi-rho_VGPD}
\end{eqnarray}
where $m_\rho$ is the $\rho$-meson mass and $\mathcal{E}(p_{\pi\pi}, \lambda_\rho)$ is the polarization vector of a massive spin-$1$ particle with momentum $p_{\pi\pi}$ and  
polarization $\lambda_\rho$. Note that the parametrization of the transition matrix element between a pseudoscalar and a vector meson involves the ``topological'' structure $\varepsilon\left(n, \mathcal{E}^*(p_{\pi\pi},\lambda_\rho), \Bar{P}, \Delta\right)$.
No such structure is possible for a transition between a pseudoscalar meson and a scalar meson.

The physical combination of unpolarized GPDs (\ref{pi-rho_VGPD}) occurring for the 
$\pi^+ \to \rho^+$ transition can be expressed as
\begin{eqnarray}
    H^{\pi^+\to\rho^+} &=& \frac{4}{9}H^{\pi\to\rho}_u + \frac{1}{9}H^{\pi\to\rho}_d 
   =\frac{5}{9}H_{S}^{\pi\to\rho} + \frac{1}{3}H_{V}^{\pi\to\rho}, \label{H_pi-rho}
\end{eqnarray}
where $H^{\pi\to\rho}_{u+d} = 2H^{\pi\to\rho}_{S}$ and $H^{\pi\to\rho}_{u-d} = 2H^{\pi\to\rho}_V$.

The first Mellin moment of the unpolarized $\pi\to\rho$ GPD is related to the electromagnetic 
FF for the $\pi\to\rho$ transition defined~\cite{Khodjamirian:1997tk} through the $\pi\to\rho$ transition matrix element of the local electromagnetic current
$J^\mu_{\mathrm{em}}(0) = \Bar{\psi}(0)\gamma^\mu\biggl(\frac{1}{6} + \frac{\tau^3}{2}\biggr)\psi(0)$:
\begin{eqnarray}
    \Braket{\rho(p_{\pi\pi}, \lambda_\rho) | J_{\mathrm{em}}^\mu(0) | \pi(k)} 
     = \varepsilon^\mu_{\; \nu\alpha\beta}\mathcal{E}^{*\nu}(p_{\pi\pi}, \lambda_\rho)p_{\pi\pi}^\alpha k^\beta \frac{F_{\rho\pi}(t)}{m_\rho}. \label{pi-rho FF}
\end{eqnarray}
This results in the following sum rules for transition GPDs (\ref{H_pi-rho}):
\begin{eqnarray}
    \int_{-1}^1 dx H^{\pi\to\rho}_S(x,\xi,t) = 3F_{\rho\pi}(t); \qquad
    \int_{-1}^1 dx H^{\pi\to\rho}_V(x,\xi,t) = 0. \label{pi-rho_1st}
\end{eqnarray}

Similarly, the polarized $\pi\to\rho$ GPDs are defined through the matrix element of the axial vector light-cone quark operators  (\ref{Def_OSV5}):
\begin{eqnarray}
       &&  \frac{1}{2} \int \frac{d \lambda}{ 2 \pi} e^{i \lambda x (\bar{P} \cdot n)} \big\langle\rho^b(p_{\pi\pi}, \lambda_\rho) \big|
    \begin{Bmatrix}
        \hat{O}_{5S} \\ \hat{O}^c_{5V} \\
    \end{Bmatrix}
    \biggl(-\frac{\lambda n}{2}, \frac{\lambda n}{2}\biggr)\big| \pi^a(k) \big\rangle  \nonumber \\
    && \mbox{} = \frac{1}{f_\pi}\biggl[(\mathcal{E}^*\cdot \Delta)
    \begin{Bmatrix}
        \delta^{ab}\tilde{H}^{\pi\to\rho}_{1S}(x,\xi,t) \\
        i\varepsilon^{abc}\tilde{H}^{\pi\to\rho}_{1V}(x,\xi,t)\\
    \end{Bmatrix}
    + m_\rho^2(\mathcal{E}^*\cdot n)
    \begin{Bmatrix}
        \delta^{ab}\Tilde{H}^{\pi\to\rho}_{2S}(x,\xi,t) \\
        i\varepsilon^{abc}\Tilde{H}^{\pi\to\rho}_{2V}(x,\xi,t) \\
    \end{Bmatrix}
    \biggr], \label{pi-rho_AGPD}
\end{eqnarray}
for the isoscalar and the isovector polarized $\pi\to\rho$ GPDs, $\tilde{H}^{\pi\to\rho}_{1,2 \, S}$ and $\tilde{H}^{\pi\to\rho}_{1,2 \, V}$, respectively. 
The pion decay constant and the $\rho$-meson mass factors are introduced to render the $\pi \to \rho$ GPDs dimensionless.
The isospin relations for the  polarized $\pi\to\rho$ GPDs are analogous to  eq.~(\ref{H_pi-rho}).

\subsection{$e^-\pi\to e^-\gamma \rho$ BH + DVCS amplitudes}

\begin{figure}[t]
   \centering
   \includegraphics[width=0.8\columnwidth]{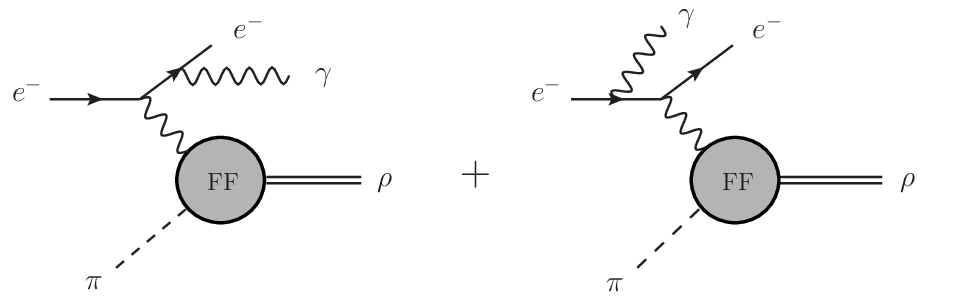}
   \caption{The BH process of $e^-\pi\to e^-\gamma\rho$ in which the lepton radiates the outgoing photon. Shaded blobs represent the  $\pi\to\rho$ transition electromagnetic FF (\ref{pi-rho FF}).}
   \label{fig:BH}
\end{figure}

The amplitude of the $e^-\pi \to e^-\gamma\rho$ reaction includes the contributions from both the BH and non-diagonal $\pi \to \rho$ Virtual Compton Scattering (VCS).

The BH amplitude describes the interaction of the target pion with 
the local electromagnetic probe, while the outgoing photon 
is produced via Bremsstrahlung from the leptonic lines (see figure~\ref{fig:BH}). 
To the LO in QED, it can be calculated exactly: 
\begin{eqnarray}
    \mathcal{M}_{\mathrm{BH}}(e^-\pi\to e^-\gamma\rho) &=& \frac{e^3}{t}\varepsilon^{*\nu}(q', \lambda_\gamma) \Bar{u}(l', \lambda_e')\left[ \gamma_\nu\frac{1}{\slashed{l}'+\slashed{q}'}\gamma_\mu + \gamma_\mu\frac{1}{\slashed{l}-\slashed{q}'}\gamma_\nu\right]u(l, \lambda_e) \nonumber \\
    && \mbox{} \times \Braket{\rho(p_{\pi\pi}, \lambda_\rho) | J_{\mathrm{em}}^\mu(0) | \pi(k)},
\end{eqnarray}
where $e$  
denotes the elementary electric charge,
$\varepsilon^\nu (q', \lambda_\gamma)$ is the polarization vector of the photon with helicity $\lambda_\gamma$, 
and $u(l, \lambda_e)$ ($\Bar{u}(l', \lambda_e')$) denotes the Dirac spinor of the incoming (outgoing) electron with helicity $\lambda_e$ ($\lambda_e'$). 
We adopt the Dirac slash notation $\slashed{l} \equiv l^\mu \gamma_\mu$ and neglect the electron mass. The parametrization of the $\pi \to \rho$ transition matrix element of the local electromagnetic current is specified in eq.~(\ref{pi-rho FF}).

\begin{figure*}[h]
\centering
\includegraphics[page=1,width=0.8\textwidth]{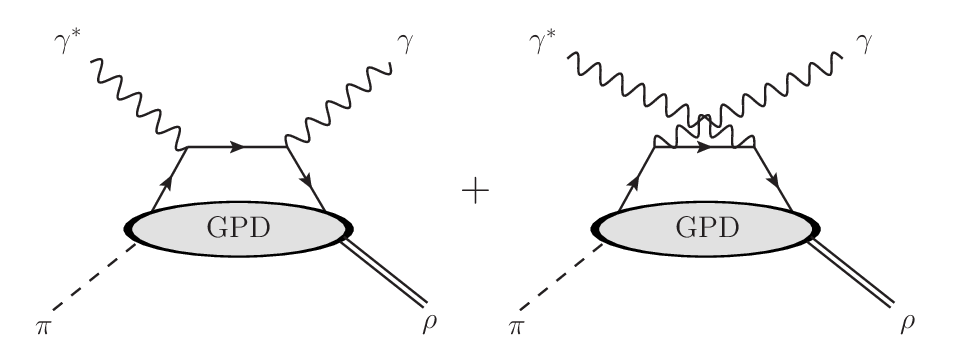}
\caption{Handbag diagrams for the non-diagonal deeply virtual Compton scattering   $\gamma^*\pi\to\gamma\rho$ contributing into the hadronic tensor (\ref{Def_Htensor_pi_rho}) to the LO in $1/Q$. }
\label{fig:handbag}
\end{figure*}

The amplitude of the non-diagonal VCS, 
\begin{eqnarray}
    \mathcal{M}_{\mathrm{VCS}}(e^-\pi\to e^-\gamma\rho)
     =\frac{ie^3}{Q^2}\varepsilon^{*}_\mu(q', \lambda_\gamma)
    \Bar{u}(l', \lambda_e')\gamma_\nu u(l, \lambda_e) {\cal T}^{\mu\nu}_{\pi\to\rho},
\end{eqnarray}
is expressed in terms of the hadronic tensor ${\cal T}^{\mu\nu}_{\pi\to\rho}$ defined as the $\pi \to \rho$ transition matrix element of the time-ordered product of electromagnetic currents:
\begin{eqnarray}
    {\cal T}^{\mu\nu}_{\pi\to\rho} = -i\int d^4x e^{iq'\cdot x} \Braket{\rho(p_{\pi\pi}, \lambda_\rho) | \mathscr{T}\left[ J_\mathrm{em}^\mu(x)J_\mathrm{em}^\nu(0) \right] | \pi(k) }.
\label{Def_Htensor_pi_rho}    
\end{eqnarray}
We compute the hadronic tensor at the LO in perturbation theory within the Bjorken limit $Q \to \infty$ and constant $x_B$,  
to the leading twist-$2$ accuracy,
taking into account the handbag diagrams depicted in figure~\ref{fig:handbag}.
This yields the expression for the hadronic tensor (\ref{Def_Htensor_pi_rho}) through  convolutions of the hard scattering kernels with appropriate combinations (cf. eq.~(\ref{H_pi-rho})) of unpolarized and polarized $\pi \to \rho$ transition GPDs, (\ref{pi-rho_VGPD}) and (\ref{pi-rho_AGPD}):
\begin{eqnarray}
    {\cal T}^{\mu\nu}_{\pi\to\rho} &=& -\frac{1}{2}g^{\mu\nu}_\perp \int_{-1}^1 dx C^+(x, \xi) \frac{\varepsilon\left(n, \mathcal{E}^*(p_{\pi\pi},\lambda_\rho),\Bar{P},\Delta\right)}{m_\rho}H^{\pi\to\rho}(x,\xi,t) \nonumber \\
    && \mbox +\frac{i}{2}\varepsilon^{\mu\nu}_\perp\int_{-1}^1 dx C^{-}(x,\xi)\frac{1}{f_\pi}\left[ (\mathcal{E}^*\cdot\Delta)\tilde{H}^{\pi\to\rho}_1(x,\xi,t) + m_\rho^2 (\mathcal{E}^*\cdot n)\tilde{H}^{\pi\to\rho}_2(x,\xi,t) \right], \nonumber \\ \label{pi-rho_hadronic_T}
\end{eqnarray}
where we employ the conventional notations
\begin{eqnarray}
    g^{\mu\nu}_\perp &=& g^{\mu\nu} - \tilde{p}^\mu n^\nu - n^\mu\tilde{p}^\nu;
    \quad
    \varepsilon^{\mu\nu}_\perp = \varepsilon^{\mu\nu\rho\sigma}\tilde{p}_\rho n_\sigma;
\end{eqnarray}
and $C^{\pm}(x,\xi)$ denote the LO hard scattering kernels 
\begin{eqnarray}
    C^{\pm}(x,\xi) = \frac{1}{x-\xi+i\epsilon} \pm \frac{1}{x+\xi-i\epsilon}, \label{hard_kernel}
\end{eqnarray}
with an infinitesimal regulating $\epsilon>0$.

\subsection{Angular structure in the $\rho(770)$ resonance region}

We now examine the angular distribution of the 
$e^-\pi\to e^-\gamma\pi\pi$ 
reaction amplitude in the decay angles of the final state 
$2\pi$ system in the vicinity of the $\rho(770)$ resonance. 
We work out the angular distribution of the cross section and point out its specific dependence on the polarization state of the intermediate $\rho$-meson.  

We also consider a description of the same reaction in terms of 
$\pi \to \pi \pi$ 
transition GPDs. We explore the decay angular dependence of 
$\pi\to\pi\pi$ GPDs in the $\rho(770)$ resonance region and specify their PW expansion in the decay angles of the $2 \pi$ system in terms of the real-valued spherical harmonics. This leads us toward a generalization for the case of an arbitrary intermediate $2 \pi$ resonance of spin-$\ell$ presented in 
section~\ref{Sub_sec_Expansion_in_pion_decay_PWs} 
resulting  in construction of the double PW expansion of 
$\pi \to \pi \pi$ 
transition GPDs in the decay angles of the final state 
$2 \pi$ system.

In the vicinity of the $\rho$(770) resonance, the $e^-\pi\to e^-\gamma\pi\pi$ process is primarily driven  by the contribution of the $\pi\to\rho$ transition, followed by the $\rho\to\pi\pi$ decay.
The $\rho\to\pi\pi$ decay process is described with the help of the effective Lagrangian \cite{Dumbrajs:1983jd}
\begin{eqnarray}
    \mathcal{L}_{\rho\pi\pi} = g_{\rho\pi\pi}\varepsilon^{abc}\rho^a_\mu \pi^b \partial^\mu \pi^c,
    \label{L_rho_pipi}
\end{eqnarray}
where $g_{\rho\pi\pi}$ stands for the $\rho\pi\pi$ coupling constant. The $\rho\pi\pi$ coupling can be extracted from the $\rho\to\pi\pi$ decay width 
$\Gamma_{\rho}$ expressed as
\begin{eqnarray}
    \Gamma_{\rho } = |g_{\rho\pi\pi}|^2 \frac{m_\rho}{48\pi}\left(1-\frac{4m_\pi^2}{m_\rho^2}\right)^{\frac{3}{2}}.
\end{eqnarray}
The experimental value of $\Gamma_{\rho}\simeq 149.1~\mathrm{MeV}$~\cite{ParticleDataGroup:2022pth} yields $g_{\rho\pi\pi}\simeq 6.01$.

\begin{figure*}[t]
    \centering
    \includegraphics[page=1,width=0.45\textwidth]{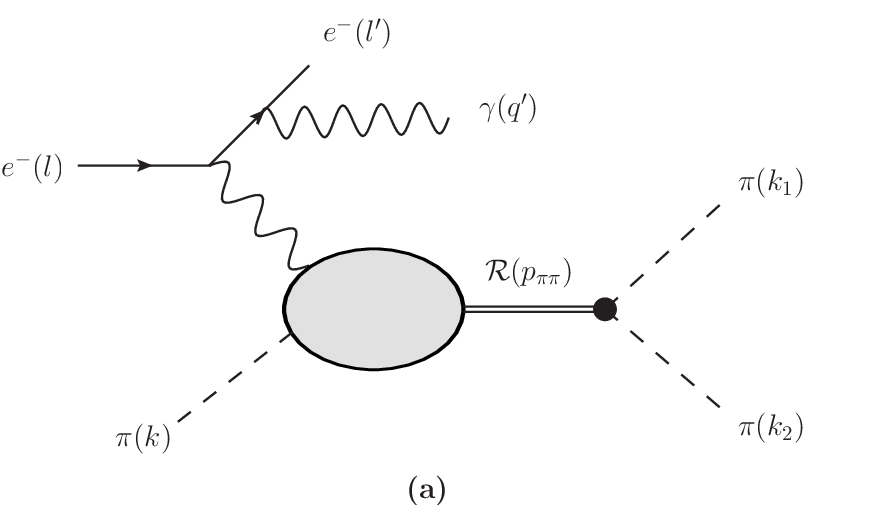} \ \ \includegraphics[page=1,width=0.45\textwidth]{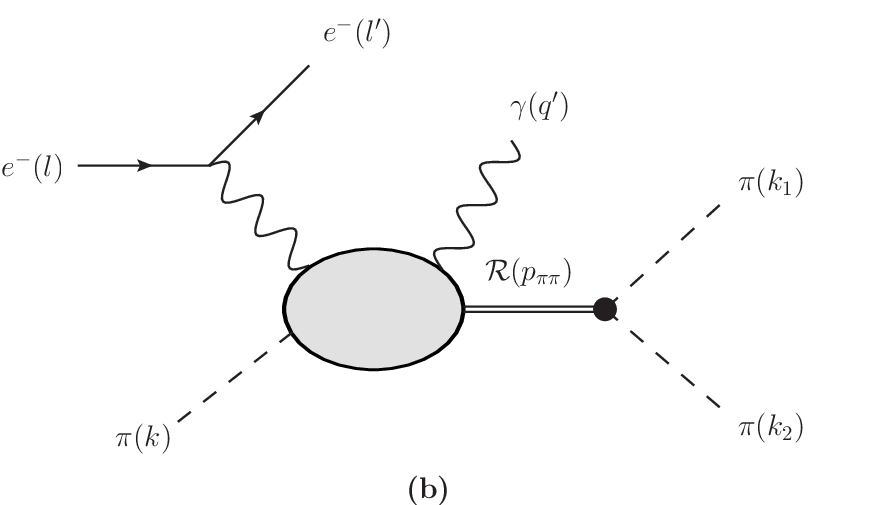}
    \caption{Feynman diagrams corresponding to the $e^-\pi\to e^-\gamma\mathcal{R}(p_{\pi \pi})\to e^-\gamma\pi\pi$ amplitude. The intermediate meson resonance $\mathcal{R}$ 
decays into two pions. The Bethe-Heitler and Deeply Virtual Compton Scattering contributions involving the $\pi\to\mathcal{R}$ transition are shown in diagrams {\bf (a)} and {\bf (b)}, respectively. For simplicity, we do not show explicitly the BH diagram with a photon emitted from the incoming leptonic line.}
\label{fig:pi-2pi}
\end{figure*}

The amplitude 
$\mathcal{M}(e^-\pi\to e^-\gamma\rho\to e^-\gamma\pi\pi)$ 
includes both the BH and the DVCS contributions, as presented in figure~\ref{fig:pi-2pi}.
We assume the $\rho(770)$ resonance propagator to be described by the  standard Breit-Wigner formula.
With these assumptions, one can present the amplitude of the $e^-\pi\to e^-\gamma\rho\to e^-\gamma\pi\pi$ process as 
\begin{eqnarray}
    \mathcal{M}(e^-\pi\to e^-\gamma\rho \to e^-\gamma\pi\pi) &=& C_\mathrm{iso}g_{\rho\pi\pi}(k_1-k_2)_\mu \frac{i\sum_{\lambda_\rho} \mathcal{E}^\mu(p_{\pi\pi}, \lambda_\rho) \mathcal{E}^{*\nu}(p_{\pi\pi}, \lambda_\rho) }{W_{\pi\pi}^2 - m_\rho^2 + im_\rho\Gamma_\rho} \nonumber \\
    && \mbox{} \times \mathcal{M}_\nu(e^-\pi\to e^-\gamma\rho), \label{Amp_pi-2pi}
    \label{Isolated_rho_contribution}
\end{eqnarray}
where $C_\mathrm{iso}$ is the isospin factor for the $\rho\to\pi\pi$ decay, for instance, 
$C_\mathrm{iso} = \pm {1}/{\sqrt{2}}$   
for
$\rho^\pm\to\pi^\pm\pi^0$.
The amplitude $\mathcal{M}^\mu(e^-\pi\to e^-\gamma\rho)$ is defined as
\begin{eqnarray}
    \mathcal{M}\left(e^-\pi\to e^-\gamma\rho(p_{\pi\pi}, \lambda_\rho)\right) = \mathcal{M}_\nu(e^-\pi\to e^-\gamma\rho) \mathcal{E}^{*\nu}(p_{\pi\pi}, \lambda_\rho).
\end{eqnarray}
The $\rho$-meson polarization sum    
in the numerator of propagator in (\ref{Isolated_rho_contribution}) is given by
\begin{eqnarray}
    \sum_s \mathcal{E}^\mu(p_{\pi\pi}, \lambda_\rho) \mathcal{E}^{*\nu}(p_{\pi\pi}, \lambda_\rho) = -g^{\mu\nu} + \frac{p_{\pi\pi}^\mu p_{\pi\pi}^\nu}{W_{\pi\pi}^2},
\end{eqnarray}
where we substitute $p_{\pi\pi}^2 = W_{\pi\pi}^2$ instead of $m_\rho^2$.

The expression for the $e^-\pi\to e^-\gamma\pi\pi$ amplitude given in eq.~(\ref{Amp_pi-2pi}) can be used to analyze the $2\pi$ decay angular structure of the $e^-\pi\to e^-\gamma\pi\pi$ cross section near the $\rho(770)$ resonance energy. To achieve this, it is instructive to explicitly compute the product of the $\rho$-meson polarization vector and the $\rho\pi\pi$ vertex function instead of using the polarization sum. We evaluate $k_1\cdot \mathcal{E}(p_{\pi\pi}, \lambda_\rho)$ in the c.m. frame of the final state pions, where the intermediate $\rho$-meson is at rest, and the polarization states 
$\lambda_\rho = 0$ 
and 
$\lambda_\rho = \pm 1$ 
correspond to the longitudinal and right and left circular polarizations of the $\rho$-meson, respectively.
Our choice of the $2\pi$ decay angles, as given in eq.~(\ref{pi-angles}), allows us to calculate this convolution as
\begin{eqnarray}
    k_1\cdot\mathcal{E}(p_{\pi\pi}, \lambda_\rho) = -\Vec{k}_1\cdot\vec{\mathcal{E}}(0, \lambda_\rho) = |\vec{k}^*_1|\sqrt{\frac{4\pi}{3}}Y^{\lambda_\rho}_{1}(\theta_\pi^*, \varphi_\pi^*), 
\end{eqnarray}
where  $Y_\ell^m(\theta_\pi^*, \varphi_\pi^*)$ denote the 
Condon-Shortley phase convention spherical harmonics
(\ref{Ylm_CondSh}).
The magnitude of the pion $3$-momentum in the $2\pi$ c.m. frame is expressed as 
$|\vec{k}_1^*| = 
\frac{W_{\pi \pi}}{2} 
\sqrt{1- \frac{4m_\pi^2}{W_{\pi \pi}^2}}. 
$

Integrating the squared amplitude over the $2\pi$ decay azimuthal angle $\varphi_\pi^*$ yields the distribution in the polar angle $\theta_\pi^*$,
\begin{eqnarray}
    &&\int_0^{2\pi}d\varphi_\pi^* \big|\mathcal{M}(e^-\pi\to e^-\gamma\rho\to e^-\gamma\pi\pi)\big|^2 \nonumber \\
    && \mbox{}
    = C_\mathrm{iso}^2 g_{\rho\pi\pi}^2\frac{W_{\pi\pi}^2-4m_\pi^2}{(W_{\pi\pi}^2 - m_\rho^2)^2 + m_\rho^2\Gamma_\rho^2}\frac{4\pi}{3} \nonumber \\ 
    && \quad\times \sum_{\lambda_\rho} \big|\mathcal{M}\left(e^-\pi\to e^-\gamma\rho(p_{\pi\pi}, \lambda_\rho)\right)\big|^2 \biggl[\frac{3}{2}\cos^2\theta_\pi^* \delta_{\lambda_\rho, 0} + \frac{3}{4}\sin^2\theta_\pi^*\left(\delta_{\lambda_\rho, 1} + \delta_{\lambda_\rho, -1}\right)\biggr]. \nonumber \\
    \label{Polar_angle_distribution}
\end{eqnarray}
This result shows that contributions to the $e^-\pi\to e^-\gamma\rho\to e^-\gamma\pi\pi$ cross section from the $\lambda_\rho = 0$ and the $\lambda_\rho = \pm 1$ polarization states of the resonance exhibit distinct angular dependencies. 
Similarly, the distribution of the cross section in the azimuthal angle $\varphi_\pi^*$ reads
\begin{eqnarray}
    &&\int_{-1}^{1}d\cos\theta_\pi^* \big|\mathcal{M}(e^-\pi\to e^-\gamma\rho\to e^-\gamma\pi\pi)\big|^2 \nonumber \\
    && \mbox{} = C_\mathrm{iso}^2 g_{\rho\pi\pi}^2\frac{W_{\pi\pi}^2-4m_\pi^2}{(W_{\pi\pi}^2 - m_\rho^2)^2 + m_\rho^2\Gamma_\rho^2}\frac{2}{3} \nonumber \\ 
    && \quad\times \biggl\{ \sum_{\lambda_\rho} \big|\mathcal{M}\left(e^-\pi\to e^-\gamma\rho(p_{\pi\pi}, \lambda_\rho)\right)\big|^2\left(\delta_{\lambda_\rho, 0} + \delta_{\lambda_\rho, 1} + \delta_{\lambda_\rho, -1}\right) \nonumber \\
    && \qquad - 2\operatorname{Re}\left[\mathcal{M}\left(e^-\pi\to e^-\gamma \rho(p_{\pi\pi},+1)\right)\mathcal{M}^*\left(e^-\pi\to e^-\gamma\rho(p_{\pi\pi}, -1)\right) e^{2 i \varphi_\pi^*} \right] \biggr\}.
    \label{Azimuthal_angle_distribution}
\end{eqnarray}
Apart from a flat distribution, we recover a trigonometric distribution arising from the interference between the amplitudes with $\lambda_\rho = \pm 1$.

It is also very instructive to consider the angular structure of the $\pi\to\pi\pi$ GPDs in the vicinity of the $\rho(770)$ resonance, where $W_{\pi\pi}\simeq m_\rho$. 
This can be performed by matching the description of the VCS amplitude in terms
of $\pi\to\pi\pi$ GPDs to that in  terms of $\pi \to \rho$ GPDs. 

In order to describe the VCS contribution into $e^-\pi\to e^-\gamma\pi\pi$
we consider the $\pi\to\pi\pi$ hadronic tensor 
\begin{eqnarray}
    {\cal T}^{\mu\nu}_{\pi\to\pi \pi} = -i\int d^4x e^{iq'\cdot x} \Braket{ \pi(k_1) \pi(k_2)| \mathscr{T}\left[ J_\mathrm{em}^\mu(x)J_\mathrm{em}^\nu(0) \right] | \pi(k) }.
\label{Def_Htensor_pi_2pi}    
\end{eqnarray}
The leading (twist-$2$) contribution following from the handbag diagrams in the Bjorken limit is expressed as
\begin{eqnarray}
{\cal T}^{\mu\nu}_{\pi\to\pi\pi} &=& -\frac{g^{\mu\nu}_\perp}{2}\frac{\varepsilon(n, \Bar{P}, \Delta, k_1)}{f_\pi^3} {\cal H}^{\pi \to \pi \pi} (\xi,t;W_{\pi\pi}^2, \theta_\pi^*, \varphi_\pi^*) \nonumber \\
    && \mbox{}+ \frac{i\varepsilon^{\mu\nu}_\perp}{2}\frac{i}{f_\pi}
         \tilde{\cal H}^{\pi \to \pi \pi} (\xi,t;W_{\pi\pi}^2, \theta_\pi^*, \varphi_\pi^*),
     \label{HTensor_pi_to-2pi}
   \end{eqnarray}
where we define the $\pi \to \pi \pi$ transition Compton FFs as 
convolutions of hard scattering kernels 
(\ref{hard_kernel}) 
with 
$\pi \to \pi \pi$ 
unpolarized and polarized transition GPDs
(\ref{Def_Hunpolarized_pi_to2pi}) and (\ref{Def_Hpolarized_pi_to2pi}):
\be
{\cal H}^{\pi \to \pi \pi} (\xi,t;W_{\pi\pi}^2, \theta_\pi^*, \varphi_\pi^*) &=&
\int_{-1}^1 dx C^+(x,\xi)H^{\pi\to\pi\pi}(x,\xi,t;W_{\pi\pi}^2, \theta_\pi^*, \varphi_\pi^*), \nn \\
\tilde{\cal H}^{\pi \to \pi \pi} (\xi,t;W_{\pi\pi}^2, \theta_\pi^*, \varphi_\pi^*) &=&
\int_{-1}^1 dx C^-(x,\xi) \tilde{H}^{\pi\to\pi\pi}(x,\xi,t;W_{\pi\pi}^2, \theta_\pi^*, \varphi_\pi^*).
  \label{Transition_FF_Compton}
\ee
For a given process, the integrands of 
eq.~(\ref{Transition_FF_Compton}) 
are written in terms of corresponding physical combinations of transition GPDs, 
{ e.g.} for the case of $\pi^+ \to  \pi^+ \pi^0$ transition see  eq.~(\ref{H_pi-2pi}).

By matching the VCS  amplitude expressed through the $\pi \to \pi \pi$ hadronic tensor
(\ref{HTensor_pi_to-2pi})
to the amplitude  (\ref{Isolated_rho_contribution}), which is presented in terms of the  $\pi \to \rho$ hadronic tensor 
(\ref{pi-rho_hadronic_T})
in the vicinity of
the $\rho(770)$, 
we obtain the contribution of the $\rho(770)$ resonance into $\pi \to \pi \pi$ GPDs.

For the unpolarized  $\pi\to\pi\pi$ GPD in the $\rho$-meson resonance region we get
\begin{eqnarray}
&&
    H^{\pi\to\pi\pi}(x,\xi,t;W_{\pi\pi}^2, \theta_\pi^*, \varphi_\pi^*)\Big|_{\rho(770)} = -2C_{\mathrm{iso}}\frac{f_{\pi}^3g_{\rho\pi\pi}}{m_\rho}\frac{1}{W_{\pi\pi}^2-m_\rho^2+im_\rho\Gamma_\rho}H^{\pi\to\rho}(x,\xi,t). \nn \\ &&
    \label{H-rho_unpol}
\end{eqnarray}
Note that the $\rho$-meson contribution into the
unpolarized $\pi\to\pi\pi$ GPD  exhibits no dependence on the $2\pi$ decay angles.
The angular dependence of the corresponding contribution into the VCS amplitude is determined by the ``topological'' pseudotensor structure $\varepsilon(n, \Bar{P}, \Delta, k_1)$ occurring in the definition of the unpolarized $\pi\to\pi\pi$ GPD (\ref{Def_Hunpolarized_pi_to2pi}).
From eq.~(\ref{GramDet_4}) one may check that 
\begin{eqnarray}
    \varepsilon(n, \Bar{P}, \Delta, k_1) \sim \sqrt{1-\cos^2\theta_\pi^*}  \sin\varphi_\pi^*.
\end{eqnarray}
The $W_{\pi\pi}^2$-dependence of eq.~(\ref{H-rho_unpol}) is solely governed by the Breit-Wigner propagator. Furthermore, the dependencies on $x$, $\xi$, and $t$ are inherited from those of the unpolarized $\pi\to\rho$ GPD. 
We also note that the GPD 
$H^{\pi\to\pi\pi}$ 
in the $\rho$-meson region
includes both the real and imaginary parts as a function of 
$W_{\pi \pi}^2$, 
that reflects instability of the $\rho$-meson with respect to the decay into $2 \pi$. 

For the polarized GPD, we obtain 
\begin{eqnarray}
    &&\tilde{H}^{\pi\to\pi\pi}(x,\xi,t;W_{\pi\pi}^2, \theta_\pi^*, \varphi_\pi^*)\Big|_{\rho(770)} = 2C_{\mathrm{iso}}g_{\rho\pi\pi}\frac{1}{W_{\pi\pi}^2-m_\rho^2+im_\rho\Gamma_\rho} \nonumber \\
    && \quad \times \biggl[ \frac{W_{\pi\pi}^2 - t - 3m_\pi^2 + 2t'}{4}\tilde{H}^{\pi\to\rho}_1(x,\xi,t) \nonumber \\
    && \qquad + \frac{m_\rho^2(1-\xi)(s-2s_1 - W_{\pi\pi}^2 + 2m_\pi^2)}{2(s-W_{\pi\pi}^2)}\tilde{H}^{\pi\to\rho}_2(x,\xi,t) \biggr],
\label{H_pi_to_2pi_polarized_from_rho}    
\end{eqnarray}
where the $2\pi$ decay angular dependence is introduced through the invariants 
$s_1$ and $t'$.

The relations
(\ref{s_1}) 
and 
(\ref{t'}) 
enable us to work out explicitly  
the angular dependence of the polarized GPD 
(\ref{H_pi_to_2pi_polarized_from_rho}):
\begin{eqnarray}
    &&\tilde{H}^{\pi\to\pi\pi}(x,\xi,t;W_{\pi\pi}^2,\theta_\pi^*,\varphi_\pi^*)\Big|_{\rho(770)} = C_\mathrm{iso}g_{\rho\pi\pi}\frac{1}{W_{\pi\pi}^2-m_\rho^2+im_\rho\Gamma_\rho}\sqrt{1-\frac{4m_\pi^2}{W_{\pi\pi}^2}} \nonumber \\
    && \quad \times \biggl[ \left( R_{1,0} \tilde{H}^{\pi\to\rho}_1(x,\xi,t) + m_\rho^2(1-\xi)\tilde{H}^{\pi\to\rho}_2(x,\xi,t) \right)\cos\theta_\pi^* \nonumber \\
    && \qquad + R_{1,1} \tilde{H}^{\pi\to\rho}_1(x,\xi,t) \sin\theta_\pi^*\cos\varphi_\pi^*\biggr], \label{H-rho_pol}
\end{eqnarray}
with the coefficients 
$R_{1,0}(x_B,\Delta^2,W_{\pi \pi}^2)$ 
and 
$R_{1,1}(x_B,\Delta^2,W_{\pi \pi}^2)$ 
defined in 
eq.~(\ref{t'}). 

It is instructive to rewrite  the $\rho$-meson contributions into unpolarized and polarized 
$\pi \to \pi \pi$ GPDs 
(\ref{H-rho_unpol}) and (\ref{H-rho_pol}) 
in terms of the so-called real-valued spherical harmonic functions of sine and cosine types 
$\operatorname{Y}_{\ell, m}(\theta_\pi^*, \varphi_\pi^*)$ 
(\ref{Def_Real_Sph_Harmonics}) 
(see discussion of 
section~\ref{Sub_sec_Expansion_in_pion_decay_PWs}).
\bi
\item
Polarized $\pi \to \pi \pi$ GPD turns to involve $\operatorname{Y}_{1, 0}(\theta_\pi^*, \varphi_\pi^*) \sim \cos \theta_\pi^*$
and
 $\operatorname{Y}_{1, 1}(\theta_\pi^*, \varphi_\pi^*) \sim \sin \theta_\pi^* \cos \varphi_\pi^*$.
Also note that the contribution of the constant term 
$\sim \operatorname{Y}_{0, 0}(\theta_\pi^*, \varphi_\pi^*)$  
cancels exactly,%
\footnote{We employ the spin sum in the numerator of the $\rho$-meson propagator analytically continued off-shell replacing 
$m_\rho^2$ by $W_{\pi \pi}^2$.}
leaving only the contribution of spherical harmonics with 
$\ell = 1$.

\item The unpolarized 
$\pi \to \pi \pi$ 
GPD obtains a contribution proportional to 
$\frac{1}{\sin \varphi_\pi^* \sin \theta_\pi^* } \operatorname{Y}_{1, -1}(\theta_\pi^*, \varphi_\pi^*) \sim {\rm const}$.
\ei

By repeating a similar exercise with the contribution of the spin-$2$ $f_2(1270)$ 
meson, one may check that the polarized 
$\pi \to \pi \pi$ 
GPD obtains contributions from the 
$\ell=2$ 
spherical harmonics
$\operatorname{Y}_{\ell=2 \, m=0}(\theta_\pi^*, \varphi_\pi^*)$, 
$\operatorname{Y}_{\ell=2, \,  m=-1}(\theta_\pi^*, \varphi_\pi^*)$ 
and 
$\operatorname{Y}_{\ell=2, \, m=-2}(\theta_\pi^*, \varphi_\pi^*)$.
The unpolarized 
$\pi \to \pi \pi$ 
GPD gets contributions proportional to 
$\frac{1}{\sin \varphi_\pi^* \sin \theta_\pi^* } \operatorname{Y}_{\ell=2, \,  m=-1} (\theta_\pi^*, \varphi_\pi^*)$
and 
$\frac{1}{\sin \varphi_\pi^* \sin \theta_\pi^* } \operatorname{Y}_{\ell=2, \,  m=-2} (\theta_\pi^*, \varphi_\pi^*)$.
This suggest the possibility of generalization for a  spin-$\ell$  $2\pi$-resonance, that is addressed in the next section.

\section{\texorpdfstring{\bm{$\pi\to\pi\pi$} transition GPDs 
in the resonance region}{\pi\to\pi\pi transition GPDs in the resonance region}}
\label{Sec_Resonance_region}

\subsection{Expansion of $\pi\to\pi\pi$ GPDs in partial waves of the $2\pi$ decay angles}
\label{Sub_sec_Expansion_in_pion_decay_PWs}

The common description of a generic spin-$J$ resonance production and decay
$2 \to 3$
process
\be
a+b \to c+R_J, \ \ R_J \to d+e
\ee
is performed relying on the formalism of the density matrix (see { e.g.} \cite{Chung:186421}).
The angular distributions in the resonance rest frame decay angles
$\theta^*$
and
$\varphi^*$
are considered at the level of the cross section rather than at the level of the amplitude.
However, this kind of description makes no specific reference to the mechanism of resonance formation.

In the case of non-diagonal DVCS, in order to take the full advantage of the control over the resonance production by a non-local QCD probe, it might be convenient to consider the expansion into resonance decay angles partial waves at the level of the amplitude. 
The LO leading twist-2 generalized Compton FFs 
of non-diagonal 
$\pi \to \pi \pi$ DVCS 
(\ref{Transition_FF_Compton}) 
turn to be functions of
$\xi$,
$t=(q-q')^2$,
$W_{\pi \pi}^2$
and
$\theta_\pi^*$,
$\varphi_\pi^*$.
This enables us to consider the expansion of the corresponding Compton FFs and transition GPDs in spherical harmonics.
By trading the dependence of the transition GPDs on the $2 \pi$  decay angles for the discrete orbital angular momentum labels 
$\ell$ and $m$, 
we obtain a convenient parametrization of the transition GPDs. Our approach also facilitates the development of phenomenological models, as displayed in section~\ref{Sec_Omnes}.

We argue that at this point it is more appropriate to use the real-valued angular harmonics. The key argument consists in the most clear way of addressing the analytic properties of the non-diagonal Compton amplitude in variables 
$\xi$ and $W_{\pi \pi}^2$.
Indeed, within the collinear factorized description, the LO leading twist-2 amplitude of non-diagonal DVCS is
presented as a convolution of 
$\pi \to \pi \pi$ 
transition GPDs with the conventional hard scattering kernels,
as written in eq.~(\ref{hard_kernel}).
This provides a discontinuity of the Compton FFs along the cut 
$\xi \in [-1;\,1]$, 
which corresponds to the usual 
$s$- and $u$-channel cuts in the
symmetric variable 
$\nu = \frac{s-u}{4m_\pi}$ 
\cite{Anikin:2007yh,Diehl:2007jb}. 
For the case of the diagonal DVCS, the discontinuity along this cut is referred to as the ``imaginary part of the Compton FF'' corresponding in the LO to values of a GPD on the crossover lines 
$x=\pm \xi$.   
However, in the case of non-diagonal DVCS reaction,  transition
GPDs themselves turn to possess an imaginary part for 
$W_{\pi \pi}^2 \ge 4m_\pi^2$. 
Therefore, the usual slang terminology referring to the ``imaginary'' and ``real'' parts of the Compton FFs has to be used with some care in this case. Thus, in order to follow accurately the origin of a phase of the Compton FFs, while constructing the PW expansion, it looks natural to avoid the appearing of a trivial phase from the azimuthal angle dependence and make use of
the real-valued spherical harmonics of sine and cosine types defined for $|m| \le \ell$ as%
\footnote{We employ the modified definition of section~5.1.6 of ref.~\cite{khersonskii1988quantum} explicitly including  the $(-1)^m$  Condon–Shortley phase factor. Also, mind the factor 
$\sqrt{2}$ 
introduced for harmonics with 
$m \ne 0$ 
to preserve the normalization of the orthogonal basis.}
\be
\operatorname{Y}_{\ell, \, m}(\theta^*_\pi, \varphi^*_\pi)=
 \begin{cases}(-1)^m \sqrt{2} \sqrt{\frac{2 \ell+1}{4 \pi} \frac{(\ell-|m|) !}{(\ell+|m|) !}} P_{\ell}^{|m|}(\cos \theta_\pi^*) \sin (|m| \varphi_\pi^*),
 & \text { for }  -\ell \le m<0;
 \\ \sqrt{\frac{2 \ell+1}{4 \pi}} P_{\ell} (\cos \theta_\pi^*),
 & \text { for } m=0;
 \\ (-1)^m \sqrt{2} \sqrt{\frac{2 \ell+1}{4 \pi} \frac{(\ell-m) !}{(\ell+m) !}} P_{\ell}^m(\cos \theta_\pi^*) \cos (m \varphi_\pi^*),
  & \text { for }  0 <m  \le \ell,\end{cases}
\label{Def_Real_Sph_Harmonics}
\ee
where
$P_{\ell}^m(z)$
stand for the associated Legendre polynomials. The spherical harmonics (\ref{Def_Real_Sph_Harmonics}) 
form an orthogonal basis on the unit sphere
$\int d \Omega_\pi^* \operatorname{Y}_{\ell, m}(\theta^*_\pi, \varphi^*_\pi) \operatorname{Y}_{\ell',  m'}(\theta^*_\pi, \varphi^*_\pi)= \delta_{\ell \ell'} \delta_{mm'}$.
The relation of the real-valued spherical harmonics $\operatorname{Y}_{\ell,  m}(\theta^*_\pi, \varphi^*_\pi)$
to the familiar Condon–Shortley phase convention  spherical harmonics 
(\ref{Ylm_CondSh}) 
is specified in 
eq.~(\ref{Rel_Spher_Harm}).

The general form of the expansion of unpolarized and polarized $\pi \to \pi \pi$ transition GPDs in the spherical harmonics (\ref{Def_Real_Sph_Harmonics})
can be worked out by a generalization of the $f_0$-, $\rho$- and $f_2$-meson exchange contributions into $\pi \to \pi \pi$ GPDs:
\begin{eqnarray}
   H^{\pi\to\pi\pi}_S(x,\xi,t;W_{\pi\pi}^2, \theta_\pi^*, \varphi_\pi^*) &=& \frac{1}{ \sin\theta_\pi^* |\sin\varphi_\pi^*| }\sum_{\ell = 1}^\infty\sum_{m = -\ell}^{-1}H^{\ell,m}_{S}(x,\xi,t;W_{\pi\pi}^2)\operatorname{Y}_{\ell, m}(\theta_\pi^*, \varphi_\pi^*), \nn \\ \label{PW_expansion_GPDs_S_unpolarized} \\
   \tilde{H}^{\pi\to\pi\pi}_S(x,\xi,t;W_{\pi\pi}^2, \theta_\pi^*, \varphi_\pi^*) &=& \sum_{\ell = 0}^\infty\sum_{m = 0}^\ell \tilde{H}^{\ell,m}_{S}(x,\xi,t;W_{\pi\pi}^2)\operatorname{Y}_{\ell, m}(\theta_\pi^*, \varphi_\pi^*). \label{PW_expansion_GPDs_S}
\end{eqnarray}
At this stage the isospin contents does not make a distinction, so we provide the results for the case of the isoscalar unpolarized and
polarized GPDs 
$H^{\pi \rightarrow \pi \pi}_S$ and $\tilde{H}^{\pi \rightarrow \pi \pi}_S$. 
However, switching to the isovector operators requires only relabelling. We conclude this subsection with a few comments on the properties of the double PW expansion of 
$\pi \to \pi \pi$ transition GPDs.
\bi
\item The expansion of unpolarized GPDs goes over the sine-type harmonics, while polarized GPDs involve the cosine-type harmonics making
these GPDs odd/even functions of $\varphi_\pi^*$, in accordance with the requirements of the $P$-symmetry.

\item
The ``topological'' Lorentz structure
$i \varepsilon(n, \bar{P}, \Delta, k_1)$ 
occurring in the definition of unpolarized 
$\pi \to \pi \pi$ 
GPDs explicitly depends on the decay angles of the final state $2 \pi$ system.
The square of this structure, as can be seen from eq.~(\ref{GramDet_4}),
turns to be proportional to
$\sin^2 \theta_\pi^* \sin^2 \varphi_\pi^*$. 
To make the spherical functions appear properly in the expression for the amplitude the PW expansion of unpolarized GPD (\ref{PW_expansion_GPDs_S_unpolarized}) 
is defined with  the 
$\frac{1}{ \sin \theta_\pi^* |\sin \varphi_\pi^*|}$ 
prefactor.

\item This prefactor 
$\frac{1}{ \sin \theta_\pi^* |\sin \varphi_\pi^*|}$ 
introduces no ambiguity in
(\ref{PW_expansion_GPDs_S_unpolarized}), 
since for 
$\ell \ge 1$, $\sin (|m| \varphi_\pi^*)$ 
with 
$m=-1, \ldots -\ell$  
is linear in 
$\sin \varphi_\pi^*$;
and 
$P_\ell^{|m|}(\cos \theta_\pi^*)$ 
with 
$|m|>0$ 
always contains a power of  
$\sin \theta_\pi^*$.
\ei

\subsection{Dispersive approach and  the Omn\`{e}s representation for $\pi \to \pi \pi$  transition GPDs}
\label{Sec_Omnes}

\begin{figure}[H]
    \centering
    \includegraphics[height = 2.5cm]{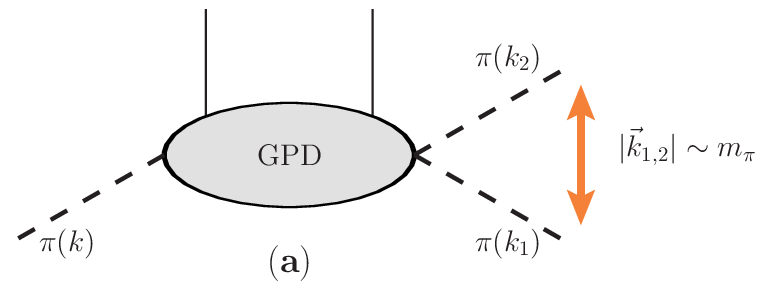} \ \ \includegraphics[height = 2.5cm]{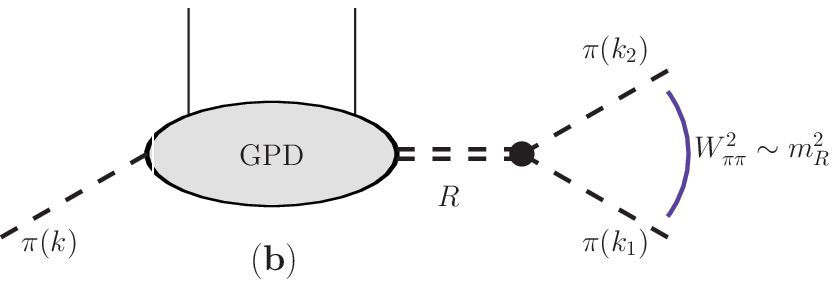}
    \caption{{\bf (a)} $\pi \rightarrow \pi \pi$ transition GPD in the chiral regime; {\bf (b)} $\pi \rightarrow \pi \pi$ transition GPD in the resonance region;  }
    \label{Fig_Ch_vs_Res}
\end{figure}

In this section we focus on the dependence of 
$\pi \to \pi \pi$ 
transition GPDs on the invariant mass 
$W_{\pi \pi}$ of the $2 \pi$ 
system. Our reasoning generally follows the line of the dispersive analysis of 
$2 \pi$ 
DAs presented in  
refs.~\cite{Polyakov:1998ze,Lehmann-Dronke:2000hlo}. 
This can be viewed as an implementation, for the case of (pseudo)scalar hadrons, of the general program outlined in ref.~\cite{Polyakov:1998sz}, aimed at establishing a link between the chiral and resonance region regimes for transition GPDs, see figure~\ref{Fig_Ch_vs_Res}.

Our principle tools are the method of dispersive analysis in the combination with the Watson-Migdal final state interaction theorem
\cite{Watson:1954uc,WOS:A1955WN97000001} 
and the approach based on the Muskhelishvili-Omn\`{e}s-type integral equations 
\cite{Omnes:1958hv,muskhelishvili2013singular}.
This framework has been successfully applied for detailed studies of  meson FFs 
\cite{Gasser:1984ux,Guerrero:1997ku}.
A broader context and a detailed historical overview of the method is given in 
refs.~\cite{Oller:2020guq,Oller2019}. 
Also, see a recent study of two-meson FFs in the unitarized chiral perturbation theory ref.~\cite{Shi:2020rkz}.

\begin{figure}[H]
    \centering
    \includegraphics[height = 3.0cm]{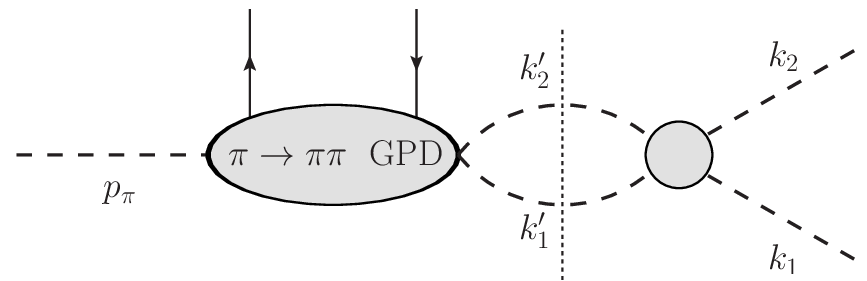}
    \caption{ Unitarity diagram for  $\pi \to \pi \pi$ transition GPDs. Dotted line denotes the cut producing the imaginary part for
    $H^{\pi \to \pi \pi}(x,\xi, t; W^2_{\pi \pi}, \theta_\pi^*, \varphi_\pi^*)$ from the $2 \pi$ intermediate state (elastic unitarity). Round gray blob denotes the $\pi \pi$-scattering amplitude.}
    \label{Fig_Unitarity}
\end{figure}

Relying on the Watson-Migdal final state interaction theorem,
we relate the imaginary part of the
$\pi \to \pi \pi$ 
transition GPDs below the inelastic 
$4 \pi$ 
threshold 
($W_{\pi \pi}^2 <16 m_\pi^2$) 
to the pion-pion scattering amplitude, see figure~\ref{Fig_Unitarity}.
In this study we do not consider the effects of inelasticity, however a generalization accounting for the effect of several coupled channels is feasible employing the formalism presented in ref.~\cite{Ropertz:2018stk}.

For definiteness we are going to consider the case of the isoscalar  $\pi \to \pi \pi$ 
GPDs defined from the non-diagonal transition matrix elements of non-local operators isoscalar vector 
$\hat{O}_S$
operator. A generalization for the cases of isoscalar pseudovector $\hat{O}_{5 S}$  
and isovector operators 
$\hat{O}_{V}$ and $\hat{O}_{5 V}$ 
is straightforward. 
The imaginary part of 
the Fourier transformed matrix element defining the isoscalar unpolarized GPD, 
associated with the discontinuity along the cut in the  $W_{\pi \pi}^2$ complex plane, can be presented as  
\be
&&
\operatorname{Im}    {\cal F}(x) \langle \pi_b(k_1) \pi_c(k_2)| \hat{O}_S \left(- \frac{\lambda n}{2}, \frac{\lambda n}{2}\,  \right) | \pi_a(p_\pi) \rangle \nn \\ && =
 \frac{1}{2!}\int d(\text{phase space}) {\cal F}(x) \langle \pi_{b'}(k_1') \pi_{c'}(k_2')| \hat{O}_S \left(- \frac{\lambda n}{2}, \frac{\lambda n}{2}\,  \right) | \pi_a(p_\pi) \rangle^*
 P^{I=1 \, b'c'}_{\; bc} \nonumber \\
&& \quad \times A_{\pi \pi}^{I=1}(k_1,k_2 | k_1',k_2' ),
\label{FInal_state_int_theorem}
\ee
where we  use a  compact notation for the Fourier transform 
$
{\cal F}(x)  \ldots  \equiv \frac{1}{2} \int \frac{d \lambda}{ 2 \pi} e^{i \lambda x (\bar{P} \cdot n)} \ldots\,,
$
and
$\frac{1}{2!}$ 
in the right-hand side (r.h.s.) of 
eq.~(\ref{FInal_state_int_theorem}) 
is the symmetry coefficient for identical pions. Note that, since for the case of isoscalar operators
the final state two-pion must be in the isospin $I=1$ state, only the $I=1$
$\pi \pi$-scattering amplitude arises in 
eq.~(\ref{FInal_state_int_theorem}). 
This is ensured by the $I=1$ projection operator 
$P^{I=1 \, b'c'}_{\; bc}$.

Therefore, the final state interaction theorem 
(\ref{FInal_state_int_theorem}) 
results in the following relation for the imaginary
part of unpolarized isoscalar 
$\pi \to \pi \pi$ 
transition GPD associated with a cut in the complex 
$W_{\pi \pi}^2$-plane:
\be
&&
\operatorname{Im} \left\{ \sin \theta_\pi^* |\sin \varphi_\pi^*| H_S^{\pi \to \pi \pi}(x,\xi,t; W^2_{\pi \pi}, \theta_\pi^*, \varphi_\pi^*) \right\} \nn \\ &&= \frac{1}{2!}\int d(2 \pi \; \text{phase space})
 \sin \theta_\pi^{'*} |\sin \varphi_\pi^{'*}|
\left(H_S^{\pi \to \pi \pi}(x,\xi, t; W_{\pi \pi}^2, \theta^{'*}_\pi, \varphi^{'*}_\pi) \right)^* \nonumber \\
&& \quad \times A_{\pi \pi}^{I=1}(k_1,k_2 | k_1',k_2' ). 
\label{Final_state_int_for_GPD}
\ee
Computing the phase space integral in 
(\ref{Final_state_int_for_GPD}) 
is most easily performed  within  the 
$2\pi$ 
c.m. system, in which
$ \vec{k}'_1 =- \vec{k}'_2$; $E'_1=E'_2=\sqrt{m_\pi^2+|\vec{k}'_1|^2}$; $W_{\pi \pi}=E'_1+E'_2$:
\be
\int d(2 \pi \; \text{phase space}) &\equiv& 
\int(2 \pi)^4 \delta^{(4)}\left(k_1+k_2-k'_1-k'_2\right) \frac{d^3 k'_1}{2 E'_1(2 \pi)^3} \frac{d^3 k'_2}{2 E'_2(2 \pi)^3} \nn \\
 &=& \int \delta(W_{\pi \pi}-2 E'_1) \frac{|\vec{k}'|^2 d|\vec{k}'| d \Omega_\pi^{'*}}{4 E_1^{'2}(2 \pi)^2}  
 = \int \frac{d\Omega_\pi^{'*}}{32 \pi^2}\sqrt{1-\frac{4 m_\pi^2 }{ W_{\pi \pi}^2}}.
\ee

To parametrize  the $\pi \pi$-scattering amplitude occurring in (\ref{Final_state_int_for_GPD})
we employ the usual form of the PW expansion, see
eq.~(1.18) of ref.~\cite{Alfaro_red_book}:
\be
A_{\pi \pi}^I= 8 \pi W_{\pi \pi} \sum_\ell (2 \ell+1)  a_\ell^I(W^2_{\pi \pi})   P_\ell\left[\cos \left(\theta_{\mathrm{cm}}\right) \right],
\ee
where 
$\theta_{\mathrm{cm}}$ 
is the angle between
$\vec{k}_1$ and $\vec{k}'_1$ 
in the 
$2\pi$ 
c.m. system.
The elastic unitarity condition
\be
\operatorname{Im} a_\ell^I(W^2_{\pi \pi})= |\vec{k}_1| |a_\ell^I(W^2_{\pi \pi})|^2
\ee
implies
\be
a_\ell^I(W^2_{\pi \pi})= \frac{1}{|\vec{k}_1|} \sin \left[\delta_\ell^I\left(W^2_{\pi \pi}\right)\right] e^{i \delta_\ell^I\left(W^2_{\pi \pi}\right)},
\ee
where 
$\delta_\ell^I\left(W^2_{\pi \pi}\right)$ 
are the scattering phases.
Therefore, the $\pi \pi$-scattering amplitude can be parametrized as
\be
A_{\pi \pi}^I=
8\pi \frac {W_{\pi \pi}}{|\vec{k}_1|} \sum_\ell(2 \ell+1) \sin \left[\delta_\ell^I\left(W^2_{\pi \pi}\right)\right] e^{i \delta_\ell^I\left(W_{\pi \pi}^2\right)} P_\ell\left[\cos \left(\theta_{\mathrm{cm}}\right)\right],
\ee
where
$\frac {W_{\pi \pi}}{|\vec{k}_1|}
=\frac{2  }{ \sqrt{1 -\frac{4m_\pi^2}{W_{\pi \pi}^2}}}$.

Now, by plugging the double PW expansion
(\ref{PW_expansion_GPDs_S})
into
(\ref{Final_state_int_for_GPD}), we obtain
\be
&&
\operatorname{Im}
\sum_{\ell=1}^\infty \sum_{m=-\ell}^{-1} H_{S}^{\ell,m}(x,\xi,t; W_{\pi \pi}^2) \operatorname{Y}_{\ell, m}\left(\theta_\pi^*, \varphi_\pi^*\right)
\nn \\&&
=  \frac{1}{2!} \frac{1}{32 \pi^2} \int_{-1}^1 d \cos \theta^{'*}_\pi \int_0^{2 \pi} d \varphi^{'*}_\pi
\sum_{\ell'=1}^\infty \sum_{m'=-\ell'}^{-1} \operatorname{Y}_{\ell',    m'}(\theta^{'*}_\pi, \varphi^{'*}_\pi)  \left( H^{\ell, m}_S (x,\xi,t; W_{\pi \pi}^2)  \right)^*\nn \\  && \quad \times
16 \pi \sum_{\ell=0}^\infty (2 \ell+1) \sin \left[\delta_\ell^{I=1} \left(W^2_{\pi \pi}\right)\right] e^{i \delta_l^{I=1}\left(W^2_{\pi \pi}\right)} P_\ell\left[\cos \left(\theta_{\mathrm{cm}}\right)\right],
\label{Im_part_for_Omenes}
\ee
where the cosine of the angle 
$\theta_{\mathrm{cm}}$ 
between
$\vec{k}_1$ and $\vec{k}'_1$ 
in the 
$2\pi$ 
c.m. system is expressed as
\be
\cos \left(\theta_{\mathrm{cm}}\right)=
\cos \theta_\pi^*  \cos \theta^{'*}_\pi +\sin \theta_\pi^*  \sin \theta^{'*}_\pi  \cos( {\varphi}^{'*}_\pi -\varphi_\pi^*).
\ee

To perform the integration over the angles in 
(\ref{Im_part_for_Omenes})
we  substitute the explicit expressions for the real-valued spherical harmonics
$\operatorname{Y}_{\ell',  m'}$
(\ref{Def_Real_Sph_Harmonics}) 
into 
(\ref{Im_part_for_Omenes}),
shift the integration variable
\be
\tilde{\varphi}_\pi={\varphi}^{'*}_\pi -\varphi_\pi^*,
\ee
and make use of the multiplication formula 
(\ref{Mult_flormula_sin}).
Next, we perform the 
$\cos \theta^{'*}$ 
integral employing
\be
\int_{-1}^1 P_{\ell'}^{|m'|} (\cos \theta^{'*}_\pi) P_{\ell}^{|m'|} (\cos \theta^{'*}_\pi) d \cos \theta^{'*}_\pi=\frac{2( \ell'+|m'|) !}{(2 \ell'+1)(\ell'-|m'|) !} \delta_{\ell, \ell'},
\ee
which allows to lift the $\ell'$ summation in the r.h.s. of (\ref{Im_part_for_Omenes}).
Finally, equating coefficients at spherical harmonics in the left- and right-hand sides of 
(\ref{Im_part_for_Omenes})
yields the following equation for the imaginary part of the expansion coefficients 
$H^{\ell, m}_S$:
\be
\operatorname{Im} H^{\ell, m}_S(x,\xi,t; W^2_{\pi \pi}) &=& \sin \left[\delta_\ell^{I=1}\left(W_{\pi \pi}^2 \right)\right] e^{i \delta_\ell^{I=1}\left(W^2_{\pi \pi} \right)} \left(  H^{\ell, m}_S (x,\xi,t; W^2_{\pi \pi}) \right)^* \nn \\
&=& \tan \left[\delta_\ell^{I=1}\left(W^2_{\pi \pi} \right)\right] \operatorname{Re} H^{\ell, m}_S(x,\xi,t; W^2_{\pi \pi}).
\ee

This relation can be transformed into the familiar Muskhelishvili-Omn\`{e}s-type
$N$-subtracted dispersion relation~\cite{Omnes:1958hv}:
\be
H^{\ell, m}_S (x,\xi,t; W^2_{\pi \pi}) &=&
\sum_{k=0}^{N-1} \frac{W^{2 k}_{\pi \pi}}{k !} \frac{d^k}{dW_{\pi \pi}^{2 k}} H^{\ell, m}_S (x,\xi,t; W^2_{\pi \pi}=0) \nonumber \\
&& \mbox + \frac{W^{2 N}_{\pi \pi}}{\pi} \int_{4 m_\pi^2}^{\infty}d\omega \frac{\tan \left(\delta_\ell^{I=1}( \omega)\right)
\operatorname{Re}\left\{  H^{\ell, m}_S(x,\xi,t; s) \right\}
}{\omega^N\left(\omega-W^2_{\pi \pi}-i \epsilon\right)}.
\ee
Assuming one subtraction 
($N=1$) 
suffices to ensure the convergence at infinity, the solution 
of this dispersion relation takes the form:
\be
&&
H^{\ell, m}_S(x,\xi,t; W^2_{\pi \pi})
=
H^{\ell, m}_S(x,\xi,t; W^2_{\pi \pi}=0) \, \Omega_\ell^{I=1}(W_{\pi \pi}^2),
\label{N=1subtracted_Omnes_representation}
\ee
where the so-called  Omn\`{e}s functions
\be
\Omega_\ell^{I}( W_{\pi \pi}^2)=\exp \left(\frac{ W_{\pi \pi}^2}{\pi} \int_{4 m_\pi^2}^{\infty} \frac{d \omega }{ \omega} \frac{\delta^I_\ell \left(\omega \right)}{\omega-W_{\pi \pi}^2-i \epsilon }\right) \label{Omnes_function}
\ee
are constructed to match the phase of  the $\pi \pi$-scattering amplitude above the $2\pi$ threshold, and to remain real otherwise.

Generally, this opens a possibility for constraining the PWs $\{H,
\, \tilde{H}\}^{\ell, m}(x,\xi,t;W^2_{\pi \pi})$,
in terms of well-known $\pi \pi$-scattering phases 
$\delta_\ell^I\left( \omega \right)$  
and a few unknown subtraction ``constants'' 
$\{H,
\, \tilde{H}\}^{\ell, m}  (x,\xi,t;W_{\pi \pi}^2=0)$,
which specify the near-threshold behavior of 
$\pi \to \pi \pi$ GPDs.

A way to constrain  the near-threshold behavior of $\pi \to \pi \pi$ GPDs relying on their chiral properties was presented in section~\ref{Sub_sec_Threshold}.
However, the strict threshold limit turns to be insufficient to fix all harmonics of the PW expansions 
(\ref{PW_expansion_GPDs_S_unpolarized}) and (\ref{PW_expansion_GPDs_S}). 
Particularly,
unpolarized $\pi \to \pi \pi$ GPDs just vanish
in this limit. Moreover, as can be seen from the prescriptions (\ref{Threshold_theta})
and 
(\ref{Threshold_phi}),  
some of the spherical harmonics 
$\operatorname{Y}_{\ell, \, m}(\theta^*_\pi, \varphi^*_\pi) \Big|_{\rm th}$   vanish at the threshold; therefore, corresponding PWs remain unconstrained. More deep insight at the near-threshold behavior of 
$\pi \to \pi \pi$ 
transition GPDs can be obtained by the systematical analysis employing 
{ e.g.} the methods of refs.~\cite{Kivel:2002ia,Diehl:2005rn}.

\subsection{Froissart-Gribov projection for $\pi \to \pi \pi$ transition GPDs}
\label{Sec:FG}

The FG projections \cite{Froissart:1961ux,Gribov:1961fm}
arise within the dispersive analysis of the diagonal DVCS
amplitudes expanded in the PWs of the $t$-channel
\cite{Kumericki:2008di,Muller:2014wxa}. 
In the case of the DVCS on a scalar target  
$\gamma^* h \to \gamma h$,  
the FG projections are formulated for the generalized FFs 
$F_J(t)$ 
occurring in the expansion of corresponding Compton FFs with respect to the cross channel  SO$(3)$ 
PWs (for definiteness we consider the case of the charge even combination of unpolarized GPDs):
\be
\mathcal{H}_{+}(\cos \theta_t, t)=
\sum_{\substack{J=0 \\ \text { even }}}^\infty
F_J(t) P_J(\cos \theta_t).
\label{Def_PW_expansion_cross_channel}
\ee
Here 
$\theta_t$ 
refers to the c.m. scattering angle of the cross-conjugated  
$\gamma^* \gamma \to \bar{h} h$
process. Neglecting the finite hadron mass corrections,
\be
 \cos  \theta_t  \sim   \frac{1}{\xi} + O(1/Q^2).  
\ee
With help of the fixed-$t$ dispersion relation for the DVCS amplitude 
\cite{Anikin:2007yh,Diehl:2007jb}
analytically continued to the cross channel, the  spin-$J$
generalized FFs 
$F_J(t)$  
can be expressed through the known absorptive part of the Compton 
amplitude  proportional to the GPD on the crossover line 
$x=\xi$, 
$
\im \mathcal{H}_{+}(x, t)= \pi H_{+}(x,x, t),
$
as convolutions with the second kind Legendre functions ${\cal Q}_J(z)$ of even $J \ge 0$:
\be
F_J(t)=2(2J+1) \int_0^1 d x\left(\frac{\mathcal{Q}_J(1 / x)}{x^2}-\frac{1}{x} \delta_{J,0}\right) H_{+}(x, x, t)+4 D(t) \delta_{J,0} \,.
\label{FJ}
\ee
Here $D(t)$  is the subtraction constant  ($D$-term FF)
given by the sum of coefficients of the Gegenbauer expansion of the $D$-term
\cite{Polyakov:1999gs}, 
and the second kind Legendre functions 
${\cal Q}_J(z)$
with  integer
$J \ge 0$ are defined in the complex-$z$ plane with the cut along the segment $[-1;\,1]$
(see { e.g.} ref.~\cite{Yanke1}):
\be
\mathcal{Q}_J\left(z\right)=\frac{1}{2} \int_{-1}^1 d z' P_J(z') \frac{1}{z-z'}.
\label{Neumann_representation}
\ee
 
The FG projections 
(\ref{FJ}) 
are argued to provide a tool for expanding the non-local QCD string-like probes created by the hard subprocess of a hard exclusive reaction
into a tower of local probes of spin-$J$. 
Therefore, in addition to the usual electromagnetic (and axial) FFs accessible with the elementary spin $J=1$ electroweak probes 
($\gamma, \,W^\pm, \, Z^0$), 
one may study  hadron's responses on the cross channel excitations of arbitrary spin-$J$ induced by the non-local QCD string-like quark (and gluon) 
operators on the light-cone.
In particular, the created spin-$2$ probe allows to access gravitational FFs of hadrons defined from hadronic matrix elements of QCD 
energy-momentum tensor, which are currently in focus of intensive theoretical and experimental studies 
\cite{Polyakov:2018zvc,Burkert:2018bqq}.

We now consider a generalization of the FG projection framework for the case of non-diagonal hard exclusive
processes. In the case of non-diagonal DVCS the expansion of the transition Compton FFs in the cross channel SO$(3)$ PWs is performed
referring to the cross channel counterpart reaction
 \be
 \gamma^*(q)+ \gamma(-q') \to \pi(-k)+R(p_{\pi \pi}) \to \pi(-k)+\pi(k_1)+ \pi(k_2),
 \ee
with the Mandelstam variable $t=(q-q')^2$ playing the role of the invariant c.m. energy and the $t$-channel scattering angle $\theta_t$
defined between $\vec{q}$ and $\vec{p}_{\pi \pi} \equiv \vec{k}_1+\vec{k}_2$
in the $\gamma^*(q) \gamma(-q')$ c.m. frame.

Applying the general method for determining
the invariant amplitudes for the binary scattering problem suggested in
refs.~\cite{Munczek1963, Alfaro_red_book}, see appendix \ref{App_Inv_Ampl_PW},
it is straightforward to specify the basis of the cross channel SO$(3)$ PWs relevant for a given $\pi \to \pi \pi$ transition GPD.
\bi
\item  Polarized $\pi \to \pi \pi$ transition GPDs 
(\ref{Def_Hpolarized_pi_to2pi}) 
are directly expanded in the Legendre polynomials of $\cos \theta_t$. Thus, from a perspective of
the cross channel SO($3$) PW expansion, polarized $\pi \to \pi \pi$ transition GPDs turn  to be completely analogous
to the case of a diagonal GPD of a spinless hadron.  

\item
The case of unpolarized $\pi \to \pi \pi$ transition GPDs 
(\ref{Def_Hunpolarized_pi_to2pi})
is slightly more tricky. 
One may check that it has to be expanded in derivatives of the Legendre polynomials of $\cos \theta_t$.
This can be seen as a consequence of choice of the ``topological'' tensor  structure 
\be
\varepsilon(n, \bar{P}, \Delta, k_1) 
\label{Tensor_structure_top}
\ee 
occurring in the definition of unpolarized $\pi \to \pi \pi$ transition GPDs. 
The tensor structure 
(\ref{Tensor_structure_top}), involving the light-cone vector $n$~(\ref{Lc_vector_covariant}),
can be constructed with the 
help of the gradient operator $\partial/ \partial \bar{P}^T_\mu$ at constant $\bar{P}^2$, $\Delta^2$ and $\bar{P} \cdot \Delta$
(\ref{GradientP}):
\be
\varepsilon(\mu, \bar{P}, \Delta, k_1) \frac{\partial}{\partial \bar{P}^T_\mu},
\ee
where the gradient operator, acting on the corresponding invariant generalized FF, produces 
$n^\mu$: 
$$
\frac{\partial}{\partial \bar{P}^T_\mu } \varphi(t,\cos \theta_t ) \sim n^\mu \frac{d}{d \cos \theta_t}\varphi(t, \cos \theta_t).
$$
Thus, the cross channel SO$(3)$ PW expansion of unpolarized $\pi \to \pi \pi$ transition GPDs looks
similar to that of the magnetic combination $H^{(M)} \equiv H+E$ of  spin-$\frac{1}{2}$ hadron GPDs, addressed in ref.~\cite{Semenov-Tian-Shansky:2023ysr}, and involves derivatives of the Legendre 
polynomials.
\ei

We introduce shorthand notations for different isospin content $H^{\ell, m} \equiv \{H^{\ell, m}_{S}, \, H^{\ell, m}_{VI=0,1,2} \}$; and, respectively,
$\tilde{H}^{\ell, m} \equiv \{\tilde{H}^{\ell, m}_{S}, \, \tilde{H}^{\ell, m}_{VI=0,1,2} \}$ for 
the coefficients of the PW expansions (\ref{PW_expansion_GPDs_S_unpolarized}), (\ref{PW_expansion_GPDs_S}), and their isovector counterparts.
The corresponding FG projections are then denoted by $F^{\ell, m}_J \equiv \{F^{\ell, m}_{J \, S}, \, F^{\ell, m}_{J \, VI=0,1,2} \}$
and
$\tilde{F}^{\ell, m}_J \equiv \{\tilde{F}^{\ell, m}_{J \, S}, \, \tilde{F}^{\ell, m}_{J \, VI=0,1,2} \}$.
\bi
\item For the polarized case, for $J \ge 0$ and  $0 \le m \le \ell$ the FG projection takes the form
\be
\tilde{F}_{J }  ^{\ell m}(t, W_{\pi\pi}^2)
= 2 \int_0^1 dx  \tilde{H}_{}^{\ell, m}(x,x,t;W_{\pi \pi}^2)    (2J+1) \frac{{\cal Q}_J (1/x)}{x^2},
\label{FG_projection_Polarized}
\ee
where we assume that the dispersion relation for the corresponding transition Compton FF does not require a subtraction.

\item
The FG projections with $J \ge 1$ and $-\ell \le m <0$ of the unpolarized Compton transition FFs then yield the FFs:
\be
F_J^{\ell m}(t,W_{\pi\pi}^2)
= 2 \int_0^1 dx  H^{\ell, m}(x,x,t; W_{\pi \pi}^2)    \frac{ (2J+1)}{ J(J+1)} \frac{(-1)}{x}   \sqrt{\frac{1}{x^2}-1}  \,
{\cal Q}_J^1(1/x )\,,
\label{FG_projection_Unpolarized}
\ee
where ${\cal Q}_J^1(1/x )$ stand for the associated Legendre functions of the second kind,
$
 {\cal Q}_J^1(z)=\left(z^2-1\right)^{\frac{1}{2}} \frac{d   {\cal Q}_J(z)}{d z }.
$

\ei
Following the line of analysis of section~3 of ref.~\cite{Semenov-Tian-Shansky:2023ysr}, once neglecting the effect of final hadron mass corrections 
({i.e.} $\cos \theta_t \sim \frac{1}{\xi}$),
the FFs (\ref{FG_projection_Unpolarized}) and (\ref{FG_projection_Polarized}) encode transitions from a single pion state to a two-pion state 
of angular momentum $\ell$ with angular momentum projection $m$ induced by a cross channel exchange of spin-$J$. The FG projection represents quantities that, at least in principle,
can be computed directly from measurable observable quantities, transition GPDs on the crossover line $x=\xi$, which are accessible through the BSA in non-diagonal DVCS.

However, this interpretation becomes more tricky once trying to account for the corrections due to finite hadron masses. As demonstrated in ref.~\cite{Semenov-Tian-Shansky:2023ysr},
in the diagonal DVCS case, the finite mass 
corrections 
due to a refined relation between the skewness variable $\xi$ and the cosine of the $t$-channel scattering angle
$ \cos \theta_t \sim   \frac{1}{\xi \beta}$ with $\beta= \sqrt{1 - \frac{4m^2}{t}}$ result 
in mixing of the spin-$J$ contribution into the FG projection $F_J$ with those of higher spins. 
It comes from the necessity to re-expand the cross channel SO$(3)$ PW expansion in $P_l(\cos \theta_t)$ in terms of $P_l(\frac{1}{\xi})$. 
The effect of this mixing can be put under control making some specific assumptions on the relevant difference between the conformal spin and cross channel angular momentum, 
ensuring that the cross channel spin-$J$ exchange still makes the dominant contribution to the FG projection $F_J$.
Accounting for the finite hadron mass corrections in the non-diagonal case looks even more tricky.  
Indeed, to the leading accuracy in $1/Q^2$, from eq.~(\ref{costhetat}) we get:
\be
 \cos  \theta_t \sim \frac{1}{\xi} \frac{t+\xi(W_{\pi \pi}^2-m_\pi^2)}{ \lambda^{1/2}(t,W_{\pi \pi}^2,m_\pi^2)} + {\cal O}\biggl(\frac{1}{Q^2}\biggr). 
\ee
The mixing between the cross channel PWs looks rather involved even in the chiral limit $m_\pi=0$, however, it can be put under control with the help of the iterative
procedure addressed in ref.~\cite{Semenov-Tian-Shansky:2023ysr}.

\section{Numerical results}
\label{sec:results}

In this section, in order to illustrate the application of our formalism,  we 
estimate the cross section of
hard exclusive electroproduction of a photon off a pion
with a non-diagonal transition from a pion into a two-pion system in the 
$\rho$-meson resonance region ($p_{\pi \pi}^2 \equiv  W_{\pi \pi}^2 \simeq m_\rho^2$):
\be
e^-(l) + \pi^+(p_\pi) \to e^-(l') + \gamma(q')+ \rho(p_{\pi \pi}, \lambda_\rho) \to
e^-(l')+ \gamma(q') +
\pi^+(k_1) + \pi^0(k_2).
\label{Reaction_2pi_rho}
\ee
We introduce phenomenological models for $\pi \to \rho$ and $\pi \to \pi \pi$ transition GPDs and 
provide estimates of contributions of the non-diagonal $\pi \to \pi \pi$ DVCS and the BH processes into the cross section 
(\ref{CS_7fold})
of the reaction (\ref{Reaction_2pi_rho}) and work out the $2\pi$ decay angular distributions in the vicinity of the 
$\rho(770)$ 
resonance. We show that the cross section turns to be sensitive to the polarization states of the produced resonance. Thus,  even in the experiments with unpolarized target, the study of angular distribution of the resonance decay products allows to get access to the information on the resonance polarization state.

Under the kinematic conditions of current experiments, the overall magnitude of the cross section is too small to realistically expect experimental access to this reaction using the approach where a quasi-real pion target is generated through a Sullivan-type process (\ref{Sullivan}).
Still, we believe our calculations are helpful for clarifying the general features 
of the framework of transition GPDs for non-diagonal hard exclusive electroproduction reactions, which will find broader application once the
approach is generalized for the physically relevant case of 
$N \to \pi N$
transitions.

Our first exercise accounts for the isolated $\rho$-meson contribution to
(\ref{Reaction_2pi_rho}) 
described within the approach of section~\ref{sec:3}. Conceptually, our phenomenological model for $\pi \to \rho$ transition GPDs is very similar to the model for nucleon-to-$\Delta$ and $N^*$ transition GPDs employed in the analysis of 
ref.~\cite{Semenov-Tian-Shansky:2023bsy}. Namely, the unpolarized 
$\pi\to\rho$ 
GPD, 
$H^{\pi\to\rho}$,
is constructed using the conventional  Radyushkin double distribution Ansatz~\cite{Radyushkin:1998es, Radyushkin:1998bz}, 
with a factorized $t$-dependence defined by the 
$\pi\to\rho$ 
transition electromagnetic FF (\ref{pi-rho FF}): 
\begin{eqnarray}
    H^{\pi\to\rho}(x,\xi,t) = \frac{5}{3}H_{DD}(x,\xi)F_{\rho\pi}(t).
    \label{H_Ansatz_unp}
\end{eqnarray}
For the  $\pi \to \rho$ electromagnetic transition FF we adopt the phenomenological parametrization of ref.~\cite{Khodjamirian:1997tk}:
\begin{eqnarray}
    F_{\rho\pi}(\Delta^2) = \frac{A^{\rho\pi}}{\Delta^4 - B^{\rho\pi}\Delta^2 + C^{\rho\pi}}, \label{pi-rho_FF_param}
\end{eqnarray}
with the parameter values $A^{\rho\pi} = 0.92~\mathrm{GeV}^4$, $B^{\rho\pi} = 3.96~\mathrm{GeV}^2$, and $C^{\rho\pi} = 2.48~\mathrm{GeV}^4$.
The double distribution part, $H_{DD}(x,\xi)$, is given  by
\begin{eqnarray}
    H_{DD}(x,\xi) = \int_{-1}^1 d\beta\int_{-1+|\beta|}^{1-|\beta|}d\alpha \delta(x-\beta-\alpha \xi) h^{(b)}(\beta, \alpha) q(\beta). \label{RDDA}
\end{eqnarray}
Here the profile function $h^{(b)}(\beta, \alpha)$  \cite{Musatov:1999xp},
\begin{eqnarray}
    h^{(b)}(\beta, \alpha) &=& \frac{1}{2^{2b+1}}\frac{\Gamma(2b+2)}{[\Gamma(b+1)]^2}\frac{[(1-|\beta|)^2-\alpha^2]^b}{(1-|\beta|)^{2b+1}},
\end{eqnarray}
depends on the parameter $b$, which we set to $b=1$.  The forward distribution $q(x)$ is chosen to be of the valence quark type  
\begin{eqnarray}
    q(x) = {\cal N} x^r(1-x)^p\theta(x),
    \label{Parametrization_q(x)}
\end{eqnarray}
with $r = -0.5$, and $p = 2$, the normalization $\cal N$ is chosen
so that the first Mellin moment of $q(x)$ is normalized to $1$. Therefore, the overall normalization of the GPD (\ref{H_Ansatz_unp}) is fixed by the sum rule (\ref{pi-rho_1st}).

\begin{figure}[t]
   \centering
   \includegraphics[width=0.30\columnwidth]{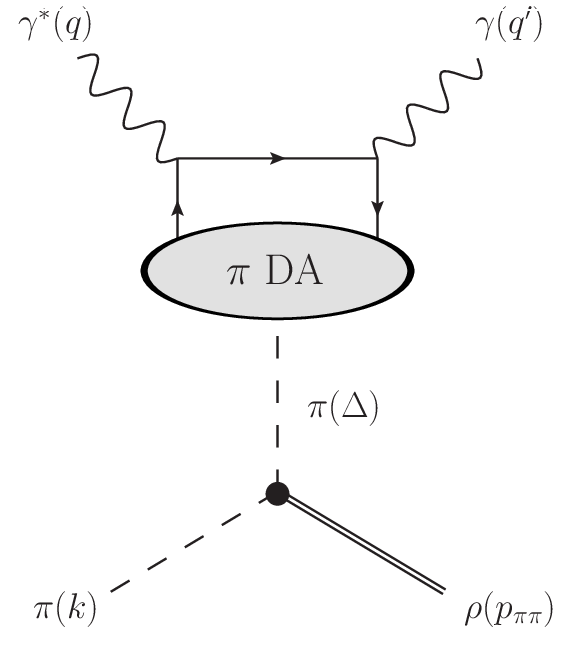}
   \caption{Cross channel pion exchange model for the polarized $\pi\to\rho$ GPDs. Upper blob stands for the twist-$2$ pion DA; $\rho\pi\pi$ interaction is described with the help of the effective vertex from (\ref{L_rho_pipi}). 
   }
   \label{fig:pi-pole}
\end{figure}

In order to build a model for the polarized $\pi\to\rho$ GPDs, we assume that for $t \sim \Delta^2_{\min}$ the hadronic matrix elements,  eq.~(\ref{pi-rho_AGPD}), are dominated by the contribution of the pion exchange in the cross channel, see figure~\ref{fig:pi-pole}.
This allows us to construct the isovector polarized $\pi\to\rho$ GPD, $\tilde{H}_{1V}^{\pi\to\rho}$, using the $\rho\pi\pi$ effective coupling constant $g_{\rho\pi\pi}$ and the twist-2 pion DA $\phi_\pi(y)$
\cite{Lepage:1980fj} defined through the matrix element of the axial isovector light-cone operator between vacuum and one-pion state. In our study, we 
employ the asymptotic form of the pion DA keeping only the $n=0$ term in the 
Gegenbauer expansion:
\begin{eqnarray}
\phi_\pi(y)= \frac{3}{4} (1-y^2) \sum_{\substack{n=0\\ \text{even}}}
a_n C_n^{\frac{3}{2}}(y); \quad a_0=1.
\end{eqnarray}  
Finally, for the polarized $\pi \to \rho$ GPD $\tilde{H}_{1V}$  we obtain
\begin{eqnarray}
    \tilde{H}_{1V}^{\pi\to\rho}(x,\xi,t) &=& \frac{4f_\pi^2g_{\rho\pi\pi}}{m_\pi^2 - t}\theta(\xi-|x|)\phi_\pi\biggl(\frac{x}{\xi}\biggr).
    \label{Htilde_model_pi_rho}
\end{eqnarray}
The support domain of the GPD (\ref{Htilde_model_pi_rho}) is restricted to the 
Efremov-Radyushkin-Brodsky-Lepage region, $-\xi \leq x \leq \xi$.
GPDs $\tilde{H}^{\pi\to\rho}_{1,2 S}$ and $\tilde{H}^{\pi\to\rho}_{2V}$ are  neglected within this approach.

\begin{figure}[t]
    \centering
    \subfigure{
        \includegraphics[width=0.45\linewidth]{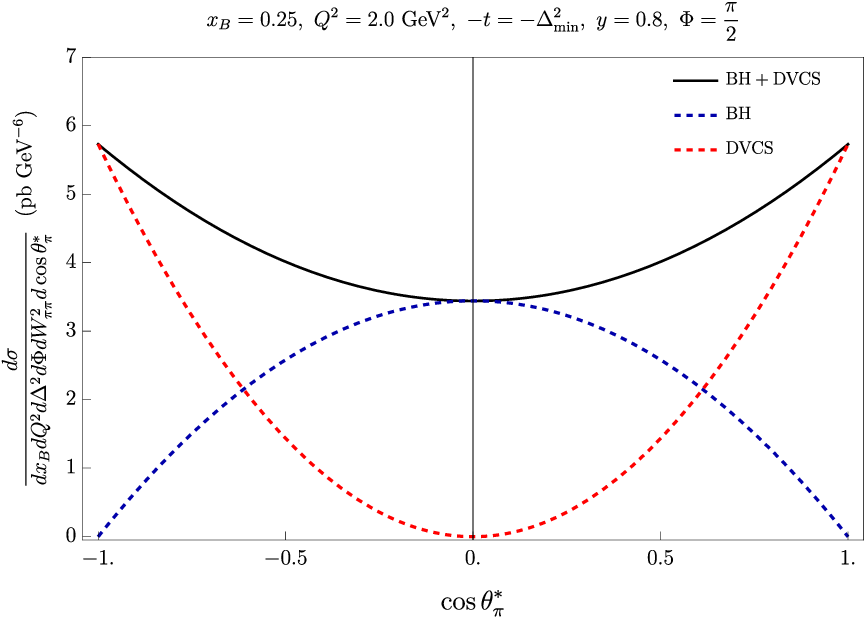}
    }
        \ \
    \subfigure{
        \includegraphics[width=0.45\linewidth]{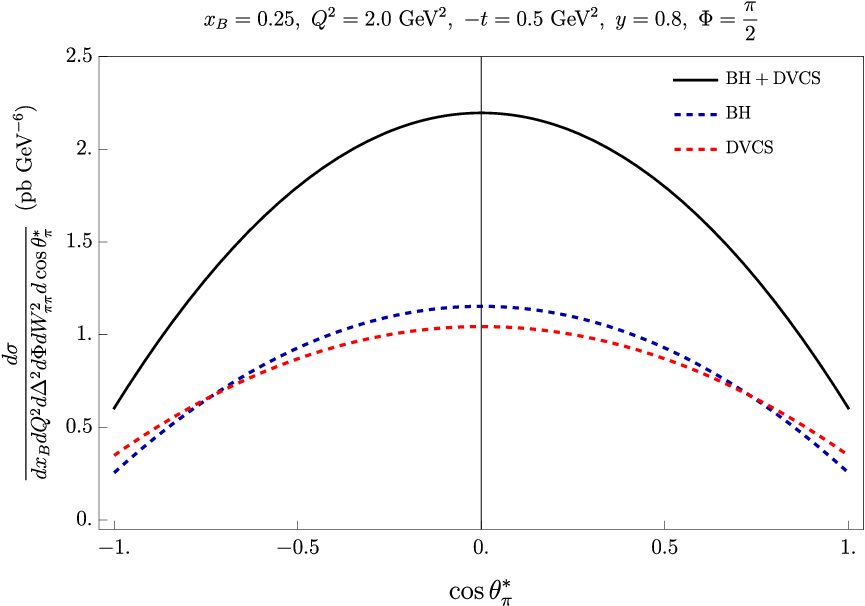}
    }
    \subfigure{
        \includegraphics[width=0.45\linewidth]{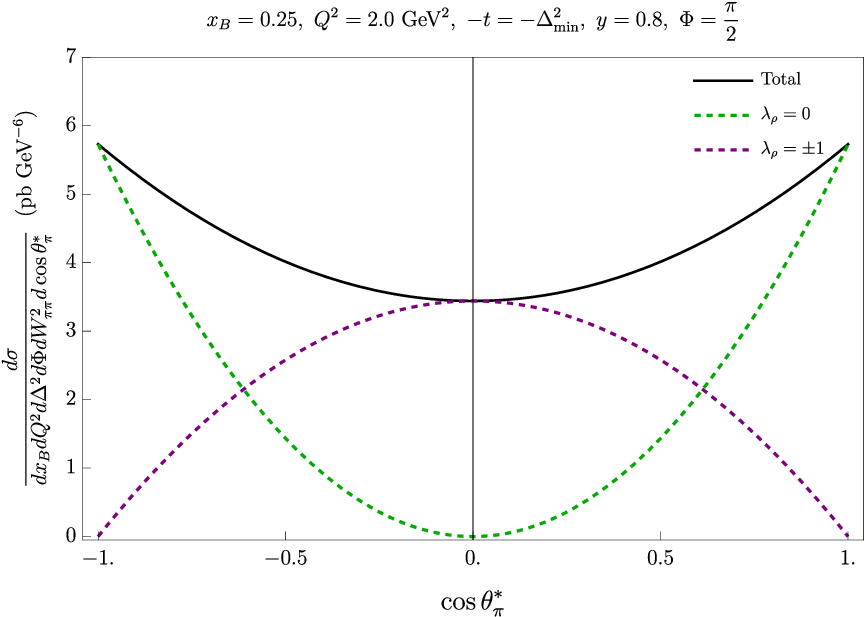}
    }
        \ \
    \subfigure{
        \includegraphics[width=0.45\linewidth]{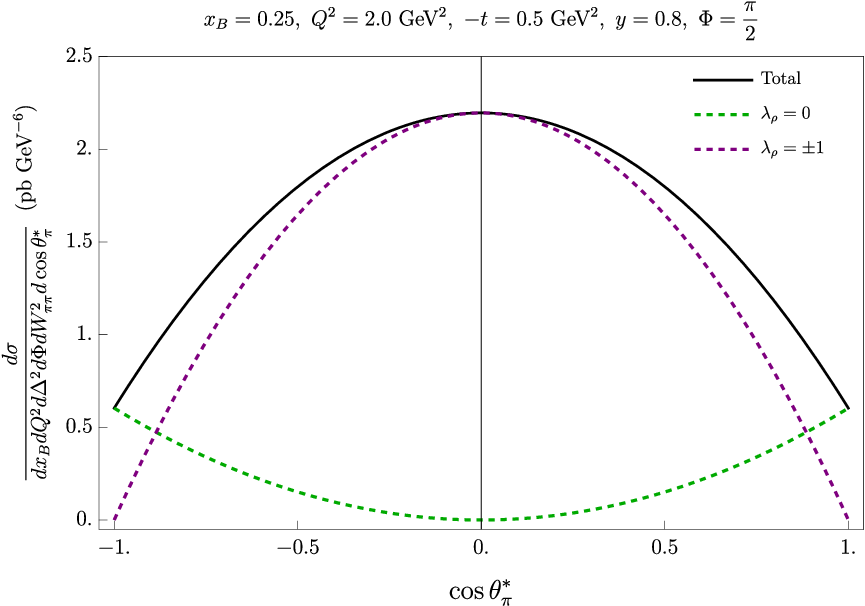}
    }
    \caption{The $2\pi$ decay angular distributions of the $e^- \pi^+ \to e^- \gamma \rho^+ \to e^- \gamma \pi^+ \pi^0$ differential cross section (\ref{CS_7fold}) integrated over $\varphi_\pi^*$. We compare the BH and DVCS cross sections (upper panels) and the contributions from the $\lambda_\rho = 0$ and the $\lambda_\rho = \pm 1$ helicity states of the produced $\rho$-meson (lower panels).}
    \label{fig:crosssection}
\end{figure}

Incorporating the isolated Breit-Wigner distribution for the $\rho(770)$ resonance, as specified in eq.~(\ref{Isolated_rho_contribution}), we
estimate the cross section (\ref{CS_7fold}) 
using the  models
(\ref{H_Ansatz_unp}) 
and 
(\ref{Htilde_model_pi_rho}) 
for the 
$\pi\to\rho$ 
GPDs and the phenomenological parametrization 
(\ref{pi-rho_FF_param}) 
for the transition FF.

In figure~\ref{fig:crosssection}, we present the differential cross section (\ref{CS_7fold}) integrated over $\varphi_\pi^*$ as a function of $\cos \theta_\pi^*$. Our estimates are performed within kinematics typical for the
JLab@12 GeV experiment: $x_B = 0.25$, $Q^2 = 2.0~\mathrm{GeV}^2$, $y = 0.8$.
We select $\Phi = \frac{\pi}{2}$ to maximize the cross section and set the invariant energy of the pion pair to $W^2_{\pi\pi} = m_\rho^2$.
The minimum value of the momentum transfer $-t$ in this kinematic region is given by $-\Delta^2_\mathrm{min}\simeq 0.2~\mathrm{GeV}^2$. 

In the upper panels of figure~\ref{fig:crosssection}, we show the cross sections of  the BH and DVCS subprocesses and the total cross section for 
$-t = -\Delta^2_{\min}$ 
and $-t = 0.5~\mathrm{GeV}^2$. 
It is worth noting that, to the LO in $1/Q$ expansion, the BH-DVCS interference term is suppressed by the threshold factor $\sqrt{\Delta_{\min}^2 - t}$.
We also note that the 
$\gamma^*\pi\to\gamma\rho \to \gamma \pi \pi$ DVCS cross section 
is dominated by the contribution of the cross channel pion exchange 
(\ref{Htilde_model_pi_rho}) 
to the polarized $\pi \to \rho$ GPD. The overall magnitude of the total cross section in the vicinity of the $\rho(770)$ resonance turns to be a few units of picobarn; this has to be compared to the cross section of the diagonal photon electroproduction off a pion \cite{Amrath:2008vx}, that is at least two orders of magnitude larger for the similar kinematics, and still is known to
be quite challenging to be measured with the JLab@12~GeV \cite{Chavez:2022tkf}. 

In the two lower panels of figure~\ref{fig:crosssection}, we present the angular distributions resulting from the specific polarization states of the intermediate $\rho$-meson. Equation~(\ref{Polar_angle_distribution}) implies that the
$\lambda_\rho=0$ helicity state of
$\rho(770)$ results in a distribution proportional to $\cos^2\theta_\pi^*$, whereas 
the $\lambda_\rho=\pm 1$ states
produce a $\sin^2\theta_\pi^*$ distribution. At $t = \Delta^2_\mathrm{min}$, the total cross section exhibits a convex shape, which shows the dominance of the longitudinally polarized $\rho(770)$ produced in the intermediate state.
As $-t$ increases, we find that the DVCS cross section becomes dominated by the contribution of the $\lambda_\rho = \pm 1$ helicity states, as the sign of its curvature changes to negative. On the other hand, for the BH cross section, its $\sim \sin^2\theta_\pi^*$ behavior does not change with varying $-t$. 
Furthermore, as it can be seen from the plots in the left panel of figure~\ref{fig:crosssection}, at the minimum value of $-t$, the BH cross section is dominated by the contribution of the $\lambda_\rho=\pm 1$ helicity states, while the DVCS cross section is  dominated by the contribution of the $\lambda_\rho=0$ helicity state. 

\begin{figure}[t]
    \centering
    \subfigure{
        \includegraphics[width=0.45\linewidth]{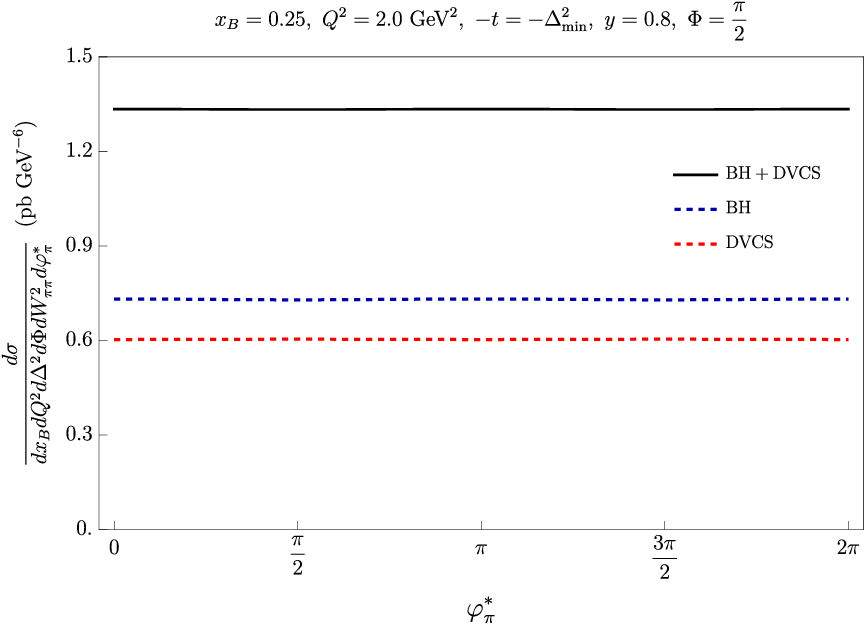}
    }
        \ \
    \subfigure{
        \includegraphics[width=0.45\linewidth]{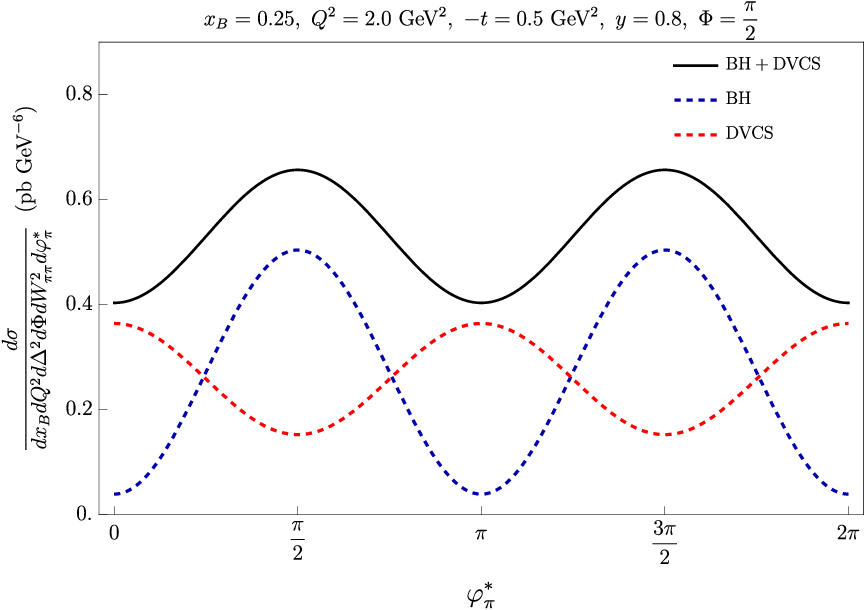}
    }
    \caption{The $2\pi$ decay angular distributions of the $e^- \pi^+ \to e^- \gamma \rho^+ \to e^- \gamma \pi^+ \pi^0$ differential cross section  (\ref{CS_7fold}) integrated over $\cos\theta_\pi^*$. We compare the BH and DVCS cross sections for $-t = -\Delta_\mathrm{min}^2$ (left panel) and $-t = 0.5~\mathrm{GeV}^2$ (right panel).}
    \label{fig:crosssection_phi}
\end{figure}

For completeness, we also present the azimuthal $2\pi$ decay angular distributions of the differential cross section for the same process in figure~\ref{fig:crosssection_phi}. As can be seen from eq.~(\ref{Azimuthal_angle_distribution}), the amplitude squared of the $e^-\pi \to e^- \gamma \rho$ process for each helicity state of $\rho(770)$ results in a flat distribution. The interference between the amplitudes for $\lambda_\rho = +1$ and $\lambda_\rho = -1$ leads to angular distributions proportional to $\cos 2\varphi_\pi^*$.
At $t = \Delta_\mathrm{min}^2$, this term is strongly suppressed for both the BH and the DVCS contributions, resulting in nearly flat distributions. 
As $-t$ increases, its $\cos 2\varphi_\pi^*$ behavior becomes more pronounced, with the BH and the DVCS cross sections displaying opposite patterns in their angular distributions.

Our second exercise presents the application of the dispersive 
technique for modelling $\pi \to \pi \pi$ 
transition GPDs. 
As we show in section~\ref{Sec_Omnes}, 
the dependence of $\pi \to \pi \pi$ GPDs on 
the invariant mass $W_{\pi \pi}$  of the final state $2\pi$ system 
for a given PW $H^{\ell ,m}$ of the expansion 
(\ref{PW_expansion_GPDs_S}) 
can be constrained in terms of known scattering phases of the
$\pi \pi$-scattering amplitude with the help of the Muskhelishvili-Omnès dispersion relation. This allows to establish a contact between transition GPDs and the $2\pi$-spectrum characteristics accessed in the $\pi \pi$-scattering reaction.

In a hypothetical ideal scenario, in which  measurements of non-diagonal
$e^- \pi \to  e^- \gamma \pi \pi$
cross section 
are supposed to be complete, the experiment is expected to provide a detailed knowledge of transition Compton FFs (\ref{Transition_FF_Compton}) as functions of
$\xi$, $t$,
$W_{\pi \pi}$, and the
$2\pi$ decay angles $\theta_\pi^*$, $\varphi_\pi^*$. This would make possible a systematic PW analysis of transition Compton FFs.

\begin{figure}[t]
    \centering
    \subfigure{
        \includegraphics[width=0.5\linewidth]{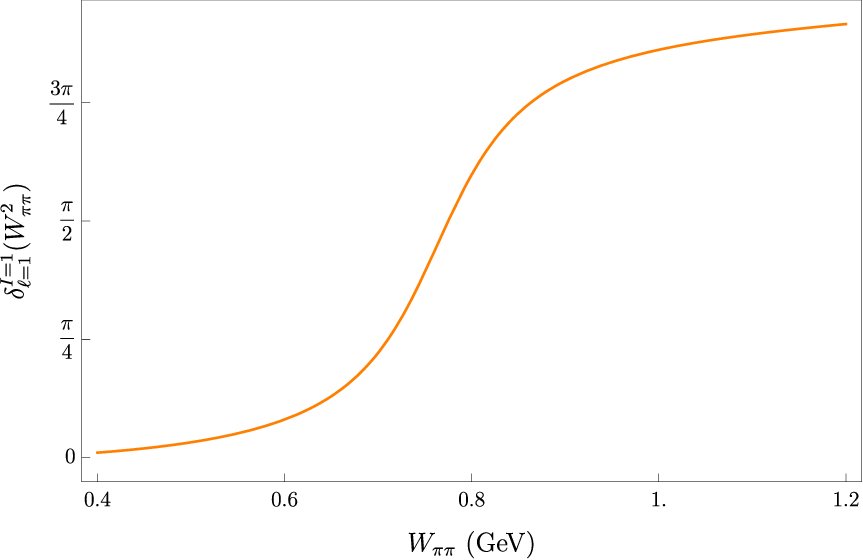}
        }
    \caption{The $\ell = 1$, $I = 1$ $\pi\pi$-scattering phase shift from ref.~\cite{Pelaez:2019eqa}.}
    \label{fig:pipi-PhaseShift}
\end{figure}

We illustrate the application of this approach considering the $\ell=1$, $m=-1$ partial wave. 
To construct a model for $H^{\ell=1 ,m=-1}$ 
we propose an Ansatz describing the near-threshold behavior  and then develop the $W_{\pi\pi}$-dependence through the $N=1$-subtracted Omnès representation (\ref{N=1subtracted_Omnes_representation}) employing the global parametrization of the isovector $P$-wave $\pi \pi$-scattering  phase 
$\delta^{I=1}_{\ell=1}$ presented in ref.~\cite{Pelaez:2019eqa} 
\footnote{For the parametrization and parameter values therein, see eq.~(22) of ref.~\cite{Pelaez:2019eqa}.}, see figure~\ref{fig:pipi-PhaseShift}.
Since the $P$-wave of $\pi \pi$-scattering is dominated by the single $\rho(770)$,
the model is expected to be equivalent to the single Breit-Wigner 
contribution of  the $\rho(770)$ resonance to unpolarized $\pi \to \pi \pi$ GPD 
(\ref{H-rho_unpol}).

\begin{figure}[t!]
    \centering
    \includegraphics[width=0.6\linewidth]{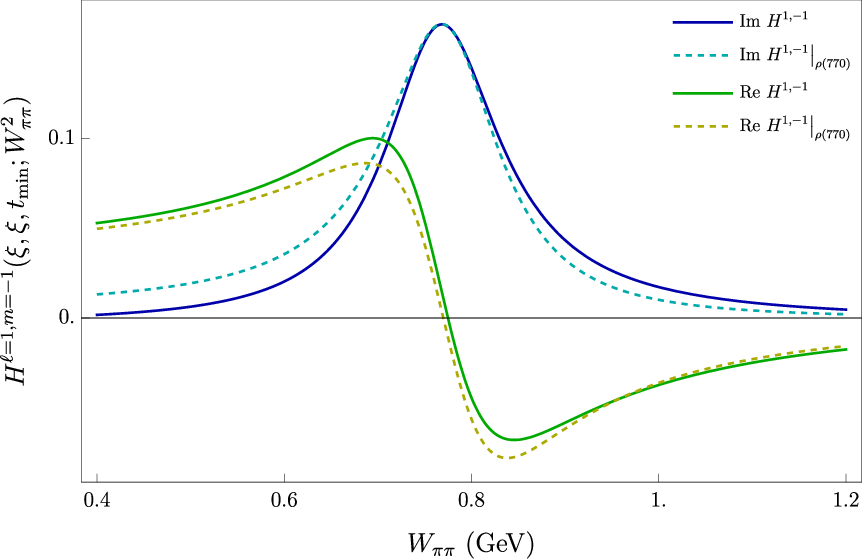}
    \caption{The imaginary and real parts of the PW $H^{\ell = 1, m = -1} (x,\xi,t,W_{\pi \pi}^2)$
    on the crossover line $x=\xi$ for $\xi$ corresponding to $x_B = 0.25$ and $t = t_{\mathrm{min}}$, within the model based on the $N=1$-subtracted Omnès 
    representation (\ref{N=1subtracted_Omnes_representation}), (\ref{Omnes_function}), compared to the contribution of the Breit-Wigner-shaped $\rho(770)$ resonance into $\pi \to \pi \pi$ transition GPD (\ref{H-rho_unpol}). 
   }
    \label{fig:PWGPD}
\end{figure}

The  near-threshold behavior of the PW 
$H^{\ell=1,m=-1}$, 
that is required as input for the representation (\ref{N=1subtracted_Omnes_representation}),
is described with the help of the Ansatz
\begin{eqnarray}
    H^{\ell=1,m=-1 }(x,\xi,t; W_{\pi\pi}^2 = 0) = N^{1, \, -1 }H_{DD}(x,\xi)\left[1-\frac{t}{A^{1, \, -1}}+\frac{t^2}{B^{1, \, -1}}\right]^{-1},
    \label{Input1m1_Omnes}
\end{eqnarray}
with $x$- and $\xi$-dependencies provided by the double distribution representation (\ref{RDDA}) with the input parton density (\ref{Parametrization_q(x)}). The form of the $t$-dependence is chosen
analogously to that of the 
$\pi\to\rho$ 
transition electromagnetic FF 
(\ref{pi-rho_FF_param}). 
The parameters 
$A^{1, \, -1}$, $B^{1, \, -1}$, 
and the overall normalization factor 
$N^{1, \, -1}$ 
are determined by matching  to the 
$\rho$-meson contribution to the unpolarized 
$\pi \to \pi \pi$ 
GPD  
(\ref{H-rho_unpol}).

In figure~\ref{fig:PWGPD}, we present the imaginary and real parts
\footnote{Note that the notion of real and imaginary are employed in the context of $W_{\pi \pi}^2$ dependence. Transition GPDs on the crossover trajectory $x=\xi$ are proportional to absorptive parts of the transition Compton FFs. They possess both real and imaginary parts as functions of $W_{\pi \pi}$. } of 
the PW expansion term $H^{\ell = 1, m = -1}$ on the crossover line $x = \xi$ 
for $\xi$ corresponding to $x_B = 0.25$ and $t = \Delta^2_\mathrm{min}$ fitted to 
the contribution of $\rho(770)$  (\ref{H-rho_unpol}) at the resonance position
$W_{\pi \pi}=m_\rho$.
This yields the following values for the relevant parameters: $N^{1,-1} \simeq 0.02$, $A^{1,-1} \simeq 0.63~\mathrm{GeV}^2$, and $B^{1,-1} \simeq 2.5~\mathrm{GeV}^4$.
Figure~\ref{fig:PWGPD} also shows the contribution of the Breit-Wigner-shaped $\rho(770)$ resonance into the 
$\pi \to \pi \pi$ 
transition GPD 
(\ref{H-rho_unpol})
with use of the Ansatz 
(\ref{H_Ansatz_unp}) for the
unpolarized $\pi \to \rho$ transition GPD. By construction, the model based on the  Omnès representation is fitted to match with the Breit-Wigner model for $W_{\pi \pi}=m_\rho$.
The good coincidence in whole range of $W_{\pi \pi}$ between the two approaches is quite expected,
as $\rho(770)$ resonance shape is known to be nearly a perfect Breit-Wigner, and
$\rho(770)$ dominates the $\ell=1$, $I=1$ $\pi \pi$-scattering PW. However, the description based on the Omnès representation can also be adapted for the PWs that can not be described by an isolated Breit-Wigner shape.

\begin{figure}[t!]
    \centering
    \subfigure{
        \includegraphics[width=0.47\linewidth]{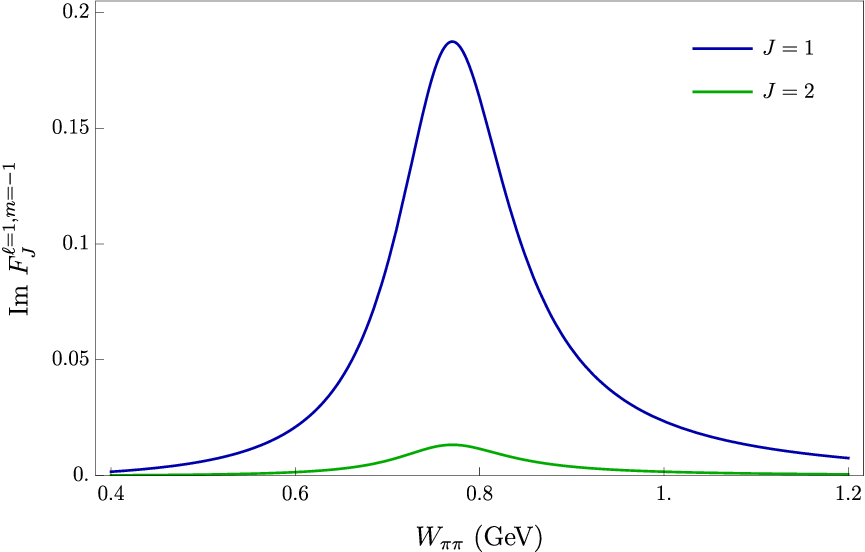}
        }
        \ \
    \subfigure{
        \includegraphics[width=0.47\linewidth]{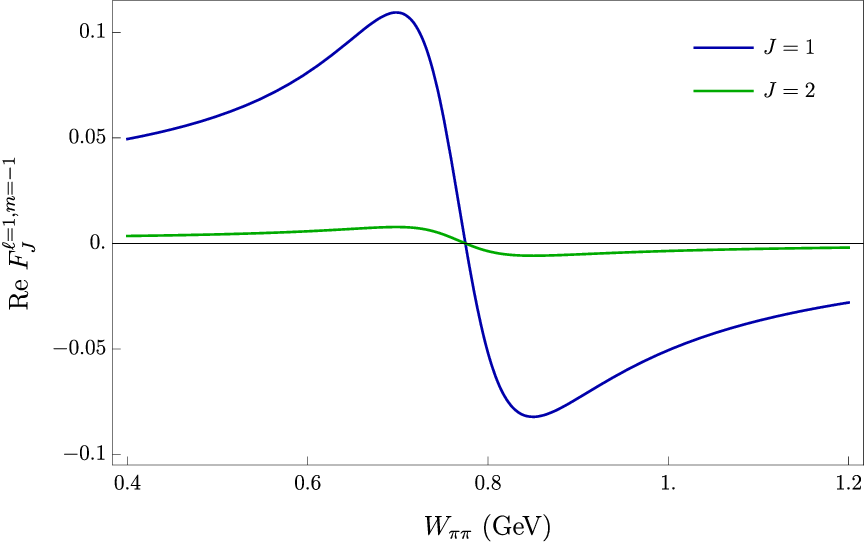}
        }
    \caption{The FG projection of the unpolarized $\pi\to\pi\pi$ Compton FF for $J = 1$ and $J = 2$ at $t = \Delta^2_\mathrm{min}$ as a function of $W_{\pi\pi}$. The left and right panels present the imaginary and real parts of $F_J^{\ell = 1, m = -1}$, respectively.}
    \label{fig:FG}
\end{figure}

The FG projections of the transition Compton FFs can be reconstructed from the known absorptive parts of the elementary Compton amplitudes. They represent observable quantities, which encode hadron's response to the excitation by means of the cross channel spin-$J$ local probes induced by the non-local quark-antiquark string-like probe created at the hard stage of the DVCS process.

We present the FG projections of the unpolarized $\pi\to\pi\pi$ Compton FF~(\ref{FG_projection_Unpolarized}) corresponding to the $J = 1$ and $J = 2$ local probes. We focus on the $\ell = 1, m = -1$ PW of the final state $2\pi$ system that is dominated by the $\rho$-meson resonance. For simplicity, 
we neglect the effect of mixing of the cross channel SO$(3)$ PWs due to final masses of the initial and final state hadrons. We use the 
model based on the $N=1$-subtracted
Omnès   representations (\ref{N=1subtracted_Omnes_representation}) and (\ref{Omnes_function}) with the near-threshold behavior described by eq.~(\ref{Input1m1_Omnes}).
Figure~\ref{fig:FG} shows the imaginary and real parts of the $J=1$ and $J=2$ 
FG projections of the unpolarized $\ell = 1, m = -1$ Compton FF as functions of $W_{\pi \pi}$. 
Quite as expected, both 
$F_{J = 1}^{1,-1}$ 
and 
$F_{J = 2}^{1,-1}$ 
inherit the 
$W_{\pi\pi}$-dependencies from the imaginary and real parts of the unpolarized $\pi\to\pi\pi$ GPD on the crossover line $x=\xi$.
The signal induced by the spin-$1$ cross channel probe  dominates over that of the spin-$2$ probe, suggesting that the graviproduction of the $\rho$-meson is suppressed compared to the production by means of the vector probe.

Here we limited our study to the channel dominated by the well isolated $\rho(770)$ resonance. Therefore, we obviously could not reveal anything beyond the
$\rho$-meson, which was included by construction. However, when applied to real data, the approach based on the FG projections can be used in hadron spectroscopy, enabling the study of resonance excitations through probes with selected quantum numbers This strategy 
can be especially helpful in the search for exotic states. Moreover, the FG projections 
of the Compton FFs can be investigated using lattice QCD methods, see e.g. \cite{Constantinou:2020pek} for a recent review.
Lattice results can be directly compared with data employing a system of sum rules similar to those worked out in ref.~\cite{Semenov-Tian-Shansky:2023ysr} for the diagonal DVCS case.

\section{Summary and outlook}
\label{sec:summary}
Transition GPDs accessed in non-diagonal hard exclusive reactions emerge as a promising tool to study the internal structure of excited hadron states and  scrutinise mechanisms of resonance formation.
In this work we addressed a toy model of a non-diagonal hard exclusive reaction involving only spinless hadrons $ \gamma^*\pi  \to  \gamma \pi \pi$ in the generalized Bjorken regime.
In order to provide a unified description of this process in the complete $2\pi$-resonance region we proposed a description in terms of the $\pi \to \pi \pi$ transition GPDs, which extend the concept of transition GPDs between a pion and isolated $2\pi$-resonances ($f_0(500)$, $\rho(770)$, $f_2(1270)$, etc.). 

We introduced the definitions of the unpolarized and polarized $\pi \to \pi \pi$  transition GPDs and specified their dependence on the kinematical variables describing the final state $2 \pi$ system. We reviewed the basic properties of the $\pi \to \pi \pi$ GPDs such as the polynomiality property of the Mellin moments, symmetry properties and worked out the chiral constraints for $\pi \to \pi \pi$ GPDs at the $2 \pi$-production threshold. We also constructed the PW expansion of $\pi \to \pi \pi$ GPDs in the real-valued spherical harmonics of the polar and azimuthal pion decay angles $\theta_\pi^*$ and  $\varphi_\pi^*$ defined in the helicity frame within $2\pi$ c.m. system. The angular distribution of the non-diagonal DVCS cross section in these angles carries valuable information on the nature of the produced intermediate $2\pi$-resonance states. This allows to make a contact with conventional tools of the PW analysis employed in resonance spectroscopy. 

Contrary to the case of usual GPDs, which are required to be real due to time reversal invariance and hermiticity, $\pi \to \pi \pi$ transition GPDs, as functions of $W_{\pi \pi}^2$, develop an imaginary part above the $2 \pi$ production threshold $W_{\pi \pi}> 2 m_\pi$. 
Relying on the methods of the dispersive analysis and the Watson-Migdal finite state interaction theorem, we constructed a representation of the
$\pi \to \pi \pi$ GPDs in terms of scattering phases of $\pi \pi$-scattering based on the Omnès solution of  the $N$-subtracted dispersion relation for the transition matrix elements defining $\pi \to \pi \pi$ GPDs. This representation connects transition GPDs, probed with hard exclusive scattering process, to the parameters of the low energy hadronic scattering  reactions characterizing the $\pi \pi$ spectrum.

Accounting for the contribution of the well isolated $\rho(770)$ fairly described by the Breit-Wigner approximation, we addressed $e^- \pi \to {e^{-}} \gamma \pi \pi$ in the generalized Bjorken kinematics in the vicinity of the $\rho(770)$ resonance. We estimated the cross section of the non-diagonal $\pi \to \pi \pi$ DVCS and the background Bethe-Heitler reaction involving $\pi \to \rho$ transition. For this purpose, we constructed a simple phenomenological model for the unpolarized $\pi  \to \rho$ transition GPD normalized to the known $\pi  \to \rho$ transition electromagnetic form factor; for the polarized $\pi \to \rho$ transition GPD $\tilde{H}_1^{\pi \to \rho}$ we adapted a simple model accounting for the pion exchange contribution in the cross channel. 

The estimated cross section rates were found to be too small to be accessed with the contemporary experiments on the deeply virtual Compton scattering off the pion target provided by the Sullivan-type reaction. 
However, the study of the angular distributions of cross sections in pion decay angles $\theta_\pi^*$ and  $\varphi_\pi^*$ allow us to illustrate several features of the formalism of transition GPDs, which may be generally applied for the physically relevant case of non-diagonal 
hard exclusive reactions involving $N \to \pi N$ transitions. The shape of the cross sections integrated over  $\varphi_\pi^*$ 
(or over $\cos \theta_\pi^*$) 
turns to be sensitive to the polarization state of the produced intermediate resonance. Moreover, the DVCS and BH cross sections exhibit clearly distinct behavior in the pion decay angles, which may aid in distinguishing between these contributions.

We also illustrated the application of the Omnès dispersive framework for the $\pi \to \pi \pi$ 
transition GPDs.
For the 
$I=1$, $\ell=1$, $m=-1$ 
partial wave of the unpolarized 
$\pi \to \pi \pi$ 
GPD, that is dominated by the contribution of the isolated 
$\rho(770)$ 
resonance, we constructed the Ansatz in terms of the 
$N=1$ 
subtracted Omnès representation employing the known $P$-wave 
$\pi \pi$ scattering phase. The model was fitted to match with the contribution of the Breit-Wigner shaped 
$\rho(770)$ 
resonance at 
$W_{\pi \pi}= m_\rho$. 
It showed a good coincidence in a broad range in $W_{\pi \pi}$. Although being just a toy exercise, it demonstrates the potential of the dispersive approach to constrain $\pi \to \pi \pi$ transition GPDs employing the low energy 
$\pi \pi$-scattering data and admits a straightforward generalization for the $\pi N$ case. 

Finally, we addressed the use of the Froissart-Gribov projection of the 
$\pi \to \pi \pi$ 
transition Compton form factors. The FG projections allow to single out the effect of the cross channel excitation of a selected spin-$J$ induced by the non-local QCD string-like light-cone quark-antiquark probe created at the hard stage of the DVCS process. Exciting hadronic resonances with probes of arbitrary $J$ may bring new tools for the resonance spectroscopy currently limited with resonance production in hadronic reactions and by elementary $J=1$ electroweak probes 
($\gamma$, $W^\pm$, $Z^0$). 
This framework can be particularly helpful in search of exotic resonances, especially those, which can not be excited from the ground state with help of probes with quantum numbers of electroweak bosons. We construct the FG projections for the unpolarized and polarized Compton transition FFs 
expanded in PWs of the produced $\pi \pi$-system.
We acknowledge the issue with mixing of the cross channel 
SO$(3)$ 
partial waves, which still needs to be resolved. We illustrate the application of the FG projection framework for the 
$\ell=1$, $m=1$ 
PW of unpolarized 
$\pi \to \pi \pi$ 
transition Compton FF.

A natural extension of the present study will be a generalization for the case of hard exclusive non-diagonal DVCS (and DVMP) reaction with the $N \to \pi N$ transition in the complete $\pi N$ resonance region. A unified description of these process in terms of $N \to \pi N$ transition GPDs
briefly introduced in refs.~\cite{Polyakov:2006dd} can be helpful in view of upcoming experimental data from the JLab@12 GeV \cite{CLAS:2023akb} and future planned experiments with the CEBAF accelerator at JLab \cite{Arrington:2021alx,Accardi:2023chb}, 
and the EIC 
\cite{Burkert:2022hjz}, 
and the EicC 
\cite{Anderle:2021wcy}.
Our results can also be useful in the context of the broad research program on studies of baryon resonances with non-diagonal hard exclusive reactions summarized in ref.~\cite{Diehl:2024bmd}.

\acknowledgments

This work was supported by Basic Science Research Program through the National Research Foundation of Korea (NRF) funded by the Ministry of Education  RS-2023-00238703; and under Grant
No. NRF-2018R1A6A1A06024970 (Basic Science Research Program).
We sincerely thank Chueng-Ryong Ji and Ho-Meoyng Choi for their insightful discussions on various aspects of this study during seminars, as well as their valuable comments on the manuscript. Additionally, we would like to thank Stefan Diehl, June-Young Kim, Bernard Pire, Hyeon-Dong Son, Lech Szymanowski, Marc Vanderhaeghen, Christian Weiss, and Ho-Yeon Won for enlightening discussions and correspondence.

\newblock

\appendix
\section{Supplementary to kinematics}

\subsection{Gram determinants}
\label{App_A}
Our notations for the Gram determinants follow appendix A of ref.~\cite{Byckling1973}.
\begin{itemize}
    \item $G_n$ denotes a generic $n \times n$ Gram determinant of 
    vectors $p_1, \cdots, p_n$; $q_1, \cdots, q_n$:
    \begin{eqnarray}
        G_n \begin{pmatrix}
            p_1, \cdots, p_n \\ q_1, \cdots, q_n
        \end{pmatrix}
        = \begin{vmatrix}
            p_1 \cdot q_1 & p_1\cdot q_2 & \cdots & p_1 \cdot q_n \\
            \vdots & \vdots & & \vdots \\
            p_n \cdot q_1 & p_n \cdot q_2 & \cdots & p_n\cdot q_n
        \end{vmatrix}. \label{GramDetG}
    \end{eqnarray}
    \item $\Delta_n$ denotes a symmetric $n \times n$ Gram determinant of 
    vectors $p_1, \cdots, p_n$:
    \begin{eqnarray}
        \Delta_n(p_1, \cdots, p_n) =
        \begin{vmatrix}
            p_1^2 & p_1\cdot p_2 & \cdots & p_1 \cdot p_n \\
            \vdots & \vdots & & \vdots \\
            p_n \cdot p_1 & p_n \cdot p_2 & \cdots & p_n^2
        \end{vmatrix}.        \label{GramDetDelta}
    \end{eqnarray}
  
\end{itemize}

We also specify the useful kinematic functions occurring in the description of the $2 \to 2$ and $2 \to 3$ scattering kinematics. 
\bi
\item The Källén  triangular function $\lambda(x,y,z)$ is defined as 
\be
\lambda(x,y,z) \equiv x^2+y^2+z^2 -2xy - 2xz - 2yz.
\label{TriangLambda}
\ee

  \item The tetrahedral function $G(x,y,z,u,v,w)$ is expressed through the Cayley determinant:
    \begin{eqnarray}
        G(x,y,z,u,v,w) = -\frac{1}{2}
        \begin{vmatrix}
            0 & 1 & 1 & 1 & 1 \\
            1 & 0 & v & x & z \\
            1 & v & 0 & u & y \\
            1 & x & u & 0 & w \\
            1 & z & y & w & 0
        \end{vmatrix}. \label{CayleyDet}
    \end{eqnarray}

\ei

Using eqs.~(\ref{cos_phi_pi}) and~(\ref{s_1}), we express the invariant $t'$ in terms of the helicity frame decay angles: 
\begin{eqnarray}
    t' &=& -\frac{W_{\pi\pi}^2-\Delta^2-3m_\pi^2}{2} \nonumber \\
    && \mbox{} -\cos\theta_\pi^* \frac{\sqrt{W_{\pi\pi}^2-4m_\pi^2} \big[m_\pi^2 (W_{\pi\pi}^2 - s) + 2Q^2 W_{\pi\pi}^2 + \Delta^2 s + s W_{\pi\pi}^2 - W_{\pi\pi}^4 +\Delta^2 W_{\pi\pi}^2\big] }{2W_{\pi\pi}(s-W_{\pi\pi}^2)} \nonumber \\
    && \mbox{} + \cos\varphi_\pi^* \sqrt{1-\cos^2\theta_\pi^*}\frac{\sqrt{W_{\pi\pi}^2-4m_\pi^2}}{s-W_{\pi\pi}^2}\sqrt{-G\big(2s-W_{\pi\pi}^2, -Q^2, -\Delta^2-2Q^2, s, 0, m_\pi^2\big)}.
    \label{tprime_through_angles}
\end{eqnarray}
We note that the angular dependencies of each term occurring in the r.h.s. of (\ref{tprime_through_angles})
correspond to the real-valued spherical harmonics (\ref{Def_Real_Sph_Harmonics}), specifically 
$\operatorname{Y}_{\ell = 0, m = 0}(\theta_\pi^*, \varphi_\pi^*), \operatorname{Y}_{\ell = 1, m = 0}(\theta_\pi^*, \varphi_\pi^*)$, 
and $\operatorname{Y}_{\ell = 1, m = 1}(\theta_\pi^*, \varphi_\pi^*)$.
In the chiral limit ($m_\pi = 0$), to the LO in $1/Q$ we get:
\begin{eqnarray}
    t' &=& -\frac{1}{2(1-x_B)}\biggl\{ (1-x_B)(W_{\pi\pi}^2-\Delta^2)+\cos\theta_\pi^*\big[W_{\pi\pi}^2(1+x_B) + \Delta^2(1-x_B) \big] \nonumber \\
    && \mbox{} -2\cos\varphi_\pi^*\sqrt{1-\cos^2\theta_\pi^*} W_{\pi\pi}\sqrt{-\Delta^2(1-x_B) - x_B W_{\pi\pi}^2}\biggr\} + \mathcal{O}(1/Q^2).
\end{eqnarray}

\subsection{Sudakov decomposition of momenta}
\label{App_Sudakov}

We construct the Sudakov expansions of momenta of the non-diagonal 
$\pi \to \pi \pi$ 
DVCS 
(\ref{DVCS_pi_to_2pi})
analogously to the case of 
$N\to\Delta$ 
DVCS described in section~4.4 of 
ref.~\cite{Goeke:2001tz}. 
We introduce two lightlike vectors, 
$\Tilde{p}^\mu = \Lambda (1, 0, 0, 1)$ 
and 
$n^\mu = \frac{1}{2\Lambda}(1, 0, 0, -1)$, 
satisfying 
$\Tilde{p}\cdot n = 1$ 
for arbitrary 
$\Lambda$. 
The momenta of the reaction 
(\ref{DVCS_pi_to_2pi}) 
admit the following expansion:
\begin{eqnarray}
    \bar{P} &=& \Tilde{p} + \frac{\bar{M}^2}{2}n, \nonumber \\
    q &=& -2\xi'\Tilde{p} + \frac{Q^2}{4\xi'}n, \nonumber \\
    \Delta &=& -2\xi \Tilde{p} + \left(\bar{M}^2 \xi + \frac{W_{\pi\pi}^2-m_\pi^2}{2}\right)n + \Delta_\perp, \nonumber \\
    p_{\pi\pi} &=& (1-\xi)\Tilde{p} + \left[ \frac{\Bar{M}^2}{2}(1+\xi) + \frac{W_{\pi\pi}^2-m_\pi^2}{4} \right]n + \frac{\Delta_\perp}{2}, \label{Sudakov_expansion}
\end{eqnarray}
with
\begin{eqnarray}
    \xi' &=& -\frac{\Bar{P}\cdot q}{2\Bar{M}^2}\left[ 1-\sqrt{1+\frac{Q^2 \Bar{M}^2}{(\Bar{P}\cdot q)^2}} \right], \nonumber \\
    \xi &=& \xi'\left[ \frac{Q^2-\Delta^2-2\xi'(W_{\pi\pi}^2-m_\pi^2)}{Q^2 + 4\xi'^2 \Bar{M}^2} \right].
\end{eqnarray}
Here 
$\Bar{P} \equiv \frac{k + p_{\pi\pi}}{2}$ 
is the average hadron momentum with 
$\Bar{P}^2 \equiv \Bar{M}^2 = \frac{m_\pi^2 + W_{\pi\pi}^2}{2} - \frac{\Delta^2}{4}$, 
$\Delta_\perp$ 
is the transverse component of the momentum transfer, $\tilde{p}\cdot\Delta_\perp = n\cdot\Delta_\perp = 0$, 
and 
$\xi$ 
is the skewness variable. The longitudinal fractions of the virtual photon momentum and the momentum transfer, 
$\xi'$ 
and 
$\xi$, 
are related in the Bjorken kinematic limit by
\begin{eqnarray}
    \xi \simeq \xi' \to \frac{x_B}{2-x_B}. 
    \label{skewedness}
\end{eqnarray}

It is helpful to rewrite the light-cone vector $n$ through the momentum vectors, to the leading accuracy in $1/Q$-expansion,
\begin{eqnarray}
    n^\kappa = \frac{q^\kappa + q'^\kappa}{\Bar{P}\cdot(q+q')} + \frac{4\xi^2}{Q^2}\Bar{P}^\kappa + \mathcal{O}(1/Q^2) \equiv \frac{\bar{\Sigma}^\kappa}{\Bar{P}\cdot \bar{\Sigma}} + \frac{4\xi^2}{Q^2}\Bar{P}^\kappa + \mathcal{O}(1/Q^2),
    \label{Lc_vector_covariant}
\end{eqnarray}
where 
$\bar{\Sigma}= \frac{1}{2}(q+q')$ 
is the average photon momentum.

\subsection{Isospin projection operators}
\label{App_projectors}

The isospin projection operators for the $I = 0, 1, 2$ states of the produced pions, which are used to define the isovector $\pi\to\pi\pi$ GPDs in eqs.~(\ref{Def_Hunpolarized_pi_to2pi}) and~(\ref{Def_Hpolarized_pi_to2pi}), are given by:
\begin{eqnarray}
    P^{I = 0, bc}_{da} &=& \frac{1}{3}\delta_{ad}\delta^{bc}, \nonumber \\
    P^{I = 1, bc}_{da} &=& \frac{1}{2}\left(\delta^c_a\delta^b_d  - \delta_a^b\delta^c_d \right), \nonumber \\
    P^{I = 2, bc}_{da} &=& \frac{1}{2}\left( \delta^b_a\delta^c_d + \delta^c_a\delta^b_d - \frac{2}{3}\delta_{ad}\delta^{bc} \right),
\end{eqnarray}
where their sum yields the identity operator $\delta_a^c \delta^b_d$, ensuring the completeness property of these projectors. 

\section{Useful properties of spherical harmonics}
\label{App_Sperical_stuff}

The Condon–Shortley phase convention  spherical harmonics
$Y_{\ell}^m$
are defined as
\be
Y_{\ell}^m(\theta, \varphi)=(-1)^m \sqrt{\frac{(2 \ell+1)}{4 \pi} \frac{(\ell-m) !}{(\ell+m) !}} P_{\ell}^m(\cos \theta) e^{i m \varphi},
\label{Ylm_CondSh}
\ee
where
$P_{\ell}^m(z)$
stand for the associated Legendre polynomials.

The relation  between the usual   spherical harmonics
$Y_{\ell}^m$
(\ref{Ylm_CondSh})
and the real-valued spherical harmonics $\operatorname{Y}_{\ell}^m$ defined in (\ref{Def_Real_Sph_Harmonics}) reads:
\be
{Y}_{\ell}^m(\theta, \varphi)= \begin{cases}\frac{1}{\sqrt{2}}\left(\operatorname{Y}_{\ell,|m|}-i \operatorname{Y}_{\ell,-|m|}\right)(\theta, \varphi), & \text { if } m<0; \\ \operatorname{Y}_{\ell,\,  0}(\theta, \varphi), & \text { if } m=0; \\ \frac{(-1)^m}{\sqrt{2}}\left(\operatorname{Y}_{\ell, \, |m|}+i \operatorname{Y}_{\ell, \, -|m|}\right)(\theta, \varphi), & \text { if } m>0.\end{cases}
\label{Rel_Spher_Harm}
\ee

The following multiplication formulas  are valid for the Legendre polynomials, see Chapter.~3, $\S$~4 of ref.~\cite{Vilenkin_book}.
\bi
\item For the cosine harmonics with $m'\ge 0$;  $m' \le \ell$:
\be
&&
\int_0^{2 \pi}  d \tilde{\varphi}  \cos(m' \tilde{\varphi}  +m' \varphi ) P_\ell \left( \cos \theta   \cos \theta'  +\sin \theta   \sin \theta'  \cos \tilde{\varphi}  \right) \nn \\ &&= P_\ell^{m'}(\cos \theta )   P_\ell^{m'}(\cos \theta'  ) \cos(m' \varphi  ) 2 \pi \frac{(\ell-m')!}{(\ell+m')!}.
\label{Mult_flormula_cos}
\ee
\item For the sine harmonics with $m'\ge 0$;  $m' \le \ell$:
\be
&&
\int_0^{2 \pi} d \tilde{\varphi}  \sin \left(m^{\prime} \tilde{\varphi} +m^{\prime} \varphi \right) P_l\left(\cos \theta  \cos \theta^{\prime}+\sin \theta  \sin \theta^{\prime} \cos \tilde{\varphi} \right) \nn \\ &&
  =P_l^{m^{\prime}}\left(\cos \theta \right) P_l^{m^{\prime}}\left(\cos \theta^{\prime}\right) \sin \left(m^{\prime} \varphi \right) 2 \pi \frac{\left(l-m^{\prime}\right)!}{\left(l+m^{\prime}\right)!}.
\label{Mult_flormula_sin}
\ee
\ei

\section{Method for determining the invariant amplitudes of a binary scattering
process}
\label{App_Inv_Ampl_PW}

In this appendix we give a brief summary of the general method for determining
the invariant amplitudes for the binary scattering problem suggested in
\cite{Munczek1963, 
Alfaro_red_book}%
\footnote{For our purpose it is sufficient to consider the case of integer spin of
intermediate states. The half-integer spin case is discussed in
\cite{Rebbi:68}.}.

\begin{figure}[h]
    \centering
    \includegraphics[height = 3.2cm]{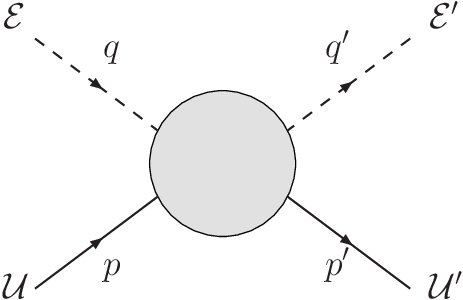}
    \caption{Binary scattering kinematics.}
    \label{Fig_binary_kinematics}
\end{figure}

Let us consider a binary scattering process with the usual kinematics
presented in
figure~\ref{Fig_binary_kinematics}.
$\mathcal{U}$, $\mathcal{E}$
($\mathcal{U}'$, $\mathcal{E}'$)
stand for polarization tensors of incoming (outgoing) states.
We introduce the following conventional kinematical quantities:
\be
\bar{P}_\mu &=& \frac{1}{2}(p'+p)_\mu, \qquad
\bar{\Sigma}_\mu=\frac{1}{2}(q'+q)_\mu, \qquad  \nu_t= \frac{1}{4} (s-u)= (\bar{P} \cdot \bar{\Sigma}), \nonumber \\
\Delta_\mu &=& (p'-p)_\mu=(q-q')_\mu, \nonumber \\
s &=& (p+q)^2, \qquad \qquad t=\Delta^2, \qquad \qquad \qquad u=(p-q')^2, \nonumber \\
\hat{\bar{P}}_\mu &=& \bar{P}_\mu- \frac{(\bar{P} \cdot \Delta)}{t} \Delta_\mu, 
\qquad
\hat{\bar{\Sigma}}_\mu=\bar{\Sigma}_\mu- \frac{(\bar{\Sigma} \cdot \Delta)}{t} \Delta_\mu.
\label{Kin_Var}
\ee
The following expression can be derived for the
$t$-channel center-of-mass scattering angle
$\cos{ 
\theta_t
}$ defined as the angle between $\vec{q} $ and $\vec{\bar{p}}' \equiv-\vec{p}'$ :
\be
\cos{ 
\theta_t
}= \frac{(\hat{\bar{P}} \cdot \hat{\bar{\Sigma}})}{ (\hat{\bar{P}}^2)^{1/2}
(\hat{\bar{\Sigma}}^2)^{1/2}}=
\frac{\nu_t+ \frac{1}{4t}(p^2-p'^2)(q^2-q'^2)}{(\hat{\bar{P}}^2)^{1/2}
(\hat{\bar{\Sigma}}^2)^{1/2}},
\label{costhetat}
\ee
where
\be
&& \hat{\bar{P}}^2=\frac{1}{2}(p^2+p'^2)- \frac{1}{4}\, t -
\frac{1}{4t}(p^2-p'^2)^2 , \nonumber \\ &&
\hat{\bar{\Sigma}}^2=\frac{1}{2}(q^2+q'^2)- \frac{1}{4}\, t -
\frac{1}{4t}(q^2-q'^2)^2. 
\ee
We label by `u' the set of all variables referring to the `upper'
particles with momenta
$q$ and $q'$
and by `l' the set of those referring the `lower'
particles with momenta
$p$ and $p'$.

In order to single out the contribution of a state of the angular momentum
$J$
in the $t$-channel to the various amplitudes, one has to
consider their behavior under rotations in the 
$t$-channel
c.m. system.
We denote by
$R_u(\phi_u)$
the rotations (characterized by the Euler angles
$\phi_u$)
of the kinematical variables of the `u' set and
by
$R_l(\phi_l)$
(characterized by the Euler angles
$\phi_l$)
those of the `l' set. An amplitude corresponding to a given
$J$
in the $t$-channel transforms as a
$(2J+1)$-dimensional
representation under the operation
$R_u$
or
$R_l$.

Let us assume that the variables of the `u' and `l' sets are
combined into factors including the complete dependence on polarization,
and these factors are separately invariant under 
$R_u( \phi_u)$
and
$R_l( \phi_l)$
operations.
In this case, the coefficients at these factors in
the expansion of the scattering amplitude
have a remarkable property that a spin-$J$ contribution
transforms like a
$J$
representation of rotation group under
$R_u$, $R_l$
operations.
Such factorization is achieved by a separate combination
of quantities that belong to the `u' and `l' sets.
In order to obtain the correct number of invariants one
has to use, in addition to polarization functions and momenta,
the gradients
$\partial/ \partial \bar{P}_\mu$ ($\partial/ \partial \bar{\Sigma}_\mu$ )
for the `l' (`u) sets. Note that these gradients are to be taken at constant
$\Delta$
to ensure momentum conservation.
Thus, one can decompose the amplitude as
\be
M= \sum_{m, \,n}
\alpha_m^{(u)} %
\alpha_n^{(l)} \varphi_{mn}(t, \cos{ \theta_t}),
\ee
where
$\alpha_m^{(l)}$
are scalar combinations of the polarization tensors (and $\gamma$-matrices, if necessary)
with momenta and gradients of the side `l', and correspondingly
$\alpha_m^{(u)}$
are scalars build from the `up' side quantities.
$\varphi_{mn}$
are the
Lorentz invariant functions of the momenta.
A spin-$J$ contribution into
$\varphi_{mn}$
has the form
\be
\varphi_{mn}^{(J)}(t, \cos{ \theta_t})= P_J(\cos{ \theta_t}) \,  \varphi_{mn}^{(J)}(t).
\ee
Hence, the following partial wave expansion, analogous to that of the spinless case, can be written for
$\varphi_{mn}(t, \cos{\theta_t})$:
\be
\varphi_{mn}(t, \cos{\theta_t})=
\sum_J P_J(\cos{ \theta_t}) \,\varphi_{mn}^{(J)}(t).
\label{scalarlike_ampl}
\ee
The amplitudes
$ \varphi_{mn}^{(J)}(t)$
are called ``scalarlike'', since for these amplitudes
the connection between the total angular momentum and the Legendre polynomials is
completely the same as in the case of scalar particles scattering.

As an example of applying this formalism, let us consider the
binary pion-nucleon scattering.
The only invariant that appears at the `u' (pion) side is just $1$.
The invariants of the `l' (nucleon) side are to be constructed
out of the Dirac spinors
$\bar{U}(p')$, $U(p)$,
momentum vectors
$p'$, $p$,
$\gamma$-matrices and the gradient operator
$\frac{\partial}{\partial {\bar{P}}_\mu}$
at constant
$\Delta$.
Note, that in order to ensure that the nucleons are on-shell,
the gradient operator must leave invariant
${\bar{P}}^2,\, \Delta^2$
and
$(\bar{P}\cdot \Delta)$.
Such gradient is denoted by
$\partial/ \partial \bar{P}^T_\mu$.
It can be constructed as follows:
\be
\frac{\partial}{\partial \bar{P}_\mu^T}=I_{\mu \nu} \frac{\partial}{\partial
\bar{P}_\nu},
\label{GradientP}
\ee
where
\be
I_{\mu \nu}= g_{\mu \nu}+ \frac{1}{(\bar{P}\cdot \Delta)^2-\bar{P}^2 \Delta^2}
\left[
\bar{P}^2 \Delta_\mu \Delta_\nu+
\Delta^2 \bar{P}_\mu \bar{P}_\nu
-(\bar{P}\cdot \Delta) (\bar{P}_\mu \Delta_\nu+ \Delta_\mu \bar{P}_\nu)
\right].
\label{PTder}
\ee
Thus, the invariants related to nucleons can be chosen as
\be
\alpha_1=\bar{U}(p') U(p), \ \ \ \alpha_2=\bar{U}(p') \gamma^\mu I_{\mu
\nu} \frac{\partial}{\partial \bar{P}_\nu}...U(p) \,.
\label{Nucleon_low_invariants}
\ee
We introduce two corresponding ``scalarlike amplitudes''
(\ref{scalarlike_ampl}):
$\varphi_1(t, \cos{\theta_t})$, $\varphi_2(t, \cos{\theta_t})$.
The relevant matrix element can then be written as
\be
M=
\bar{U}(p')
\left(
\varphi_1+
\gamma^\mu I_{\mu
\nu} \frac{\partial}{\partial \bar{P}_\nu}
\varphi_2
\right)
U(p)=
\bar{U}(p')
\left[A(\nu_t,t)+ \bar{\Sigma}_\mu \gamma^\mu B(\nu_t, t)
\right]
U(p),
\label{PiN_amplitude}
\ee
where
\be
&& A= \varphi_1+ \frac{m_N \nu_t}{\frac{1}{4}t-m_N^2}
\frac{\partial}{\partial \nu_t} \varphi_2 \, ,
\nonumber\\ &&
B= \frac{\partial}{\partial \nu_t} \varphi_2 \,.
\label{Invariant_FF_PiN}
\ee
We can now identify the combinations of the commonly used invariant amplitudes of the pion-nucleon scattering, 
$A$ and 
$B$, that are suitable for the partial wave expansion:
\be
&& A- \frac{m_N \nu_t}{\frac{1}{4}t-m_N^2}B= \sum_J \varphi_1^{(J)}(t)
P_J( \cos{\theta_t}),
\nonumber\\ &&
B= \sum_J \varphi_2^{(J)}(t) \frac{\partial}{\partial \nu_t}
P_J(\cos{\theta_t})=\sum_J \varphi_2^{(J)}(t) \frac{1}{(\hat{P}^2)^{1/2} (\hat{\Sigma}^2)^{1/2} }
P_J'(\cos{\theta_t}).
\label{Comb_Nucleon}
\ee

\bibliography{main}

\end{document}